\documentclass{article}

\usepackage{references}
\usepackage{hyperref} 
\usepackage[numbers]{natbib}
\usepackage{graphicx}
\usepackage{bm}
\usepackage{equationarray}
\usepackage{empheq}
\usepackage[normalem]{ulem}
\usepackage[usenames]{xcolor}

\newcommand{\be}{\begin{equation}}
\newcommand{\ee}{\end{equation}}
\newcommand{\ba}{\begin{eqnarray}}
\newcommand{\ea}{\end{eqnarray}}

\def\simge{\mathrel{\rlap{\raise 0.511ex
     \hbox{$>$}}{\lower 0.511ex \hbox{$\sim$}}}}
\def\simle{\mathrel{\rlap{\raise 0.511ex
      \hbox{$<$}}{\lower 0.511ex \hbox{$\sim$}}}}
      
\newcommand{\Msol}{\mbox{$\mathrm{M}_{\odot}$}}
\newcommand{\rhonuc}{\mbox{$\rho_\mathrm{nuc}$}}
\newcommand{\eq}[1]{Eq.~(\ref{#1})}
\newcommand{\eqs}[1]{Eqs.~(\ref{#1})}
\newcommand{\fig}[1]{Fig.~\ref{#1}}
\newcommand{\Singlet}{\mbox{$^1$S$_0$}}
\newcommand{\triplet}{\mbox{$^3$P$_2$}}
\newcommand{\Triplet}{\mbox{$^3$P-F$_2$}}

\begin{document}



\centerline{\Huge \bf Stellar Superfluids}
\bigskip
\bigskip


\centerline{Dany Page$^1$, James M. Lattimer$^2$, Madappa Prakash$^3$, and Andrew W. Steiner$^4$}

\bigskip

\centerline{\em $^1$ Instituto de Astronom\'ia, Universidad Nacional Aut\'onoma de M\'exico,}
\centerline{\em Mexico, DF 04510, Mexico}
\centerline{\em $^2$ Department of Physics and Astronomy, State University of New York at Stony Brook,}
\centerline{\em Stony Brook, NY 11794-3800, USA}
\centerline{\em $^3$ Department of Physics and Astronomy, Ohio University,}
\centerline{\em Athens, OH 45701-2979, USA}
\centerline{\em $^4$ Institute for Nuclear Theory, University of Washington,}
\centerline{\em Seattle, WA 98195, USA}


\bigskip \bigskip
\centerline{\bf Abstract} 
\smallskip
Neutron stars provide a fertile environment for exploring
superfluidity under extreme conditions.  It is not surprising that
Cooper pairing occurs in dense matter since nucleon pairing is
observed in nuclei as energy differences between even-even and
odd-even nuclei.  Since superfluids and superconductors in neutron
stars profoundly affect neutrino emissivities and specific heats,
their presence can be observed in the thermal evolution of neutron
stars.  An ever-growing number of cooling neutron stars, now amounting
to 13 thermal sources, and several additional objects from which upper
limits to temperatures can be ascertained, can now be used to
discriminate among theoretical scenarios and even to dramatically
restrict properties of nucleon pairing at high densities.  In
addition, observations of pulsars, including their spin-downs and
glitch histories, 
additionally support the conjecture that superfluidity and superconductivity 
are ubiquitous within, and important to our understanding of, neutron stars.

\bigskip
\section{Introduction}

In this contribution, we describe the roles of neutron superfluidity
and proton superconductivity in the astrophysical setting of neutron
stars, drawing upon lessons learned from similar phenomena occuring in
laboratory nuclei.  We will focus on both the thermal evolution (i.e.,
cooling) as well as the dynamical evolution (i.e., spin-down) of
neutron stars.  In the former, pairing dramatically affects neutrino
emission processes and the specific heat of dense matter.  In the
latter, pairing may be responsible for the observed anomalous values
of the braking index and the glitch phenomenon.  We will also briefly
describe the possibility of pairing in the presence of hyperons and
deconfined quarks.

In fermionic systems, superfluidity and superconductivity occur due to
the pairing of neutral and charged fermions, respectively.  The {\em
  Cooper Theorem} \cite{Cooper:1956qf} states that, in a system of
degenerate fermions the Fermi surface is unstable due to the formation
of ``pairs'' if there is an attractive interaction in some
spin-angular momentum channel between the two interacting particles.
The essence of the BCS theory \cite{Bardeen:1957dq} is that as a
result of this instability there is a collective reorganization of
particles at energies around the Fermi energy and the appearance of a
gap in the quasi-particle spectrum.  This reorganization manifests
itself in the formation of ``Cooper pairs''.  At high-enough
temperature, the energy gap disappears and the system reverts to its
normal state.

To begin, in Sec.~\ref{sec:nuclei}, we motivate the existence of
pairing in neutron stars by examining the pairing phenomenon in
laboratory nuclei.  We then summarize the relevant properties of
neutron stars, including their interior compositions and properties of
their crusts in Sec.~\ref{sec:ns}.  We describe in
Sec.~\ref{Sec:Pairing} how pairing in dense matter is achieved and,
in Sec.~\ref{Sec:Quarks},
we present a brief description of expectations of pairing in
deconfined quark matter that may be present in the inner core of the
most massive neutron stars.
Initially, neutron stars cool primarily due to neutrino emission in
their interiors before surface photon cooling takes over later in
their lives.  
In Sec.~\ref{Sec:Neutrinos}, we summarize the various
neutrino processes that can occur in dense matter.  Theoretically,
these processes can proceed either very rapidly (enhanced neutrino
emission) or relatively slowly.  We also describe an important
secondary process that greatly influences the interpretation of
observations -- neutrino pair emission from the pair breaking and
formation (PBF) of Cooper pairs.  This process is triggered as the
ambient temperature, decreasing because of cooling, approaches the
critical temperatures for superfluidity and superconductivity.  The
discussion in this section shows how the existence of superfluidity or
superconductivity dramatically influences neutrino emissions, leading
to both the quenching of enhanced neutrino emission and bursts of
neutrino emission due to the PBF processes.  

We present in Sec.~\ref{Sec:Cooling} simplified analytical models of 
neutron star cooling in order to gain physical insight. 
These analytical models are complemented by detailed numerical
simulations which include general relativity and state-of-the-art
microphysics, such as the dense matter equation of state, thermal
conductivities, neutrino emissivities, and specific heats.  We also
summarize in this section the abundant observational data, consisting
of estimates of surface temperatures and ages, which collectively
describe the thermal evolution of neutron stars.  We will show in
Sec.~\ref{Sec:Minimal} that the bulk of this data supports the
so-called ``Minimal Cooling Paradigm'', which supposes that no
drastically enhanced neutrino emission processes occur, or if they do,
they are quickly quenched by superfluidity or superconductivity.
Nevertheless, a few sources suggesting enhanced cooling are observed,
and we discuss the implications.

In Sec.~\ref{Sec:CasA}, we describe an outstanding recent development
in which the first real-time cooling of any isolated neutron star --
the young neutron star in the supernova remnant of Cassiopeia A -- is
observed.  The observed rate of cooling is more than 10 times faster
than expected, unless both neutron superfluidity and proton
superconductivity are present in the star's core.  These observations
provide the first direct evidence for superfluidity or superconductivity in
the interior of a neutron star, and can be verified by
continued observations of the neutron star in Cassiopeia A.

Section \ref{Sec:glitch} is devoted to other observations of neutron
stars and their dynamical evolution that may also indicate the
presence of superfluidity.  One concerns the deceleration observed in
the spin-down of pulsars, which could be due to superfluidity in their
cores, while the other is related to sporadic spin jumps, commonly
known as glitches, thought to stem from superfluidity in neutron star
crusts.

Summarizing discussions and conclusions are contained in Sec.~\ref{Sec:Conclusions}.

\section{Pairing in Nuclei\label{sec:nuclei}}

\begin{figure}[hbt]
\begin{center}
\includegraphics[width=.48\textwidth]{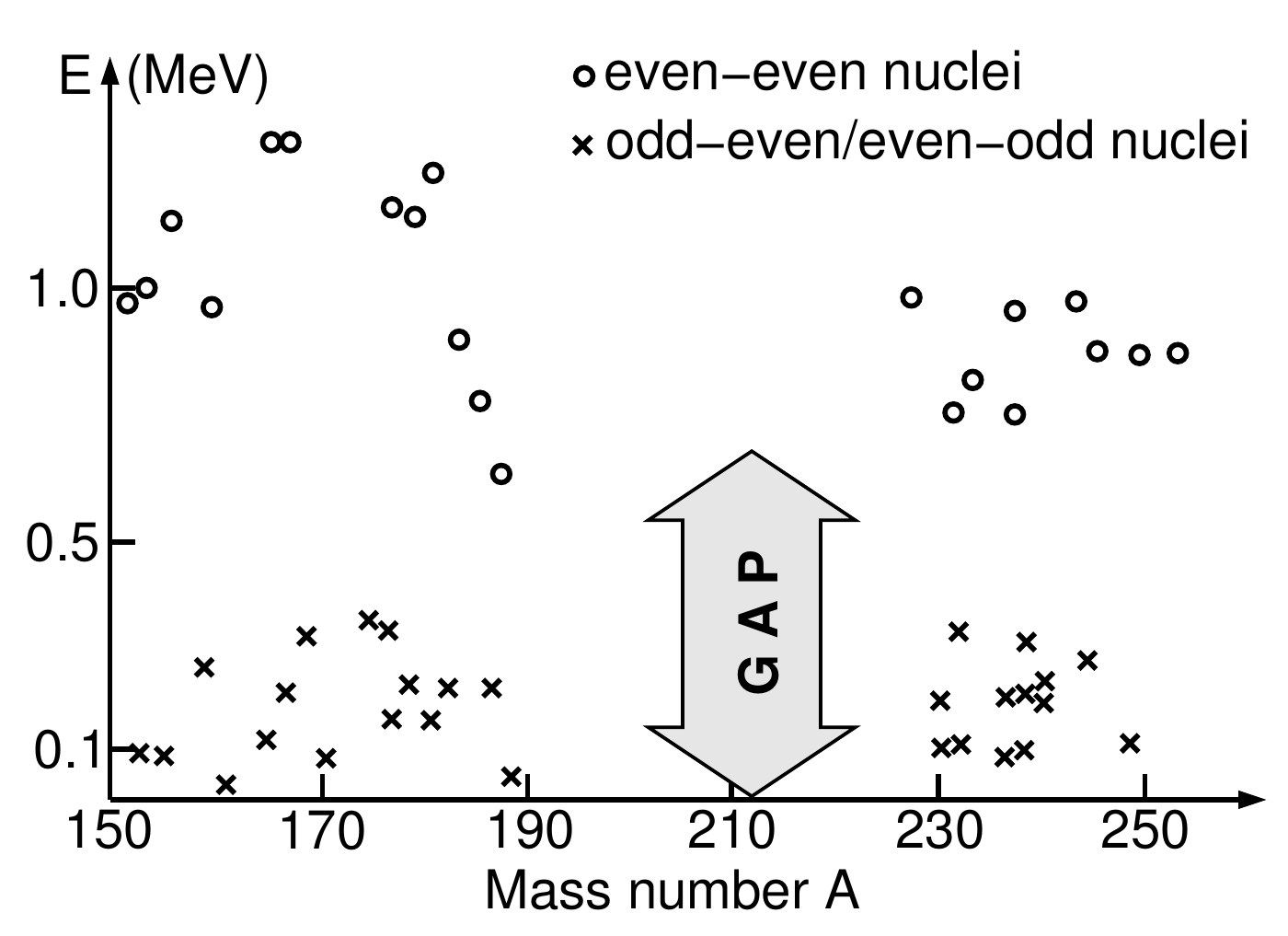}
\includegraphics[width=.48\textwidth, trim=1.2cm 1cm 1cm 1.5cm,clip]{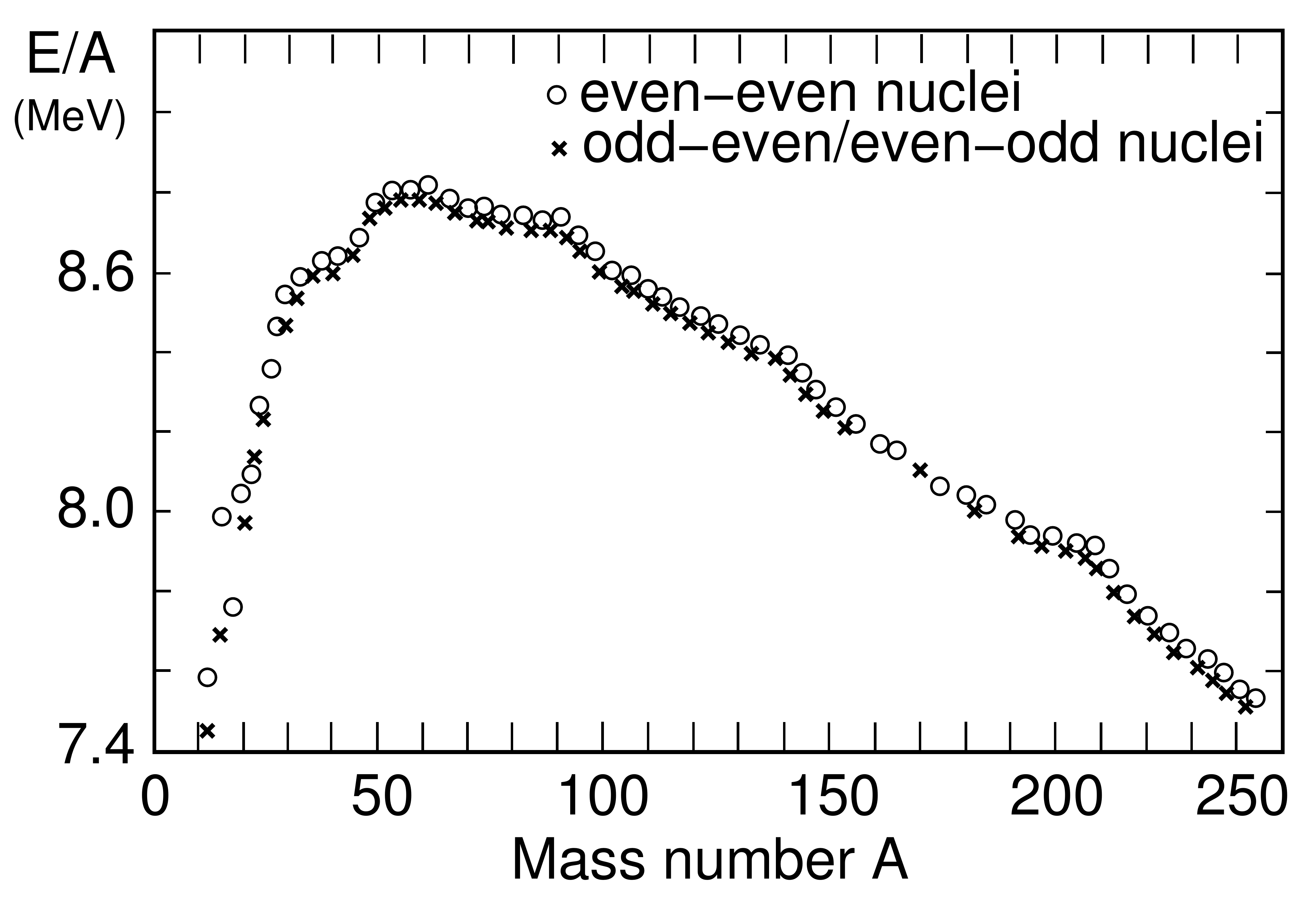}
\end{center}
\caption{Left panel: Lowest excitation levels of nuclei (adapted from \cite{Bohr:1958fk}).
Right panel: Binding energy per nucleon for the most beta-stable isobars
(adapted from \cite{Segre:1965uq}).}
\label{Fig:Gap-Nuclei}
\end{figure}

Soon after the development of the BCS theory, Bohr, Mottelson \& Pines
\cite{Bohr:1958fk} pointed out that excitation energies of nuclei exhibit
a gap, as shown in the left panel of \fig{Fig:Gap-Nuclei}.  A
nucleon in even-even nuclei, whether neutron or proton, clearly
requires a minimum energy for excitation.  This energy was interpreted
as being the binding energy of a Cooper pair which must break to
produce an excitation.  In contrast, odd-even nuclei do not show such
a gap, and this is due to the fact that they have one unpaired nucleon
which can be easily excited.  The right panel of
\fig{Fig:Gap-Nuclei} shows that pairing also manifests itself in
the binding energies of nuclei, even-even nuclei being slightly more
bound than odd-even or odd-odd nuclei\footnote{Notice that, as a
  result of pairing, the only stable odd-odd nuclei are $^2$H($Z=1,N=1$),
  $^6$Li(3,3), $^{10}$B(5,5), and $^{14}$N(7,7). All heavier odd-odd
  nuclei are beta-unstable and decay into an even-even nuclei.}.

\begin{figure*}[hbt]
\begin{center}
\includegraphics[width=.80\textwidth]{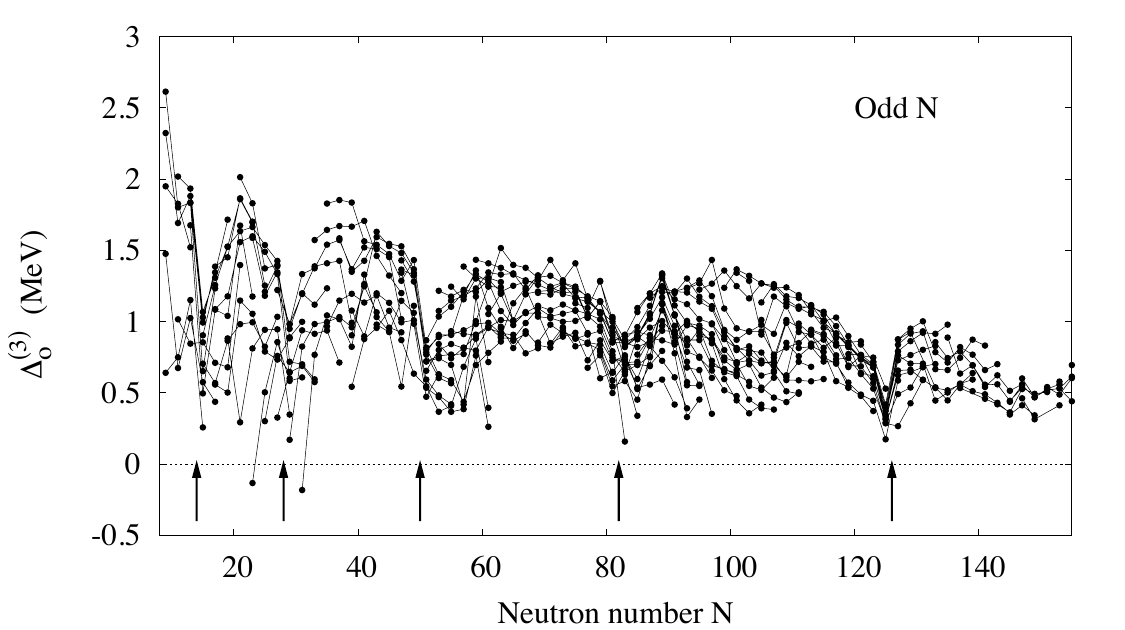}
\includegraphics[width=.80\textwidth]{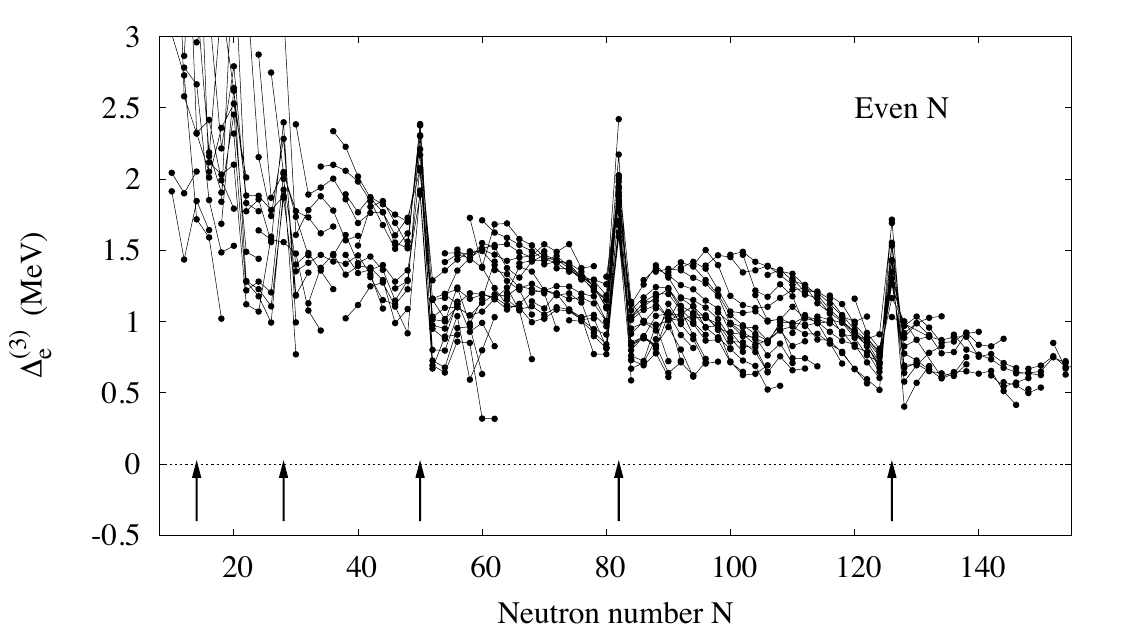}
\end{center}
\caption{Upper panel: Odd-N pairing energies. 
Lower panel: Even-N pairing energies. Figure adapted from \cite{Bertsch:2013bh}.}
\label{Fig:Ngaps-Nuclei}
\end{figure*}

The systematics of neutron pairing energies in nuclei, defined through 
\be
\Delta_{Z,N}=\pm\frac 12 (E_{Z,N+ 1} - 2E_{Z,N} + E_{Z,N-1}) \, ,
\ee
where $E_{Z,N}$ is the binding energy for charge $Z$ and neutron number $N$, and $+(-)$ 
refers to  odd-N and even-N nuclei, are shown  in \fig{Fig:Ngaps-Nuclei}.  
A few key facts to note are: 
 \begin{enumerate}
 \item Pairing energies range from about 3 to 0.5 MeV, decreasing in magnitude with increasing neutron numbers; 
 their behavior with the mass number $A=N+Z$ is well fit by~\cite{Bertsch:2013bh}
 \ba\label{deltanz}
\Delta_{N,Z} &=& {24}/{A} +  0.82 \pm 0.27\,, \quad {\rm for~}N{\rm~odd}\,, \\
\Delta_{N,Z} &=& {41}/{A} +  0.94 \pm 0.31\,, \quad {\rm for~}N{\rm~even}\,.
 \ea 
 \item Dips (peaks) occur adjacent to (at) the sequence of magic numbers \\ $N=14,28,50,82$ and 126 for $N$ odd (even).
 \end{enumerate}
Systematics of proton pairing energies for odd- and even-$Z$ as a
function of proton number $Z$ (see Fig. 2 and Table 1 of
\cite{Bertsch:2013bh}) show similar qualitative behavior, but the
magic number effects are less pronounced.  If the
pairing energy was to be extrapolated to infinite matter using
\eq{deltanz} and $A\rightarrow\infty$, the gap would vanish.
As gaps in infinite matter are predicted to be finite, a saturation
penomenon is at play.  Extended matter in $^3$He, albeit for
different reasons, also exhibits finite gaps.

In addition to the excitation spectra and binding energies of nuclei,
the pairing phenomenon plays important roles in the dynamical
properties of nuclei such as their rotational inertia and the large
amplitude collective motion encountered in fissioning
nuclei. Tunnelling effects in spontaneously fissioning nuclei receive
an enhancement factor $2\Delta^2/G^2$, amounting to an order of
magnitude or more ($G$ is a typical pairing interaction matrix element
between neighboring mean-field configurations).

Pairing effects are also evident in nuclear reactions.  For example,
thermal neutrons of energy only $\sim 0.025~{\rm eV}$ are needed to
cause the fission of $^{235}_{\;\, 92}{\rm U}$ (which results in the
even-even compound nucleus $^{236}_{\;\, 92}{\rm U}$), whereas fast
neutrons of higher energy $\sim 1~{\rm MeV}$ are needed to induce
fission of $^{238}_{\;\, 92}{\rm U}$ (the compound nucleus in this
case is the even-odd nucleus $^{239}_{\;\, 92}{\rm U}$).  What is
interesting is that this phenomenon was appreciated well before the
BCS theory was formulated, and it lies at the root of building nuclear
reactors and purifying naturally-occurring uranium to contain more of the 235
isotope than the 238 isotope.

Besides BCS pairing, the pairing energies shown in
\fig{Fig:Ngaps-Nuclei} receive contributions from other sources
since nuclear sizes are much smaller than the coherence length of the
pairing field. The odd-even staggering is caused by a combination of
effects including the pair-wise filling of the orbitals, diagonal
matrix elements of the two-body interaction, three-nucleon
interactions, the bunching of single particle levels near the Fermi
energy, and the softness of nuclei with respect to quadrupolar
deformations.  The global description of the pairing phenomenon in
nuclei is based on the Hartree-Fock-Bogoliubov approximation and
recent accounts may be found in
Refs.~\cite{Bertsch:2013bh,Brink:2005kx,Broglia:2013vn}.  The basic
cause for pairing in nuclei is, however, easy to identify. The nuclear
interaction between identical nucleons is strongly attractive in the
spin $S=0$ channel, the di-neutron being nearly bound. The even
stronger attraction between neutrons and protons in the spin $S=1$
channel produces bound deuterons, but its effects are
mitigated in heavy nuclei due to the imbalance of neutrons and
protons and attendant many-body effects.  In any case, with proton
and neutron pairing energies on the order of an MeV, nuclei represent
the highest temperature superconductors and superfluid objects in the
laboratory.  It is interesting that gaps of similar order-of-magnitude
are expected for nucleon pairing in neutron stars.

\section{Neutron Stars\label{sec:ns}}

Given that nucleon pairing is important in nuclei, we should expect
that pairing will also occur within neutron stars, as was originally
pointed out by Migdal in 1959 \cite{Migdal:1959bh}.  Although matter
within neutron stars may be heated to more than $10^{11}$~K during
birth, and may remain warmer than $10^8$~K for hundreds of thousands of years, the
nucleons are generally extremely degenerate.  Furthermore, given the
high ambient densities, the critical temperatures for pairing to occur
are large, $\sim10^8-10^{10}$~K.  The onset of pairing is expected to
take place in some parts of a neutron star's interior within minutes
to thousands of years after birth, and is expected to lead to
alteration of several important properties of matter.  While pairing
will not affect the pressure-density relation
significantly\footnote{At asymptotically-high densities where deconfined
  quark matter is thought to exist, pairing gaps could be of order 100~MeV, 
  in which case the EOS is moderately affected by the pairing
  phenomenon.} and, therefore, the overall structure of neutron stars,
the specific heat of dense matter and the emissivity of neutrinos are
dramatically influenced.  Both emissivites and specific heats
are altered at and below the critical temperature, and when the
temperature falls well below the pairing critical temperature, both
vanish exponentially.  This has important consequences for the thermal
evolution of neutron stars that will be described in several subsequent
sections of this chapter.

Superfluidity can also be important in the dynamical evolution of neutron
stars.  It has long been suspected that the so-called ``glitch''
phenomenon observed in pulsars is due to the existence of superfluids
within neutron star crusts and perhaps their outer cores.
Superfluidity within neutron stars might also significantly contribute
to the anomalous braking indices, which are related to the observed
long-term deceleration of the spin-down of pulsars.  These phenomena
will be discussed in Sec.~\ref{Sec:glitch}.

Neutron stars contain the densest form of cold matter observable in
the Universe, in excess of several times the central densities of
nuclei (which is often referred to as the nuclear saturation density
$\rhonuc \simeq 2.7 \times10^{14}$~g~cm$^{-3}$).  Note that $\rhonuc$
corresponds to the density where cold matter with a proton fraction
$x_p = 1/2$ has zero pressure.  While larger mass-energy densities are
transiently reached in relativistic heavy ion collisions, the
resulting matter is extremely ``hot".  Black holes contain a much
denser form of matter, but their interiors are not observable.  It has
long been believed that neutron stars can only form in the aftermath
of the gravitational collapse of a massive star \cite{Baade:1934ly},
commonly known as gravitational-collapse supernovae to distinguish
them from thermonuclear explosions of white dwarfs leading to Type Ia
supernovae.  However, even the collapse of stars with
masses greater than about $25 M_\odot$ are thought to produce briefly existing
proto-neutron stars before they collapse further into black
holes \cite{Woosley:2002cr}.  Nevertheless, the vast majority of
gravitational-collapse supernovae, due to the preponderance of
lower-mass stars, will produce stable neutron stars.

Two simple arguments can convince us that neutron stars, very small
and very dense, can exist.  First, consider the fastest known radio
pulsar, Terzan 5 ad (AKA PSR J1748-2446ad) \cite{Hessels:2006fk}, and posit
that the observed period of its pulses, $P=1.39$ ms, is its rotational
period.  (Pulses produced by binaries or oscillations of neutron stars
are ruled out because nearly all pulsars are observed to be slowing down, 
while orbital and vibrational frequencies increase as energy is
lost.)  Using {\em causality}, that is, imposing that the rotation velocity at
its equator is smaller than the speed of light $c$, one obtains
\be
v_\mathrm{eq} = \Omega R = \frac{2\pi R}{P} < c
\;\;\; \Rightarrow \;\;\; 
R <  \frac{c P}{2 \pi} = 65 \,\mathrm{km} \; .
\ee
This value of 65 km for the radius $R$ is but a strict upper limit; detailed theoretical models and observations
indicate radii on the order of 12 km.
Secondly,  
assuming that the star is {\em bound by gravity}, we can 
require that the gravitational
acceleration $g_\mathrm{eq} $ at the equator is larger than the centrifugal acceleration 
$a_\mathrm{eq}$ and obtain
\ba
g_\mathrm{eq} = \frac{GM}{R^2} >
a_\mathrm{eq} = \Omega^2 R = \frac{4\pi^2 R}{P^2}
\;\;\; \mathrm{or} \;\;\;
\frac{M}{R^3} > \frac{4\pi^2}{GP^2}
\nonumber
\\
\;\;\; \Rightarrow \;\;\; 
\overline{\rho} = \frac{M}{\frac{4}{3}\pi R^3}
> 8 \times 10^{13} \; \mathrm{g \, cm}^{-3} \; .
\ea
Obviously, Newtonian gravity is not accurate in this case, but we can
nevertheless conclude that the central density of these stars is
comparable to, or likely larger than, $\rhonuc$.  Theoretical models
show that densities up to $10 \rhonuc$ \cite{Lattimer:2005fk} are
possibly reachable.  In short, a neutron star is a gigantic, and
compressed, nucleus the size of a city.

\subsection{The Neutron Star Interior}
\label{Sec:NS_interior}

A ``pure neutron star", as  originally conceived by Baade \& Zwicky \cite{Baade:1934ly}
and Oppenheimer \& Volkoff \cite{Oppenheimer:1939uq}, cannot really exist.
Neutrons in a ball should decay into protons through
\be
n \rightarrow p + e^- + \overline{\nu}_e \; .
\label{Eq:DU1}
\ee
This decay is possible for free neutrons since $m_n > m_p + m_e$, where
the masses denote rest masses. However, given the large densities
expected within the neutron star interior, the relevant quantities are
not the masses, but instead the chemical potentials $\mu_i$ ($i$
denoting the species) of the participants.  The matter is degenerate
as typical Fermi energies are on the order of $10-100$ MeV, whereas
the temperature drops below a few MeV within seconds after the birth
of the neutron star \cite{Burrows:1986kx}.  Starting with a ball of
nearly degenerate neutrons, the decay of \eq{Eq:DU1} will generate a
degenerate sea of protons, electrons and anti-neutrinos.  The
interaction mean free paths of anti-neutrinos (and neutrinos) far
exceed the size of the star.  In a neutron star, these can be assumed
to immediately vacate the star, implying that $\mu_\nu=0$.  Thus, this
reaction will ultimately result in the $\beta$-equilibrium condition
\be
\mu_n = \mu_p + \mu_{e} \,,
\label{Eq:Chem}
\ee
which is equivalent to the energy minimization condition $\partial\varepsilon/\partial x_p=0$, 
where $\varepsilon$ is the energy density and the proton fraction is $x_p$.
At finite, but small temperatures, the inverse reaction
\be
p + e^-  \rightarrow n + \nu_e
\label{Eq:DU2}
\ee
also occurs since not all particles are in their lowest energy states
at all times.  

A neutron star, however, is not born from the collapse of a ``ball of
neutrons", but rather from the collapse of the iron core of a massive
star.  At densities typical of pre-collapse configurations,
$\rho\sim10^6$ g cm$^{-3}$ and $x_p\simge0.4$.  During collapse, the
reaction \eq{Eq:DU2} therefore initially dominates over \eq{Eq:DU1} in
order to reduce the proton fraction. As the density increases,
however, the neutrino mean free paths become smaller than the
collapsing core's size, so $\nu_e$ temporarily become trapped within
the core.  In this case, $\mu_{\nu_e}>0$, altering $\beta$-equilibrium
and permits $x_p$ to remain relatively large.  Only after neutrinos
are able to diffuse away, a time of approximately 10
seconds \cite{Burrows:1986kx}, will the final $\beta$-equilibrium
condition \eq{Eq:Chem} be achieved.

\begin{figure}
\begin{center}
\includegraphics[width=.95\textwidth]{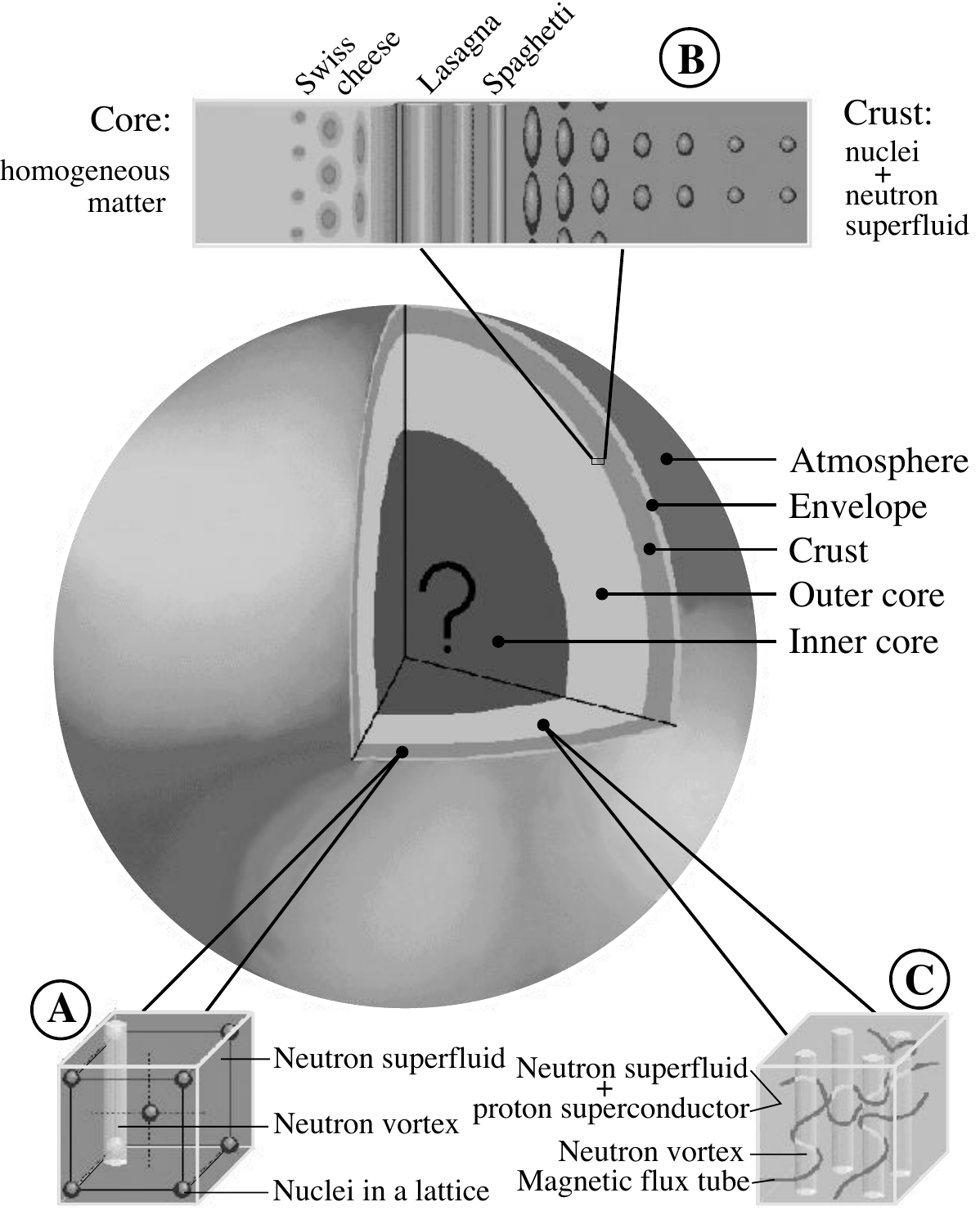}
\end{center}
\caption{Schematic illustration of the structure of a neutron star; 
figure taken  from \cite{Page:2006vn}.
The outermost layers of a neutron star, the atmophere, envelope, 
and crust are described
in Sec.~\ref{Sec:Crust}. Superfluidity in
the crust is schematically represented in inset ``A'', 
and a diagram of the pasta phases in the crust
is shown in inset ``B''. The core is separated into the 
outer core, which has the structure given in inset ``C'',
and the inner core whose nature is currently unknown.
}
\label{Fig:NStar}
\end{figure}

\eq{Eq:Chem} allows one to determine the composition of cold dense
nucleonic matter.  Near $\rhonuc$, a rough approximation to the
difference of nucleon chemical potentials is 
\be\label{muhat}
\mu_n-\mu_p\simeq4S_v\sqrt{\rho/\rhonuc}\left(1-2x_p\right), 
\ee 
where $S_v\simeq31$ MeV is the bulk nuclear symmetry energy parameter.
In dense neutron star matter, electrons are
relativistic and degenerate, so charge neutrality implies equality
between electron and proton number densities and
\be\label{mue} 
\mu_e=\hbar c\left(3\pi^2n_Bx_p\right)^{1/3},  
\ee
where $n_B=n_n+n_p=\rho/m_B$ is the baryon density and $m_B$ is the baryon mass.
  Therefore,
\eq{Eq:Chem} implies
\be\label{betaeq}
x_p=\left({4S_v\over\hbar c}\right)^3 {m_B\over 3\pi^2\rhonuc}\left({\rho\over\rhonuc}\right)^{1/2} \left(1-2x_p\right)^3.
\ee
This is a cubic equation for $x_p(\rho)$, and at the density
$\rhonuc$ the solution is $x_{p,\mathrm{nuc}}\simeq0.034$.  Neutron stars are
therefore composed predominantly, but not completely, of neutrons.
\eq{betaeq} predicts that $x_p$ increases
roughly with $\sqrt{\rho}$ near $\rhonuc$.

Notice that once $\mu_e > m_\mu \simeq 105$ MeV, muons will appear
and be stably present with the conditions of beta equilibrium
\be\label{mu}
\mu_\mu = \mu_e
\ee
and charge neutrality
\be\label{nmu}
n_p=n_\mu+n_e.
\ee
  The condition for the appearance of muons is fulfilled when the
  density is slightly above $\rhonuc$.  However, even though $n_p$ and
  $n_e$ are no longer equal, the trend that $x_p$ slowly increases with
  density is not altered by the presence of muons.  Furthermore, in
  all processes we describe below, there will always be the
  possibility to replace electrons by muons when the density is large
  enough for them to appear.

At all but the highest densities, nucleons 
can be regarded as non-relativistic in neutron star matter, but muons can be either relativistic or 
non-relativistic depending on their abundance.  For nucleons, 
the generalization of the simple approximation \eq{muhat} is~{\footnote{
Relativistic expressions for $\mu_n$ and $\mu_p$ also exist, but are omitted here in the interest of simplicity.}}:
\be
\mu_n = m_n + \frac{p_{Fn}^2}{2 m_n} + V_n \, ,\qquad 
\mu_p = m_p + \frac{p_{Fp}^2}{2 m_p} + V_p \, .
\label{Eq:muN}
\ee
For muons, 
\be
\mu_\mu = \sqrt{m_\mu^2c^4 + p_{F\mu}^2 c^2} \,,
\label{Eq:muL}
\ee
where $p_{Fi}=\hbar k_{Fi}$ is the Fermi momentum of species $i$, and
$V_n$ and $V_p$ are the mean-field energies of $n$ and $p$.  The Fermi
momenta are related to the particle densities by
$k_{Fi}^3=3\pi^2n_i$.  For the leptons, $V_e$ and $V_\mu$ are
negligibly small.  With a knowledge of $V_n$ and $V_p$, the two
$\beta$-equilibrium relations \eqs{Eq:Chem} and (\ref{mu}) can
be solved.  With four chemical potentials and two equations, a unique solution
is obtained by imposing charge neutrality, \eq{mu}, and fixing
$n_B$.  With the particle densities and chemical potentials 
known, one can calculate any thermodynamic potential, in particular
the pressure $P$ and energy density $\epsilon = \rho c^2$.  Varying
the value of $n_B$ gives us the {\em equation of state} (EOS): $P(\rho)$.  Given an EOS, an integration of the
Tolman-Oppenheimer-Volkoff (TOV,  \cite{Oppenheimer:1939uq}) equations of hydrostatic equilibrium
provides us with a well defined model of a neutron star.

The potentials $V_n$ and $V_p$ in \eq{Eq:muN} turn out to be rapidly growing functions of density,
and one can anticipate that eventually reactions such as
\be
p + e^- \rightarrow \Lambda + \overline{\nu}_e
\;\;\;\;\;\;\;\; \mathrm{and/or} \;\;\;\;\;\;\;\;
n + e^- \rightarrow \Sigma^- + \overline{\nu}_e
\label{Eq:hyperons}
\ee
may produce hyperons.
Hyperons can appear once the corresponding $\beta$-equi\-lib\-rium conditions
are satisfied, i.e., 
$\mu_n = \mu_\Lambda$ or/and $\mu_n + \mu_e = \mu_{\Sigma^-}$.
At the threshold, where $p_{F\Lambda} =0$ or $p_{F\Sigma^-} = 0$, one can expect that
$|V_{\Lambda}| \ll m_\Lambda$ and $|V_{\Sigma^-}| \ll m_{\Sigma^-}$
and thus
$\mu_\Lambda \simeq m_\Lambda$ and $\mu_{\Sigma^-} \simeq m_{\Sigma^-}$.
Since $m_\Lambda$ and $m_{\Sigma^-}$ are larger than the nucleon mass by only about 200 MeV
these hyperons\footnote{The $\Sigma^+$ is less favored as its $\beta$-equilibrium condition is
$ \mu_{\Sigma^+} = \mu_p = \mu_n - \mu_e$. Heavier baryon are even less favored, but cannot
{\em a priori} be excluded.}
are good candidates for an ``exotic" form of matter in neutron stars.
Along similar lines, 
 the lightest mesons, pions and/or kaons, may also appear stably,
and can form meson condensates.
At even larger densities, the ground state of matter is likely to
be one of deconfined quarks.
All these possibilities
depend crucially on the strong interactions terms,  $V_n$ and $V_p$.
Figure~\ref{Fig:NStar} illustrates our present understanding (or misunderstanding) of the interior of a
neutron star, with a black question mark ``{\bf ?}" in its densest part.
The outer part of the star, its {\em crust}, is described briefly in the following subsection.

When only nucleons, plus leptons as implied by charge neutrality and constrained by $\beta$-equilibrium,
 are considered, the EOS can be calculated with much more confidence than in the presence of ``exotic'' forms of matter. 
For illustrative puposes, we will generally employ the EOS of  Akmal, Pandharipande \& Ravenhall 
\cite{Akmal:1998fk} (``APR'' hereafter) in presenting our results.

Although there is no evidence that any observed neutron star or pulsar 
might actually instead be a pure quark star, theory allows this possibility.  
Such a star would be nearly completely composed of a mixture of up, down 
and strange quarks, and would differ from a neutron star in that it would be 
self-bound rather than held together by gravity.  

The reader can find a more detailed presentation and entries to the key literature in \cite{Page:2006vn}.

\subsection{The Neutron Star Envelope and Crust}
\label{Sec:Crust}

In the outer part of the star, where $\rho\simle\rhonuc/2$, a
homogeneous liquid of nucleons is mechanically unstable (known as the
spinodal instability).  Stability is, however, restored by the
formation of nuclei, or nuclear clusters.  This region, called the
{\it crust}, has a thickness of $\simle1$ km.  

Above the surface, where the pressure approximately vanishes, we might
expect the presence of an {\it atmosphere}, but there is the
possibility of having a solid surface, condensed by a
sufficiently-strong magnetic field \cite{Lai:2001uq}.  A few meters
below the surface, ions are completely pressure-ionized (the radius of
the first Bohr orbital is larger than the inter-nuclear distance when
$\rho \simge 10^4$ g cm$^{-3}$).  Matter then consists of a gas/liquid
of nuclei immersed in a quantum liquid of electrons.  When $\rho
\approx 10^6$ g cm$^{-3}$, $\mu_e$ is of the order of 1 MeV and the
electrons become relativistic.  Here, and at higher densities, Coulomb
corrections are negligible -- electrons form an almost perfect Fermi
gas.  However, Coulomb corrections to the ions are {\em not}
negligible. From a gaseous state at the surface, ions will
progressively go through a liquid state (sometimes called the {\it ocean}) and
finally crystallize, at densities between $10^2$ up to $\sim 10^{10}$
g cm$^{-3}$ depending on the temperature (within the range of
temperatures for which neutron stars are thermally detectable).
These low-density layers are commonly referred to as the {\it envelope}.  

With growing $\rho$, and the accompanying growth of $\mu_e$, it is
energetically favorable to absorb electrons into neutrons and, hence,
nuclei become progressively neutron-rich.  When $\rho \sim 4 \times
10^{11}$ g cm$^{-3}$ (the exact value depends on the assumed chemical
composition), one achieves the {\em neutron drip} point at which the
neutron density is so much larger than that of the protons that some
neutrons become unbound (i.e., $\mu_n>0$).  Matter then consists of a
crystal of nuclei immersed in a Fermi gas of electrons and a quantum
liquid of dripped neutrons.  The region containing dripped neutrons is
usually called the {\em inner crust}.  In most of this inner crust,
because of the long-range attractive nature of the nucleon-nucleon
interaction, the dripped neutrons are predicted to form a superfluid
(in a spin-singlet, zero orbital angular momentum, state $\Singlet$).
All neutron stars we observe as pulsars are rotating.  While a
superfluid cannot undergo rigid body rotation, it can simulate it
by forming an array of vortices (in the cores of which superfluidity is
destroyed).  (See, e.g., \cite{Tilley:1990ys}.)  The resulting
structure is illustrated in inset A of \fig{Fig:NStar}.

At not too-high densities, nucleons are correlated at short distances
by strong interaction and anti-correlated at larger distances by
Coulomb repulsion between the nuclei, the former producing spherical
nuclei and the latter resulting in the crystallization of the matter.
As $\rho$ approaches $\sim0.03\rhonuc$, the shapes of nuclei can
undergo drastic changes: the nuclear attraction and Coulomb repulsion
length-scales become comparable and the system is ``frustrated''.
From spherical shapes, as the density is increased, nuclei are
expected to deform, become elongated into 2D structures
(``spaghetti''), and then form 1D structures (``lasagna''), always
with denser nuclear matter surrounded by the dilute neutron
gas/superfluid which occupies an increasing portion of the volume.
When the phases achieve approximately equal volume fractions, the
geometry can invert, with dripped neutrons confined into
2D (``anti-spaghetti'' or ``ziti'') and finally 3D (``swiss cheese'')
bubbles.  The homogeneous phase, i.e. the {\em core} of the star, is
reached when $\rho \simeq 0.5-0.6 \rhonuc$.  This
``pasta'' regime is illustrated in inset B of \fig{Fig:NStar} and
is thought to resemble a liquid crystal \cite{Pethick:1998uq}.  A
compilation of the most recent progress on neutron star crust physics
can be found in the book \cite{Bertulani:2012ys}.


\section{Pairing in the Neutron Star Interior}
\label{Sec:Pairing}

\subsection{General Considerations}

\subsubsection*{Expectations from measured phase shifts}

\begin{figure*}[hbt]
\begin{center}
\includegraphics[width=.80\textwidth]{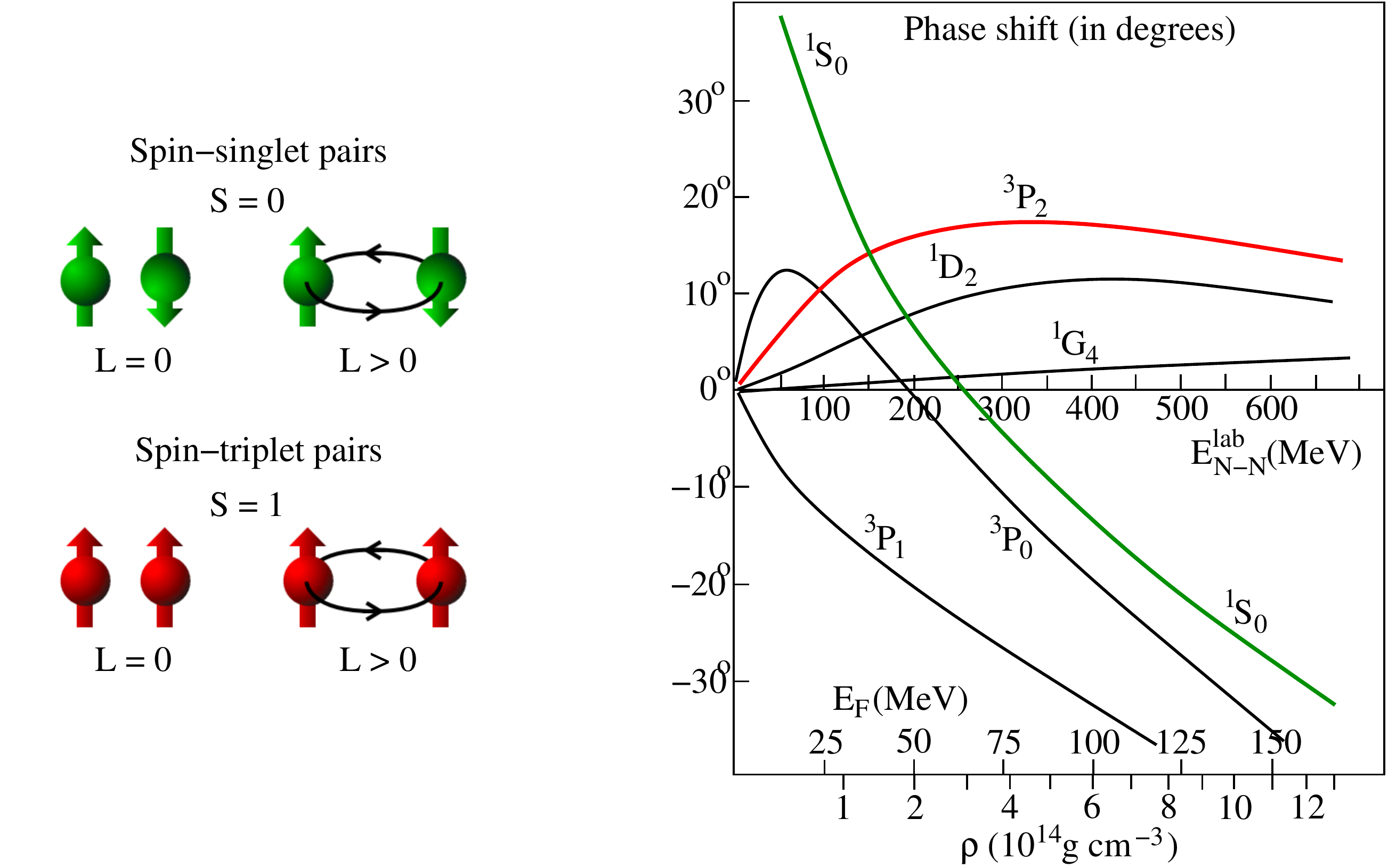}
\end{center}
\caption{
Left panel: 
Possible spin-angular momentum combinations for Cooper-pairs.
Right panel:
Phase shifts for N-N scattering as a function of the laboratory energy (middle axis)
or the neutron Fermi energy and density for a neutron star interior (lower axis).
Adapted from \cite{Tamagaki:1970uq}.}
\label{Fig:CooperPairs}
\end{figure*}

As a two-particle bound state, the Cooper pair can appear in many spin-orbital angular momentum
states (see the left panel of \fig{Fig:CooperPairs}).
In terrestrial superconducting metals, the Cooper pairs are generally in the $\Singlet$ channel, i.e., 
spin-singlets with $L=0$ orbital angular momentum, whereas in liquid $^3$He they are in spin-triplet states.
What can we expect in a neutron star?
In the right panel of \fig{Fig:CooperPairs},  we adapt a figure from one of the first works to study
neutron pairing in the neutron star core \cite{Tamagaki:1970uq} showing laboratory-measured
phase-shifts from nucleon-nucleon scattering.  A positive phase-shift implies an attractive interaction.
From  this figure, one can expect that nucleons could pair in a spin-singlet state, $\Singlet$, at low densities,
whereas a spin-triplet, $\triplet$, pairing should occur at high densities.
We emphasize that this is only a {\em presumption} (phase shifts reflect free-space interaction) as medium effects can  
strongly affect particle interactions.

\subsubsection*{The energy gap}

\begin{figure*}[htb]
\begin{center}
\includegraphics[width=.60\textwidth]{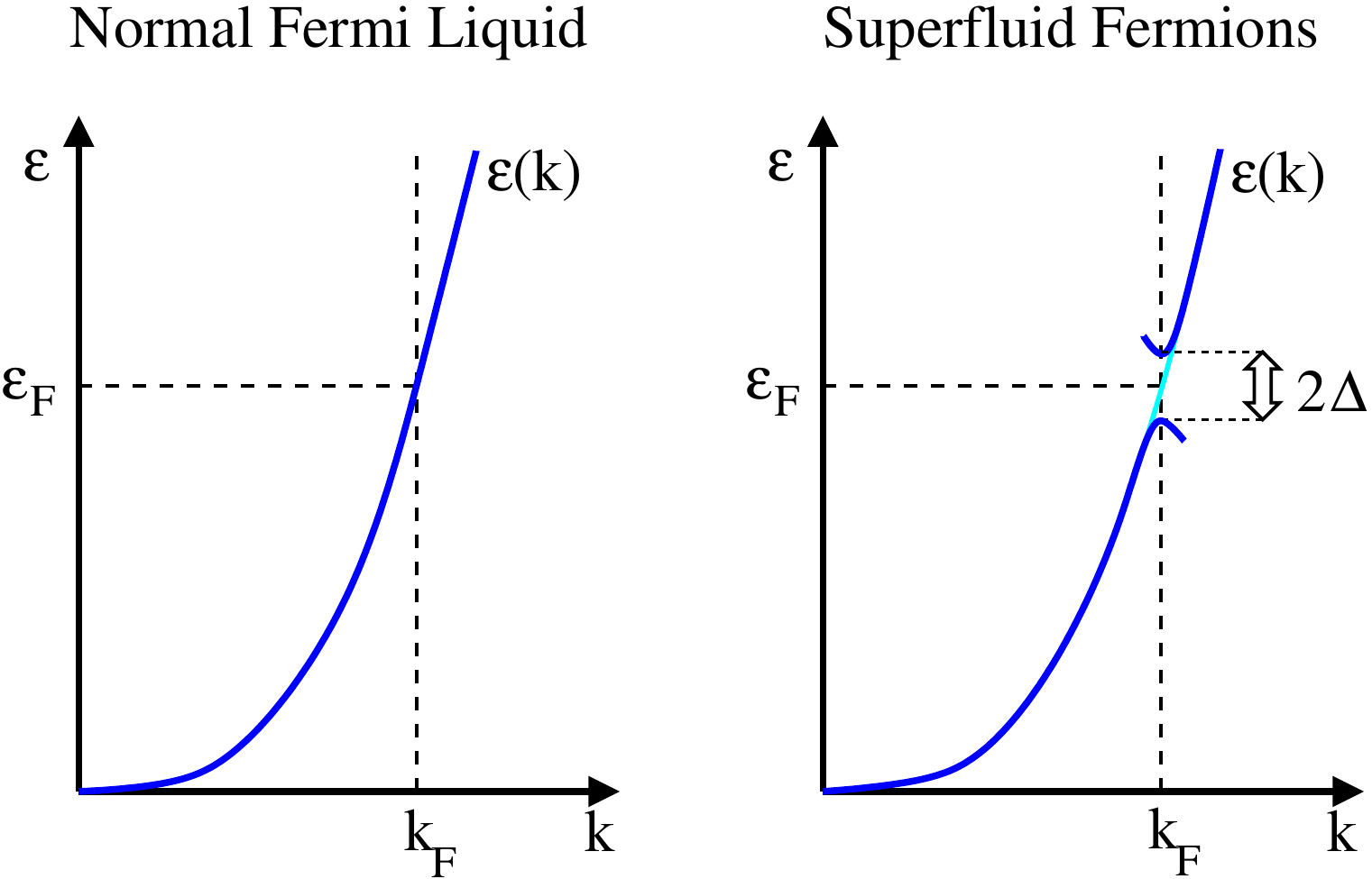}
\end{center}
\caption{Comparison of quasi-particle spectra, $\epsilon(k)$, 
for a normal and a superfluid Fermi liquid.
The reorganization of particles at $\epsilon \sim \epsilon_F$ into Cooper pairs results
in the development a gap $2 \Delta$ in the spectrum so that no particle can have
an energy between $\epsilon_F - \Delta$ and $\epsilon_F + \Delta$.
}
\label{Fig:SP_Spectrum}
\end{figure*}

In a normal Fermi system at $T=0$, all particles are in states with energies $\epsilon \le \epsilon_F$.
When $T >0$, states with energies $\epsilon \simge \epsilon_F$ can be 
occupied
(left panel of \fig{Fig:SP_Spectrum}) resulting
in a smearing of the particle distribution around $\epsilon_F$ in a range $\sim k_BT$.
It is precisely this smooth smearing of energies around $\epsilon_F$ which produces the linear $T$
dependence of $c_v$, Sec.~\ref{Sec:Cv}, and the $T^6$ or $T^8$ dependence of the neutrino
emissivities, Sec.~\ref{Sec:nuT}.

In a superfluid/superconducting Fermi system at $T=0$, all particles
are in states with energies $\epsilon \le \epsilon_F$ (actually,
$\epsilon \le \epsilon_F - \Delta$).  For nonzero temperatures that
permit the presence of Cooper pairs (and hence a gap $\Delta(T)$),
states with energy $\epsilon \ge \epsilon_F + \Delta$ can be
populated.  However, in contrast to the smooth filling of levels above
$\epsilon_F$ in the case of a normal Fermi liquid, the presence of the
$2 \Delta(T)$ gap in the spectrum implies that the occupation
probability is strongly suppressed by a Boltzmann-like factor $\sim
\exp[-2 \Delta(T)/k_BT]$.
As a result, both the specific heat of paired particles and the neutrino emissivity of all
processes in which they participate are 
strongly reduced.

\subsubsection*{The phase transition}

\begin{figure*}[t]
\begin{center}
\includegraphics[width=.95\textwidth]{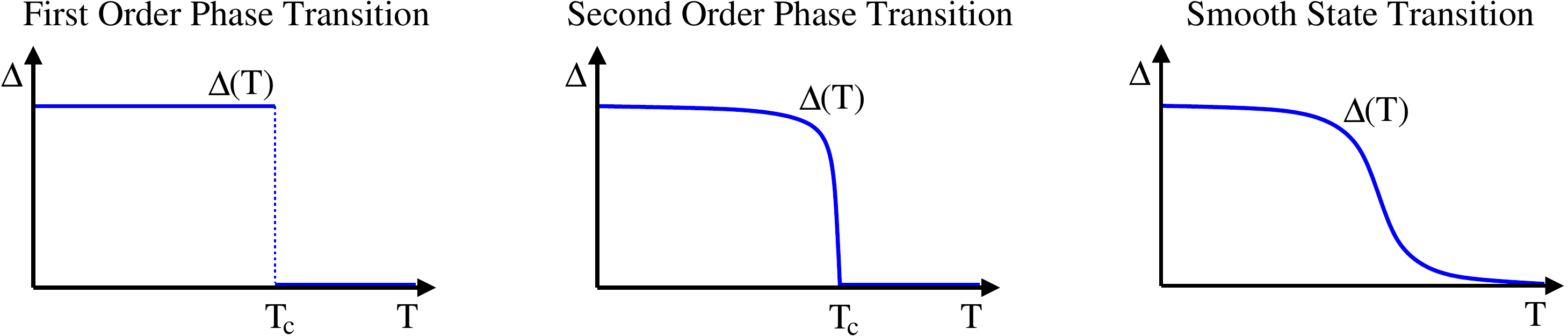}
\end{center}
\caption{Temperature evolution of the state of a system parametrized by an ``order'' parameter, $\Delta(T)$.}
\label{Fig:Order}
\end{figure*}

The transition to the superfluid/superconducting state through pairing {\em \`{a} la BCS} is usually 
a second order phase transition and the gap
$\Delta(T)$ is its order parameter (see central panel of \fig{Fig:Order}). 
Explicitly,  $\Delta(T)=0$ when $T > T_c$, the critical temperature, and, when $T$ drops below $T_c$,
$\Delta(T)$ grows rapidly but continuously, with a discontinuity in its slope at $T=T_c$.
There is no latent heat but a discontinuity in specific heat.
(Examples: superfluid $\leftrightarrow$ normal fluid; ferromagnetic $\leftrightarrow$ paramagnetic.)
In the BCS theory, which remains approximately valid for nucleons,
the relationship between the zero temperature gap and $T_c$ is
\be
\Delta(T=0) \simeq 1.75~k_B T_c \,.
\label{Eq:DTc}
\ee

In a first order phase transition there is a discontinuous change of $\Delta(T)$ at $T_c$ and the
transition occurs entirely at $T_c$ (see left panel of \fig{Fig:Order}).
There is a latent heat due to the entropy difference between the two states.
(Examples: solid $\leftrightarrow$ liquid; liquid $\leftrightarrow$ gas below the critical point.)
In a smooth state transition there is a continuous change of $\Delta(T)$ with no critical temperature
(see the right panel of \fig{Fig:Order}).
(Examples: liquid $\leftrightarrow$ gas above the critical point; atomic gas $\leftrightarrow$ plasma.)

\subsubsection*{A simple example}

A simple model can illustrate the difficulty in calculating pairing gaps.
Consider a dilute Fermi gas with a weak, attractive, interaction potential $U$.
The interaction is then entirely described by the corresponding scattering length\footnote{The 
scattering length $a$ is related to $U$ by $a = (m/4\pi \hbar^2) U_0$ with 
$U_\mathbf{k} = \int d^3r \, \exp(i \mathbf{k}\cdot \mathbf{r}) \, U(\mathbf{r}) $.}
, $a$, 
which is negative for an attractive potential.
In this case, the model has a single dimensionless parameter, $|a|k_F$, and the dilute gas corresponds to $|a|k_F \ll 1$.
Assuming the pairing interaction is just the bare interaction $U$ (which is called the
{\em BCS approximation}), the gap equation at $T=0$  can be solved analytically, giving the
{\em weak-coupling BCS-approximation} gap:
\be
\Delta(k_F)  \;  \xrightarrow{ |a| k_F \rightarrow 0} \; \Delta_{BCS}(k_F) =
\frac{8}{e^2} \left(\frac{\hbar^2 k_F^2}{2M}\right) \exp \left[ -\frac{\pi}{2|a|k_F} \right] \, .
\label{Eq:Delta_BCS}
\ee
This result is bad news: the gap depends exponentially on the pairing potential $U$.
The Cooper pairs have a size of the order of 
$\xi \sim \hbar v_F/\Delta$ (the {\em coherence length}) and thus
$\xi k_F \propto \exp [ \mathbf{+} \pi/2|a|k_F] \gg 1$ in the weak coupling limit.
There is, hence, an exponentially growing number of other particles within the pair's coherence length
when $|a| k_F \rightarrow 0$.
These particles will react and can screen, or un-screen, the interaction.
Including this medium polarization on the pairing is called {\em beyond BCS}, and in the weak
coupling limit its effect can be calculated analytically \cite{Gorkov:1961kx}, giving
\ba
\!\!\!\!
\Delta(k_F)  \; \xrightarrow{ |a| k_F \rightarrow 0} \; \Delta_{GMB}(k_F) =
\frac{1}{(4\mathrm{e})^{1/3}} \Delta_{BCS}(k_F)
\simeq 0.45 \Delta_{BCS}(k_F)
\label{Eq:Delta_GMB}
\ea
So, screening by the medium can reduce the gap by more than a factor two, even in an extremely
dilute system.

\subsection{Calculations of Pairing Gaps}
\label{Sec:Gaps}

A significant amount of work has been devoted to the calculation of pairing gaps in the neutron star environment:
see, e.g., \cite{Lombardo:2001ys,Dean:2003vn,Sedrakian2006:gk} 
or A. Schwenk's contribution to this volume for reviews.
Below we first briefly describe the Gorkov formalism
 \cite{Abrikosov:1963zr,Landau:1980ly,Lombardo:2001ys}
that will allow us to set up the stage for the 
presentation
of representative results for 
nucleon gaps. Specifically, we will address 
the neutron $\Singlet$ and $\Triplet$ and the proton $\Singlet$ and $\Triplet$ gaps
{ and briefly mention hyperon gaps.}
The effects of pairing on 
the thermal evolution of neutron stars are described in Sec.~\ref{Sec:Pairing_Effects}.
Calculations of 
pairing gaps in quark matter will be described in Sec.~\ref{Sec:Quarks}.

\subsubsection*{General formalism}

\begin{figure*}[b]
(A) $iG(1,2) = \langle T\{ \psi(1) \psi^\dagger(2) \} \rangle = $
\raisebox{-1pt}{\includegraphics[width=20pt]{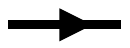}}
$=$
\raisebox{+1pt}{\includegraphics[width=20pt]{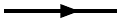}}
$+$
\raisebox{-5pt}{\includegraphics[height=15pt]{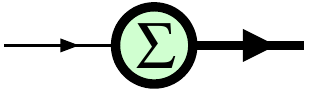}}
\smallskip
\caption{The normal state propagator and its Dyson equation.}
\label{Fig:Dyson}
\end{figure*}

Several significant effects of pairing are due to the change in the quasi-particle spectrum
that is obtained from the poles of the propagator $G$.
In \fig{Fig:Dyson}, Eq.~(A), we show the definition of $G$, the two point Green's function with one particle in and one particle out, and its Dyson equation 
which relates the
free propagator and the self energy $\Sigma$ in the case of a normal system.
The resulting quasi-particle spectrum is then
\ba
{\omega = } \epsilon(\bm k, \omega)= \frac{\hbar^2 \bm k^2}{2m} + \Sigma(\bm k, \omega) -  \epsilon_F
\simeq  \frac{\hbar^2 k_F}{m^*} (k-k_F) \,.
\label{Eq:epsilon_normal}
\ea
In obtaining the right-most result above, 
we assumed the system to be isotropic and
the spectrum is evaluated for  $k \simeq k_F$ with 
the effective mass $m^*$ 
defined through
\be
\left. \frac{\partial \epsilon(k,\omega=0)}{\partial k}\right|_{k=k_F} = \frac{\hbar^2 k_F}{m^*} \,.
\ee
The resulting spectrum $\epsilon_F + \omega$ is depicted in the left panel of \fig{Fig:SP_Spectrum}.

\begin{figure*}[t]
(B) $iF(1,2) = \langle T\{ \psi(1) \psi(2) \} \rangle \;\,\;\;\;\, = $
\raisebox{-1pt}{\includegraphics[width=20pt]{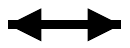}}
$= \;\;\;\; \;\;\;\;\;\;\;\; $
\raisebox{-5pt}{\includegraphics[height=15pt]{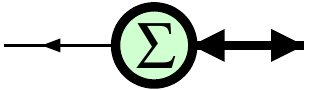}}
$+$
\raisebox{-5pt}{\includegraphics[height=15pt]{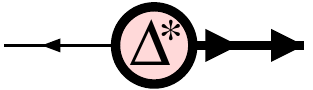}}
\smallskip
\\
(B') $iF^\dagger(1,2) = \langle T\{ \psi^\dagger(1) \psi^\dagger(2) \} \rangle = $
\raisebox{-1pt}{\includegraphics[width=20pt]{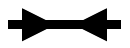}}
$= \;\;\;\; \;\;\;\;\;\;\;\; $
\raisebox{-5pt}{\includegraphics[height=15pt]{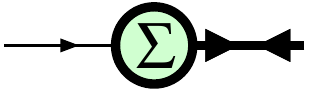}}
$+$
\raisebox{-5pt}{\includegraphics[height=15pt]{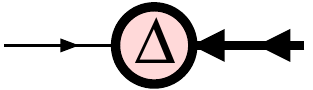}}
\smallskip
\\(C) $iG(1,2) = \langle T\{ \psi(1) \psi^\dagger(2) \} \rangle \;\;\;\; = $
\raisebox{-1pt}{\includegraphics[width=20pt]{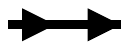}}
$=$
\raisebox{+1pt}{\includegraphics[width=20pt]{G_free.pdf}}
$+$
\raisebox{-5pt}{\includegraphics[height=15pt]{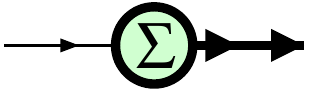}}
$+$
\raisebox{-5pt}{\includegraphics[height=15pt]{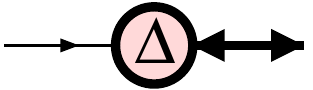}}
\bigskip
\\
(D) 
\raisebox{-5pt}{\includegraphics[height=15pt]{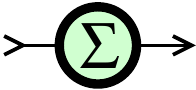}}
$=$
\raisebox{-1pt}{\includegraphics[height=27pt]{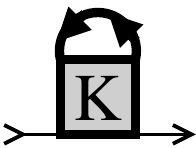}}
\hspace{3cm}
(E)
\raisebox{-5pt}{\includegraphics[height=15pt]{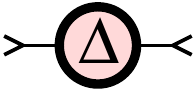}}
$=$
\raisebox{-1pt}{\includegraphics[height=27pt]{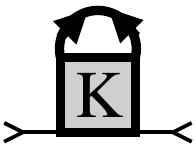}}
\caption{The Gorkov equations}
\label{Fig:Gorkov}
\end{figure*}

In the presence of a pairing instability, and the concomitant development of a condensate,
an {\it anomalous propagator} $F$ and its adjoint $F^\dagger$ can be defined, see Eq.~(B) and Eq.~(B') in \fig{Fig:Gorkov}, 
with their corresponding Gorkov equations that replace the Dyson equation. 
In addition to  
the self energy $\Sigma$, the Gorkov equations feature 
an {\it anomalous self energy}, 
or the {\it gap function} $\Delta$.
The propagator $F$ violates particle number conservation as it propagates a hole into a particle,
and vanishes in the absence of a condensate in a normal system.
The gap function is a $2 \times 2$ matrix in spin space
\ba
\hat{\Delta}(\bm k, \omega) = 
\left(\begin{array}{cc}
\Delta_{\uparrow \uparrow}(\bm k, \omega)       & \Delta_{\uparrow \downarrow}(\bm k, \omega) \\
\Delta_{\downarrow \uparrow}(\bm k, \omega)  & \Delta_{\downarrow \downarrow}(\bm k, \omega)
\end{array}\right) \,.
\label{Eq.Delta_spin}
\ea
In the case the ground state is assumed to be time-reversal invariant, 
$\hat{\Delta}(\bm k, \omega)$ (denoted $\hat{\Delta}$ for short below)  
has a unitary structure satisfying
\ba
\hat{\Delta} \hat{\Delta}^\dagger =  \hat{\Delta}^\dagger  \hat{\Delta} = \Delta^2 \, \hat{1}
\ea
where 
$\hat{1}$ is a  $2\times2$ unit matrix,
and  
\ba
\Delta^2 = \Delta^2(\bm k, \omega) = \mathrm{det} \, \hat{\Delta}(\bm k, \omega) \,.
\ea
The quantity $\Delta$ above 
will appear as the {\it  energy gap} in the quasi-particle spectrum.

The normal propagator $G$ is also modified, as depicted in Eq.~(C) in \fig{Fig:Gorkov}.
{Solving the Gorkov equations gives
$G = (\omega + \epsilon)/D$ with }
\be
D(\bm k,\omega) = \omega^2 - \epsilon(\bm k,\omega)^2 - \Delta(\bm k ,\omega)^2
\ee
and its modified poles yield a 
quasi-particle spectrum with two branches:
\ba
\omega = \pm \sqrt{\epsilon(\bm k,\omega)^2 + \Delta(\bm k,\omega)^2} \, .
\label{Eq:omega_paired}
\ea
The resulting spectrum $\epsilon_F + \omega$ is depicted in the right panel of \fig{Fig:SP_Spectrum}. 
Note that, $G$ and $\Sigma$ are also $2\times2$ matrices in spin space 
both being diagonal in structure.
The solution of the Gorkov equation for $F$ gives
$F = \hat{\Delta}^\dagger/D$.
Finally, $\Sigma$ and $\Delta$ are defined by Eq.~(D) and (E) in \fig{Fig:Gorkov}
from a kernel $K$.
Equation (D) is the gap equation and reads
\be
\Delta_{\alpha \beta}(k) = i \int \frac{d^4 k'}{(2\pi)^4} \sum_{\alpha' , \beta'} 
\frac{\langle k\alpha,-k\beta | K | k'\alpha', -k'\beta' \rangle \Delta_{\alpha' \beta'}(k')}
{D(k')} \,, 
\label{Eq:Gap_Equation}
\ee
where $k = (\bm k,\omega)$, $k' = (\bm k',\omega')$ 
and $\alpha$, $\beta$, ... denote spin indices.

Solving the full set of equations in
\fig{Fig:Gorkov} requires many approximations,
notably in the choice of the kernel $K$.
In particular, different approximate kernels $K$ are used in (D) and (E) 
as it appears in the particle-hole channel in (D), 
whereas  
in (E) it is in the particle-particle channel.
In the {\it BCS approximation}, the self energy is calculated in the normal phase
(e.g., with a Br\"uckner-Hartree-Fock or BHF scheme) with its energy dependence 
being  neglected,
i.e. $\Sigma(\bm k, \omega) \rightarrow \Sigma(\bm k)$,
and the kernel for $\Delta$ is simply taken as the bare interaction.
The latter can be a two-body force (2BF) from a potential adjusted to laboratory N-N scattering data
or that derived from an effective interaction such as a Skyrme force.
Recently, the effect of three-body forces (3BF), absent in the laboratory N-N scattering experiment,
has been considered.
Inclusion of 3BF is 
necessary to reproduce the nuclear saturation density; they are, in the bulk, repulsive and
their importance grows with increasing density.
Even at the 2BF level,  
a severe problem is encountered: to date,  
 none of the N-N interaction models 
reproduce the measured phase shifts for
$E_\mathrm{lab}$ larger than 300 MeV in the channels needed (particularly for  $\triplet$) 
for the conditions prevailing in the core of a neutron star. 
The source of this problem is  easy to identify; 
beyond 290 MeV (the threshold for pion production), 
inelastic channels begin to become dominant.

Models {\it beyond BCS} have proceeded in 
two directions. In the first approach, in the gap equation,
the bare interaction is supplemented by the inclusion of  
short-range correlations. 
In a further step, long-range correlations 
to account for medium polarization are added. 
As illustrated above with the weak-coupling result of \eq{Eq:Delta_GMB}, polarizations effects
can be significant.
In the second approach,  which goes beyond the BHF level, 
the self energy $\Sigma$ is calculated by including its
energy dependence.
Calculations of  
$\Sigma$ 
in the paired phase, see Eq.~(D) in \fig{Fig:Gorkov}, do not yet exist and are necessary.
In the following, we will neglect the energy dependence of the gap, i.e., write it as
$\hat{\Delta}(\bm k)$ instead of $\hat{\Delta}(\bm k,\omega)$.

\subsubsection*{Pairing in single spin-angular momentum channels}
\label{Sec:Gaps2}

In pairing calculations, 
the potential and the gap function are usually expanded in partial waves 
so as to focus on 
specific 
spin-angular momentum channels, $\lambda=(s,j)$.
At low $k_{Fn}$, or $k_{Fp}$, it is theoretically predicted that the preferred channel is $\lambda=(0,0)$ in $S$-wave,
i.e., the spin-singlet $\Singlet$.
At large 
Fermi momenta, the $\Singlet$ interaction becomes repulsive and the preferred channel is $\lambda=(1,2)$
in $P$ and $F$ waves (the mixing being due to the tensor interaction \cite{Takatsuka:1972bs})
, i.e., the spin-triplet $\Triplet$.
In the $\Singlet$ channel, which has also been called the ``A'' phase, the gap is spherically symmetric
and can be written as
\begin{flalign} 
& \mathrm{A \; phase} \; (\Singlet)\! : \hspace{0.1mm}  
\hat\Delta_{(0,0)}(\bm k) =
\left( \!\!\!\! \begin{array}{cc}
 0      & \Delta(k) \\
-\Delta(k)  & 0
\end{array} \!\!\!\! \right)
\;\;\;\;\;
[\Delta(k_F) = \text{energy gap}] \!\!\!\!
& 
\end{flalign}

In the $\Triplet$ channel, $\hat\Delta_\lambda$
has contributions from all possible orbital angular momenta  $l$ and their $m_j$ components, i.e.,
$\hat\Delta_\lambda = \sum_{l,m_j} \Delta^{m_j}_{l \, \lambda}(k)  \hat{G}^{m_j}_{l \, \lambda}(\hat{\bm k})$,
where the $\hat{G}^{m_j}_{l \, \lambda}(\hat{\bm k})$ are $2 \times 2$ spin matrices describing
the angular dependence of $\hat\Delta_\lambda$ which is thus {\em not} spherically symmetric.
Microscopic calculations restricted to the $\triplet$ channel \cite{Amundsen:1985nx,Baldo:1992kx}
indicate that the largest component of 
$\hat\Delta_{\lambda}$ corresponds to the $m_j =  0$ sub-channel or, possibly, the $m_j = \pm 2$ channels,  
sometimes called the ``B'' and ``C'' phases, respectively. 
For these two special cases, the energy gap $\Delta(\bm k_F)$ is given by  \cite{Amundsen:1985nx}
\begin{flalign} 
& \mathrm{B \; phase} \; (\triplet, m_j=0): \hspace{5mm}
\Delta^2(\bm k_F) = \frac{1}{2} \left[ \Delta^0_{2 \, \lambda}(k_F) \right]^2 \, \frac{1+3 \cos^2 \theta_k}{8\pi} & 
\label{Eq:Bphase}
\\
& \mathrm{C \; phase} \; (\triplet, m_j=\pm 2): \hspace{2mm}
\Delta^2(\bm k_F) = \;\;\; \left[ \Delta^2_{2 \, \lambda}(k_F) \right]^2 \;\;\; \frac{3 \sin^2 \theta_k}{8\pi} \,,& 
\label{Eq:Cphase}
\end{flalign}
where $\theta_k$ is the angle between $\bm k_F$ and the arbitrary quantization axis.
Notice that in the B phase, the gap is nodeless whereas 
in the C phase it vanishes on the equator, $\theta_k = \pi/2$.

\subsubsection*{Temperature dependence of $\Delta$ and $T_c$}

The preceding discussion 
was restricted to the zero temperature case.  
It is naturally extended to finite temperature whence
the gap becomes 
$\hat\Delta(\bm k ; T)$.
However, 
effects 
of thermal excitations are important only 
for values of  
$\bm k \simeq \bm k_F$.
We will often omit either of the arguments $\bm k_F$ or $T$ 
when 
not necessary, 
but they are always implied (as is its $\omega$ dependence).
Notice that microscopic calculations are often limited to the $T=0$ case only.

\begin{figure}[t]
\begin{center}
\includegraphics[width=.60\textwidth]{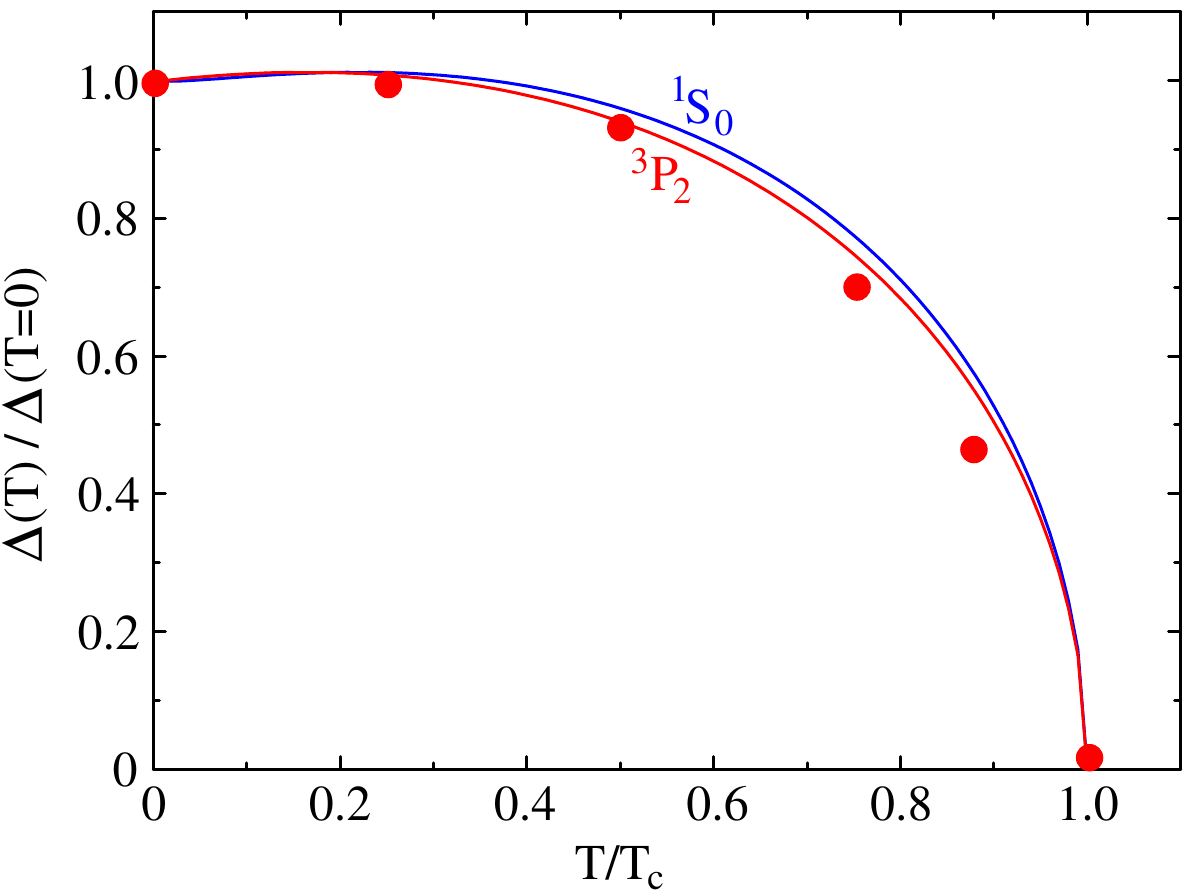}
\end{center}
\caption{
Temperature dependence of the energy gap $\Delta(k_F,T)$ for
$\Singlet$ pairing and of the angle averaged $\overline{\Delta}(k_F;T)$ for the
$\triplet$ pairing in the $m_j = 0$ case.
Continuous lines show the commonly used analytical fits of \cite{Levenfish:1994vn}
whereas 
the dots are from the calculations of \cite{Baldo:1992kx}.
$\triplet$ pairing in the $m_j = 2$ case results in values very close to the $m_j = 0$ case.
}
\label{Fig:D_T}
\end{figure}

The relationship between the 
critical temperature $T_c$ for the phase transition and the energy gap $\Delta(\bm k_F)$
is approximately given by the usual result
\be
k_B T_c \approx  0.57 \; \overline\Delta(k_F;T=0)
\ee
for all three phases A, B, and C \cite{Amundsen:1985nx,Baldo:1992kx},
where $\overline\Delta(k_F ; T)$ is obtained by angle averaging of $\Delta^2(\bm k_F,T)$ over the Fermi surface
\be
\left[ \, \overline\Delta(k_F;T) \right]^2 \equiv  \int \!\!\! \int \! \frac{d\Omega}{4\pi} \, \Delta^2(\bm k_F;T) \; .
\label{Eq:D_ave}
\ee
Obviously, $\overline\Delta(k_F;T) = \Delta(k_F;T)$ for an isotropic $\Singlet$ gap.
The temperature dependence of the energy gap $\Delta(k_F,T)$ for
$\Singlet$ pairing and of the angle averaged $\overline{\Delta}(k_F;T)$ for the
$\triplet$ pairing in the $m_j = 0$ case are shown in \fig{Fig:D_T}.

\subsubsection*{The isotropic $\Singlet$ neutron gap}

\begin{figure*}[t]
\begin{center}
\includegraphics[width=.70\textwidth]{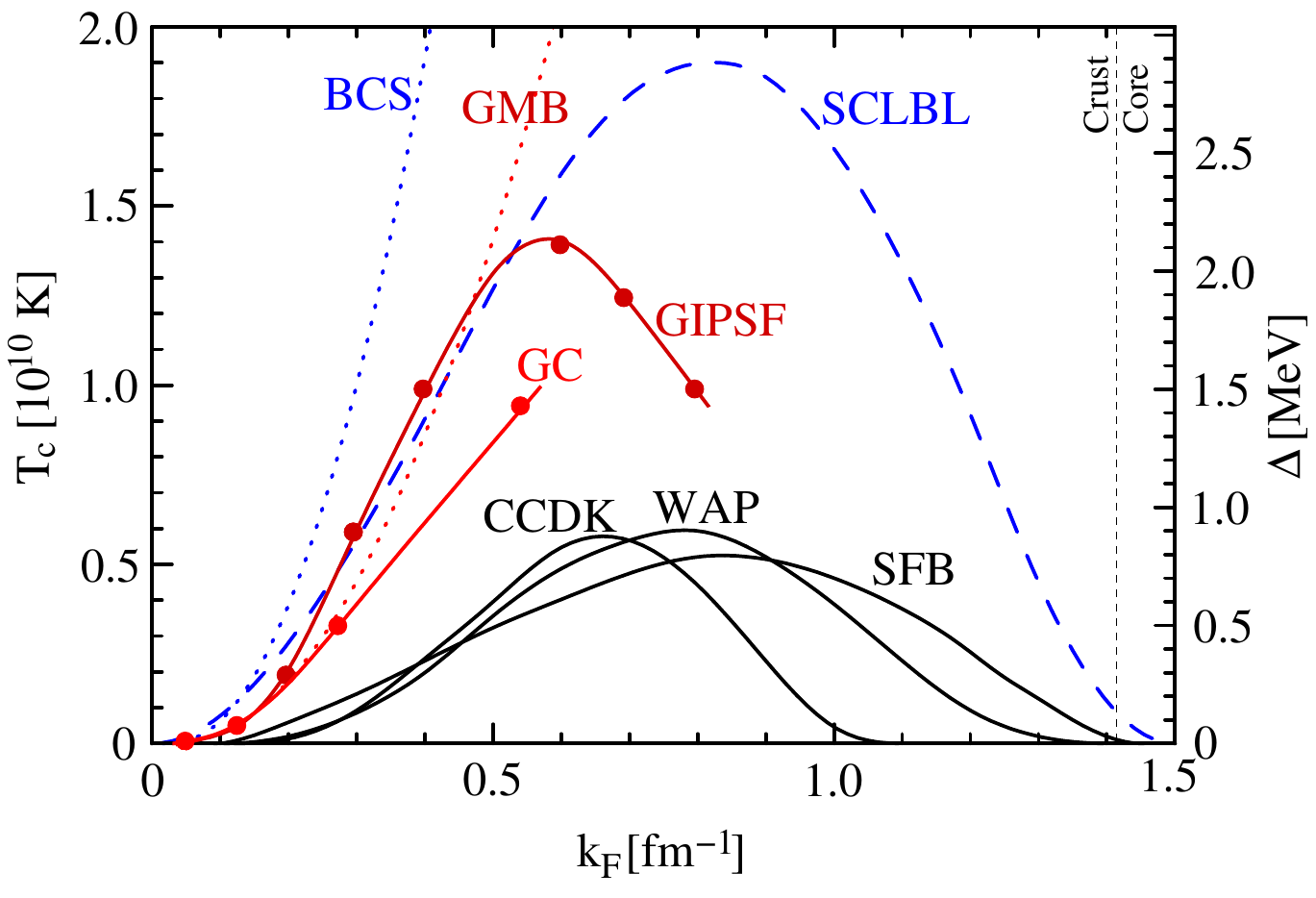}
\end{center}
\caption{
Some theoretical predictions of $T_c$ and $\Delta$, vs neutron $k_F$, for the neutron $\Singlet$ gap in uniform pure neutron matter.
The value of $k_F$ corresponding to the transition from the crust to the core is indicated.
See text for description.}
\label{Fig:Tc_n1S0}
\end{figure*}

In \fig{Fig:Tc_n1S0}, we show sets of predicted $T_c$ for the neutron $\Singlet$ pairing
in uniform pure neutron matter.
The two dotted lines marked ``BCS" and ``GMB" show the simple analytical results of
\eq{Eq:Delta_BCS} and \eq{Eq:Delta_GMB}, respectively, with $a = -18.5$ fm and
$T_c = 0.56 \,\Delta(k_F)$.
Formally, these results are 
only valid when $|a|k_F \ll 1$, i.e., $k_F \ll 0.1$ fm$^{-1}$.
The curve ``SCLBL", from \cite{Schulze:1996ys}, illustrates the results of a numerical solution of the
gap equation, using the Argone $V_{14}$ N-N potential, within the BCS approximation.
The results merge with the ``BCS" curve in the weak coupling limit $k_F \rightarrow 0$.
Also shown are results from calculations 
that take into account more sophisticated medium effects,
including medium polarization (with different schemes):
``CCDK", from \cite{Chen:1993ly}, employed a variational method within the correlated basis functions scheme, 
``WAP", from \cite{Wambach:1993zr}, employed an extension of the induced interaction scheme,
whereas 
``SFB", from \cite{Schwenk:2003ve}, went beyond \cite{Wambach:1993zr} with renormalization group methods.
In line with the simple GMB result of \eq{Eq:Delta_GMB}, these model calculations
show that
polarization has a
screening effect that quenches the gap, by a factor $\sim 3$.
These three calculations yield 
some agreement, particularly in the predicted maximum value of $T_c$,
but with a non-negligible difference in the density dependence.
The other  
two curves show 
more recent results:
``GIPSF", from \cite{Gandolfi:2009dq}, utilizes the auxiliary-field diffusion Monte Carlo technique
while ``GC", from \cite{Gezerlis:2010fk}, 
stems from a Quantum Monte Carlo calculation.
These last two models result in gaps that are intermediate between the previous models and the BCS approximation;
moreover, they converge toward the GMB value when $k_F \rightarrow 0$.

Transcribed to 
the neutron star context, the range of Fermi momenta for which
these neutron $\Singlet$ gaps are non vanishing 
corresponds mostly to the dripped neutrons in the inner crust.
The presence of nuclei, or nuclear clusters in the pasta phase, may modify 
the sizes of these gaps from their 
values in uniform matter.
The coherence length $\xi$ of the dripped neutrons is larger than
the sizes of 
nuclei, leading to proximity effects.
This issue has received some attention, see, e.g., \cite{Vigezzi:2005kx,Baldo:2005vn,Margueron:2012ys,Chamel:2012zr},
and position dependent gaps, from inside to outside of nuclei, have been calculated.
However, in most of the crust $\xi$ is
smaller than 
the internuclear distance, 
and the size of the gap far outside the nuclei
is close to its value in uniform matter. 

\subsubsection*{The isotropic $\Singlet$ proton gap}

\begin{figure}[t]
\begin{center}
\includegraphics[width=.60\textwidth]{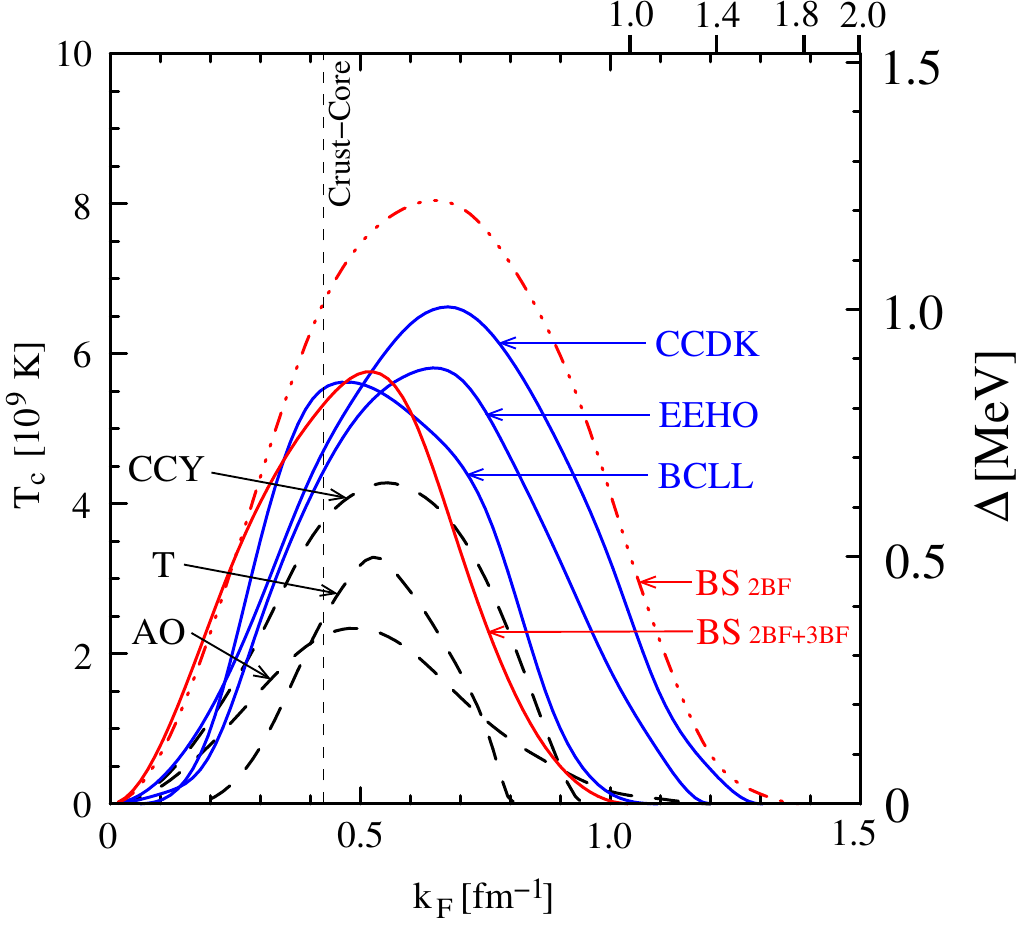}
\end{center}
\caption{
              Some theoretical predictions of $T_c$ and $\Delta$, vs proton $k_F$, for the proton $\Singlet$ gap
              in $\beta$-equilibrium uniform neutron-proton matter.
              The value of $k_F$ corresponding to the transition from the crust to the core is indicated:
              values on the right of this line correspond to the neutron star core but values on the left are
              not realized since protons in the crust are confined within nuclei which are finite size systems
              while this figure presents results for infinite matter.
              On the top margin are marked the values of the proton $k_F$ at the center of a $1.0$, $1.4$, $1.8$, and $2.0$
              $M_\odot$ star built with the APR EOS \cite{Akmal:1998fk}.
              See text for description.}
\label{Fig:Tc_p}
\end{figure}

The magnitudes of proton $\Singlet$  gaps 
are similar to those of neutrons,  
but with the important difference that, in the neutron star context in which beta equilibrium prevails, 
protons are immersed within the neutron liquid, and constitute only a small fraction
of the total baryon number
(3 to 20\% in the density range where they are expected to be superconducting).
Proton-neutron correlations cause the effective mass of the proton to be 
smaller than that of the neutron, a simple effect that reduces the size of the proton $\Singlet$ gap 
compared to that of the  neutron.  

Several theoretical predictions of $T_c$ for the proton $\Singlet$ gap are shown in \fig{Fig:Tc_p}:
``CCY" from \cite{Chao:1972kl},
``T" from \cite{Takatsuka:1973qf}, and
``AO" from \cite{Amundsen:1985ab}
that are among the first historical calculations,
whereas
``BCLL" from \cite{Baldo:1992kx},
``CCDK" from \cite{Chen:1993ly},
and ``EEHO" from \cite{Elgaroy:1996cr}
are more recent results.
All of these calculations were performed 
within the BCS approximation and very few works have gone beyond BCS
for the proton $\Singlet$ gap.
Among the latter, we show results from \cite{Baldo:2007fk}:
these authors used either only two body forces in the interaction kernel, curve ``BS$_\mathrm{2BF}$",
or two body forces supplemented by the inclusion of three body forces,  curve ``BS$_\mathrm{2BF+3BF}$"
which shows that three body forces are {\it repulsive} in the $\Singlet$ channel.
These ``BS" results also include effects of medium polarization.
Recall that for 
the $\Singlet$ pairing of neutrons in pure neutron matter,
polarization has a screening effect and quenches the gap. 
However, 
in neutron star matter, where the medium consists mostly of neutrons,
the strong $np$-correlations result in medium polarization 
inducing {\it anti-screening} \cite{Cao:2006uq} for the $\Singlet$ pairing of protons.  

\subsubsection*{The anisotropic $\Triplet$ neutron (and proton) gap}

The $\Singlet$ neutron gap vanishes 
at densities close to the crust-core transition and the
dominant pairing for neutrons in the core 
occurs in the mixed $\Triplet$ channel. 
Uncertainties in 
the actual size and the range of density 
in which this gap persists 
are, however, considerable.
As previously mentioned, a major 
source of uncertainty is the fact that even the best models of 
the N-N interaction {\em in vacuum} fail to reproduce the
measured phase shift in the $\triplet$ channel \cite{Baldo:1998zr}.
Also significant 
are the effects of the medium on the kernel and 3BF, 
even at the level of the BCS approximation.
It was found in \cite{Zhou:2004fv} that 3BF 
at the Fermi surface are 
strongly {\it attractive} in the $\Triplet$ channel in spite of being repulsive in the bulk.
Moreover, due to medium polarization a long-wavelength tensor force appears  that is not
present in the 
interaction {\em in vacuum} and results in a strong suppression of the gap
\cite{Schwenk:2004ve}.

\begin{figure}[htb]
\includegraphics[width=.99\textwidth]{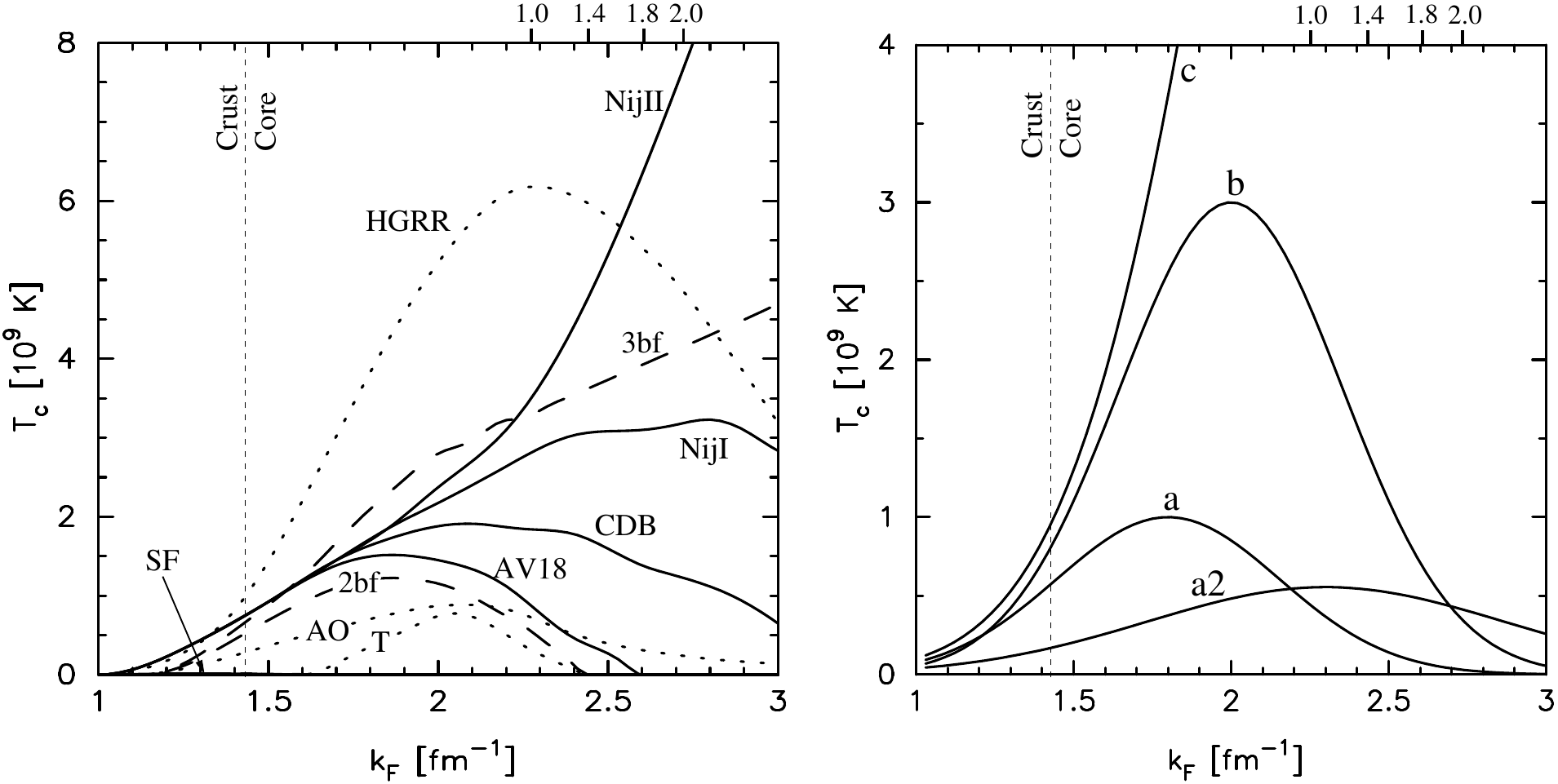}
\caption{Left panel: some theoretical predictions of $T_c$ for the neutron $\Triplet$ gap
              in uniform pure neutron and $\beta$-equilibrium matter.
              See text for description.
              Right panel: some phenomenological models of $T_c$ for the neutron $\Triplet$ gap
              used in neutron star cooling simulations.
              Models ``a'', ``b'', and  ``c'' are from \cite{Page:2004zr} and \cite{Page:2009qf}, model  ``a2'' from \cite{Page:2011ys}.
              On the top margin are marked the values of $k_{Fn}$ at the center of a $1.0$, $1.4$, $1.8$, and $2.0$
              $M_\odot$ star built with the APR EOS \cite{Akmal:1998fk}.}
\label{Fig:Tc_n}
\end{figure}

Figure~\ref{Fig:Tc_n} shows examples
of theoretical predictions of 
$T_c$ for the neutron $\Triplet$ gap.
The three dotted lines show
some of the first published models:
``HGRR'' from \cite{Hoffberg:1970hc}, ``T'' from \cite{Takatsuka:1972ij} and ``AO'' from \cite{Amundsen:1985nx}.
The four continuous lines show results of models from \cite{Baldo:1998zr} 
calculated using 
the Nijmegen II (``NijII''), Nijmegen I (``NijI''), CD-Bonn (``CDB''), and Argonne $V_{18}$ (``AV18'') potentials
(displayed values are taken from the middle panel of Figure~4 of \cite{Baldo:1998zr}).
The results of these  four models start to diverge at $k_{F n}$ above 1.8 fm$^{-1}$ and illustrate the failure of
all four N-N interactions models to fit the $\triplet$ laboratory phase-shifts above $E_\mathrm{lab} \simeq 300$ MeV.
All of these calculations were performed for pure neutron matter using the BCS approximation.

In the case of the $\Singlet$ gap, medium polarization is known to result in screening and to reduce the size of the gap.
In the case of a $\triplet$ gap, polarization with central forces is expected to result in anti-screening and to increase 
the size of the gap.
However, Schwenk \& Friman \cite{Schwenk:2004ve} showed that spin-dependent non-central forces
 do the opposite and strongly screen the coupling in the $\triplet$ channel, resulting in a $T_c$ lower than
$10^7$ K: this ``SF'' value is indicated in the figure by an arrow.

The two dashed lines in \fig{Fig:Tc_n} present results from \cite{Zuo:2008zr} where the ``2BF'' model
only considers 2-body forces (from the Argonne $V_{18}$) while the 
``3BF" model includes a meson exchange model 3-body force:
the result is a growing $\Triplet$ gap which shows no tendency to saturate at high density.
This work, for $\beta$-equilibrium matter, moreover emphasized the importance of the proton component.

Other delicate issues are the effect of the proton contaminant and the likely development of a
$\pi^0$ condensate\footnote{In the presence of a charged $\pi^-$ condensate a new Urca neutrino
emission pathway is open, see Table~\ref{Tab:Nu}. 
The development of a neutral $\pi^0$ condensate has, however, little effect on neutrino emission.}
which also strongly affects the size of the neutron  (and proton) gap(s).

In summary, 
the size and extent in density of the neutron $\Triplet$ gap in the neutron star core are poorly known.
Given these large uncertainties in 
the size of the neutron $\Triplet$ gap (about three orders of magnitude)
and the fact that neutrino emissivity is suppressed by 
an exponential Boltzmann-like factor, this gap
is often considered as a free parameter in neutron star cooling models.
The extreme sensitivity of the cooling history on the size of this gap can be utilized to 
one's advantage by inverting the problem,
as it may allows us to {\em measure} it by fitting theoretical models to observational data \cite{Page:1992nx}.
The right panel of \fig{Fig:Tc_n} presents the phenomenological neutron $\Triplet$ gaps
used in 
cooling calculations in a later section.

In the case of protons, 
their $\Triplet$ gaps have generally been overlooked 
due to their small effective masses, 
and
considered to be likely negligible \cite{Takatsuka:1997ff}.
However, in view of the strong enhancing effect of the 3BF on the neutron $\Triplet$ gap, this issue has been reconsidered 
in \cite{Zhou:2004fv} where it was shown that the proton $\Triplet$ gap can be sizable.

\subsubsection*{Hyperon gaps and nucleon gaps in hyperonic matter}

Many calculations of dense matter indicate that strangeness-bearing hyperons 
will be present in neutron star matter once the neutron chemical potential exceeds the 
rest masses of hyperons \cite{Prakash:1992vn} (see Sec.~\ref{Sec:NS_interior}).
In the likely presence of hyperons (denoted by Y) arise the issues of, first, the effect of their presence on the nucleon gaps, and second,
the possibility of hyperon pairing.
Nucleon gaps in the presence of hyperons have been studied in \cite{Zhou:2004fv} and \cite{Chen:2008mi}:
depending on the 
N-N, N-Y, and Y-Y interaction models employed, 
the nucleon gaps may be either enhanced or reduced 
by the presence of hyperons.
\\

Since the suggestion in \cite{Prakash:1996fk} and the first detailed work of \cite{Balberg:1998dz}, 
hyperon gaps have attracted some attention.
All uncertainties present in the nucleon case immediately  translate 
to the hyperon case. 
An additional problem is that very little 
is 
known about 
hyperon-hyperon interactions\footnote{Some experimental information is
available from hypernuclei~\cite{Gal05,Hashimoto06} and hadronic
atoms~\cite{Batty97}, but the data do not yet uniquely determine
the hyperon-nucleon or hyperon-hyperon interaction. Future work 
in lattice QCD~\cite{Beane12} may prove fruitful, but current results
are limited to unphysically large pion masses. },
which is generally guessed from theory
by extrapolation from the N-N interaction.
At densities not much larger 
than their 
threshold densities for appearance, hyperons have low enough concentration that they will
pair in the $\Singlet$ channel.  In both cases of $\Lambda$'s and $\Sigma^-$'s, 
the estimated gap sizes are 
similar to those of  
nucleons.
We refer the reader to \cite{Vidana:2004qa,Takatsuka:2006fu,Wang:2010kl} and references therein for details.

At very high densities, the presence of deconfined quark matter is also likely.
Quarks are expected to pair and form a color superconductor.
This subject
has developed into a field of 
its own, and we dedicate 
Sec.~\ref{Sec:Quarks} for a brief account and refer the reader to other articles in this monograph.

\subsection{Effects of Pairing for Neutron Star Cooling}
\label{Sec:Pairing_Effects}

The occurrence of pairing leads to three important effects of relevance to neutron star cooling:
\begin{description}
\item[\rm A)] 
Alteration, and possible strong suppression when $T \ll T_c$, of the specific heat $c_V$ of the paired component.
\item[\rm B)] 
Reduction, and possible strong suppression when $T \ll T_c$, of the emissivity $\epsilon_\nu$ of the neutrino processes 
the paired component is involved in.
\item[\rm C)] 
Triggering of the ``Cooper pair breaking and formation'' (PBF) with concomitant neutrino pair emission  which is  
very efficient in the case of spin-triplet pairing.
\end{description}
These effects are direct consequences of the development of the energy gap $\Delta(\bm k)$ 
and the resulting two branches in the quasi-particle spectrum, \eq{Eq:omega_paired}.
The gap severely limits the available phase space when $T \ll T_c$ and the spectrum is usually
treated in the effective mass approximation with an angle averaged gap, see \eq{Eq:D_ave},
\ba
\epsilon(\bm k) = \pm \sqrt{[\hbar v_F (k-k_F)]^2 + \overline{\Delta}(k_F,T)} \,,
\label{Eq:P_spectrum}
\ea
where $v_F \equiv \hbar k_F/m^*$.

\begin{figure*}[h]
\begin{center}
\includegraphics[width=.80\textwidth]{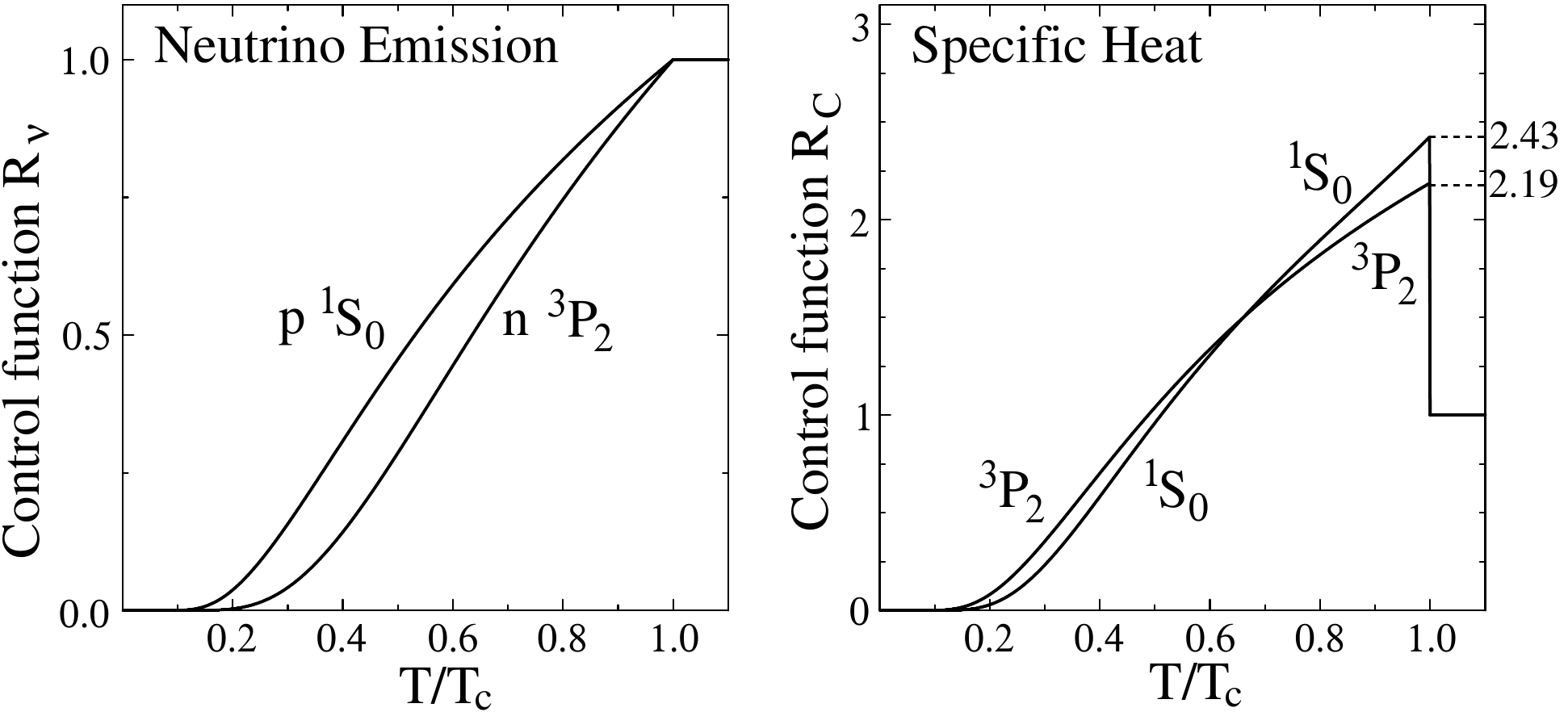}
\end{center}
\caption{Control functions for neutrino emission from the modified Urca process 
(as, e.g., $n+n \rightarrow n+n + \nu \overline\nu$)
(left panel) and the specific heat (right panel), in the presence of $\Singlet$ pairing and
$\triplet$ in the B phase (see \eq{Eq:Bphase}),
from the analytical fits of \cite{Levenfish:1994vn} and \cite{Yakovlev:1995ly}.
}
\label{Fig:Control}
\end{figure*}

In cooling calculations, these effects are introduced through ``control functions'':
\be
c_V \longrightarrow R_c \, c_V
\;\;\;\; \text{and} \;\;\;\;
\epsilon_\nu^X \longrightarrow R_X \, \epsilon_\nu^X \; .
\label{Eq:suppr}
\ee
There is a large family of such control functions for the various types of pairing and the numerous neutrino processes ``X''.
For nodeless gaps, the $R$'s are Boltzmann-like factors $\sim \exp[-2D(T)/k_B T]$ and result in a
strong suppression when $T \ll T_c$, whereas
for gaps with nodes the suppression is much milder.
Regarding the specific heat, there is a sudden increase, by a factor $\sim 2.4$ at $T=T_c$, followed by a reduction
at lower $T$.
Examples of such control functions are shown in \fig{Fig:Control}.

The effect C, neutrino emission from the formation and breaking of Cooper pairs \cite{Flowers:1976vn,Voskresenskii:1986nx},
can be interpreted as an inter-band transition (as, e.g., $n \rightarrow n + \nu \overline\nu$)
where a neutron/proton quasiparticle from the upper ($+$) branch of the spectrum of \eq{Eq:P_spectrum}
falls into a hole in the lower ($-$) branch.
Such a reaction is kinematically forbidden by the excitation spectrum of the normal phase, \eq{Eq:epsilon_normal}, 
but becomes possible in the presence of an energy-gap, \eq{Eq:P_spectrum}.
This process is described in more detail in Sec.~\ref{Sec:PBF}. 
The resulting emissivity can be significantly larger than that of the modified Urca process 
(as, e.g., $n + n \rightarrow n +n + \nu \overline\nu$) in the case of spin-triplet pairing.

\section{Superfluidity in Dense Quark Matter}
\label{Sec:Quarks}

The central densities of neutron stars can exceed the nuclear density
$\rhonuc \sim 2.7\times 10^{14}~{\rm g~cm^{-3}}$ by significant
amounts. At sufficiently high densities, a description of neutron star
interiors in terms of nucleons becomes untenable and sub-nucleonic
degrees of freedom, namely quarks, must be invoked. Interactions between quarks
is fundamentally grounded in Quantum Chromodynamics (QCD), the theory of strong
nuclear interactions. The theory has a gauge symmetry based on the Lie
group SU(3), and the associated charge is referred to as
``color''. QCD is asymptotically free: interactions between quarks
mediated by gluons become weak at short distances, or equivalently,
high densities. At low densities, strong interactions ``confine''
quarks into neutrons and protons which are color neutral. Asymptotic
freedom guarantees that, at some large density, the ground state of
zero-temperature matter will consist of nearly-free, ``deconfined''
quarks~\cite{Collins75}.

QCD has been amply tested by experiments at high energies where
asymptotic freedom has been confirmed \cite{Wilczek05}. 
Lattice-gauge calculations of
hadron masses, and of baryon-free matter at finite temperature, have
made enormous strides in recent years \cite{proc11}.  However, first-principle QCD
calculations for finite baryon density have been stymied due to the
fermion-sign problem in lattice gauge calcuations. While many guesses
are available, perturbation theory is unable to accurately predict the
density at which the deconfinement phase transition occurs. However,
it is possible that the phase transition occurs at a density lower
than the central density of some (or even all) neutron stars. In that
case, at least some neutron stars will contain deconfined quark
matter.  Such objects are referred to as hybrid quark stars.  In their
cores, up ($u$), down ($d$), and strange ($s$) quarks are the
principal degrees of freedom, the other three quark flavors (charm
($c$), bottom ($b$), and top ($t$)) being excluded because of their
large masses.  It is theoretically possible that the energy of
zero-pressure strange quark matter has a lower energy than Fe
\cite{Bodmer71,Witten84,Farhi84}, 
in which case a hybrid star would be metastable or unstable, and
nucleonic matter would spontaneously convert into strange quark
matter, creating pure quark stars that are self bound 
\cite{Alcock:1986ys,Haensel:1986zr}.
There is no experimental or observational evidence for
pure quark stars, however, and we do not consider them further in this
contribution.

\subsection{Pairing in Quark Matter}

Cooper pairing between quarks was first investigated in the late
1970s~\cite{Barrois77,Bailin79}. As all quarks are charged, pairing
between quarks is often referred to as color superconductivity because
the paired phase breaks the SU(3) gauge symmetry of QCD. First
estimates of pairing gaps in quark matter were of order 1 MeV. In this
case, neutron stars with deconfined quark matter cool very rapidly
through the quark direct Urca processes (see Table \ref{Tab:Nu}). Because some older neutron
stars are observed to be relatively warm, this naturally implies that
not all neutron stars can contain quark matter.

This situation changed drastically with the
discovery~\cite{Alford98,Rapp98} that color superconductivity implies
gaps as large as 100 MeV (see a recent review in \cite{Alford08}). 
These works suggested two possibilities: either
the so-called ``color-flavor-locked'' (CFL) phase in which all nine
combinations of flavor (up, down, strange) and color (red, green,
blue) participate in pairing, or the ``2SC'' phase where only four of
the nine combinations pair (corresponding to up and down quarks which
are either red or green).

\subsection{Theoretical Descriptions of Dense Quark Matter}

There are several formalisms which have been applied to describe color
superconducting quark matter. High-density effective theories (HDET)
were first developed in 1990s~\cite{Polchinski92,Shankar94} and then
developed further for color superconductivity in
\cite{Hong00,Schafer03}. 
To construct an effective theory, one
begins with the QCD Lagrangian, rewrites it in terms of a $1/\mu$
expansion ($\mu$ being the chemical potential), and then integrates
out hard gluons and fermionic modes corresponding to the Dirac sea.

Another commonly used alternative consists of using
Nambu--Jona-Lasinio (NJL)~\cite{Nambu61} models. The original NJL
model was a theory of strong interactions before the advent of QCD.
The four-fermion interaction of the NJL Lagrangian bears close
resemblance to that in the BCS theory of superconductivity and gives
rise to analogous effects.  Originally framed in terms of nucleon
fields, quartic interactions serve to give the nucleon its mass
through a self-energy generated by the formation of a condensate.
Modern versions involve quark fields that develop a ``quark
condensate'' which is then related to the mass of constituent
quarks~\cite{Hatsuda94}.  NJL models for color superconductivity
presume that gluonic degrees of freedom have been integrated out
resulting in point-like couplings between quarks.  For a review of the
NJL model applied to dense quark matter see 
\cite{Buballa05}.
Confinement is sometimes implemented by the addition of Polyakov loop
terms giving rise to ``PNJL'' models. HDET and NJL methods give
qualitatively similar results, but the NJL Lagrangian is a bit more
transparent, so we describe some of its details here. A chiral SU(3)
Lagrangian with superconducting quarks (adapted from~\cite{Steiner02})
is
\begin{eqnarray}
{\cal L} &=& \bar{q}_{i \alpha} \left( i \partial_\mu \gamma^{\mu}
\delta_{ij} \delta_{\alpha \beta} - m_{ij} \delta_{\alpha \beta} -
\mu_{ij,\alpha\beta} \gamma^{0} \right) q_{j \beta} \nonumber \\ && +
G_S \sum_{a=0}^{8} \left[ \left(\bar{q} \lambda_f^{a} q \right)^2 +
  \left(\bar{q} i \gamma_5 \lambda_f^a q \right)^2\right] \nonumber
\\ && + G_{\Delta} \sum_k \sum_{\gamma} \left(\bar{q}_{i \alpha}
\epsilon_{ijk} \epsilon_{\alpha\beta\gamma} q^{C}_{j \beta} \right)
\left(\bar{q}^{C}_{i^{\prime} \alpha^{\prime}}
\epsilon_{i^{\prime}j^{\prime}k^{\prime}}
\epsilon_{\alpha^{\prime}\beta^{\prime}\gamma^{\prime}} q_{j^{\prime}
  \beta^{\prime}}\right) \nonumber \\ && + G_{\Delta} \sum_k
\sum_{\gamma} \left(\bar{q}_{i \alpha} i \gamma_5 \epsilon_{ijk}
\epsilon_{\alpha\beta\gamma} q^{C}_{j \beta} \right)
\left(\bar{q}^{C}_{i^{\prime} \alpha^{\prime}} i \gamma_5
\epsilon_{i^{\prime}j^{\prime}k^{\prime}}
\epsilon_{\alpha^{\prime}\beta^{\prime}\gamma^{\prime}} q_{j^{\prime}
  \beta^{\prime}}\right) \,,
\label{eq:cfl_lag}
\end{eqnarray}
where Roman indices are for flavor and greek indices are for color,
except for $a$ which enumerates the SU(3) matrices, $m_{ij}$ is the
quark mass matrix, $\mu_{ij,\alpha\beta}$ is the chemical potential
matrix, $q$ is the quark field, $q^{C} = C \bar{q}^{T}$, and
$\epsilon$ is the Levi-Civita tensor. The first term is the Dirac
Lagrangian which describes free relativistic massive quarks at finite
density. The second term is a combination of four quark fields which
model non-superfluid quark-quark interactions and obeys the $SU(3)_L
\times SU(3)_R$ chiral symmetry present in QCD. The third and fourth
terms, which give rise to color superconductivity, are the chirally
symmetric analog of the second term in the quark-quark channel. QCD
breaks 
$U_A(1)$ symmetry, and so these four-fermion
interactions can be supplanted by six-fermion interactions in order to
do the same~\cite{tHooft76}.

The first step in obtaining the thermodynamic potential in the
mean-field approximation is to replace the quark bilinears $\bar{q}_i
q_i$ and $\bar{q}_{i \alpha} i \gamma_5 \epsilon^{ijk}
\epsilon^{\alpha \beta \gamma} q^{C}_{j \beta}$ with their
ground-state expectation values. The former is the quark condensate
associated with the breaking of chiral symmetry and the latter gives
rise to the superconducting gap, $\Delta^{k \gamma}$. Having made this
replacement, the non-constant terms in the Lagrangian take the form
$\bar{q} M q$, where $M$ is a matrix representing the inverse
propagator. This matrix can be diagonalized in the standard way to
obtain the individual quark dispersion relations and the thermodynamic
potential
\begin{eqnarray}
\Omega &=& -2 G_S \sum_{i=u,d,s} \left< \bar{q}_i q_i \right>^2
-\sum_k \sum_{\gamma} \frac{\left|\Delta^{k \gamma}\right|^2}{4
  G_{\Delta}} \nonumber \\ && - \int \frac{d^3 p}{\left(2
  \pi\right)^3} \sum_i \left[\frac{\lambda_i}{2}+T \ln
  (1+e^{-\lambda_i/T}) \right] \,,
\end{eqnarray}
where $\lambda_i$ gives the energy eigenvalues and $i$ runs over 3
flavors, 3 colors, and the Dirac indices (36 total). In this
formulation, there is a manifest parallelism between the quark
condensates, $\left<\bar{q} q\right>$, and the superconducting gaps.
The minimum of the thermodynamic potential with respect to the
superconducting gap gives the gap equation, and the minimum of the
thermodynamic potential with respect to the quark condensates gives
the ``mass gap'' equation, i.e. the equation which controls the
dependence of the dynamically generated quark masses.

The energy eigenvalues $\lambda_i$ cannot be computed analytically at
all densities, except in two limiting situations. At low densities,
where chiral symmetry is spontaneously broken, the gaps are zero. In
this case, the quark dispersion relations are $\sqrt{p^2+m^{* 2}_i}\pm
\mu$, where $m^{*}_i=m_i-4 G_S\left<\bar{q_i} q_i\right>$ are
effective masses. The corresponding quark condensates are given
by~\cite{Buballa98}
\begin{equation}
\left<\bar{q_i}{q_i}\right> = - \frac{3}{\pi^2}
\int_{p_{Fi}}^{\Lambda} p^2 dp \frac{m_i^{*}}{\sqrt{p^2+m_i^{* 2}}}
\,.
\end{equation}
At high densities, the gaps are larger than the quark masses, hence
the latter do not play significant roles and can be ignored. With
progressively increasing density color-flavor locking becomes
increasingly perfect, hence the name the CFL phase. In this phase, we
can assume flavor symmetry and 8 of 9 quarks (3 colors times 3
flavors) have the dispersion relation $\sqrt{(p-\mu)^2+\Delta^2}$,
while the remaining quark has the dispersion relation
$\sqrt{(p-\mu)^2+4 \Delta^2}$. In general, these properties are
coupled so that both the masses and the gaps appear in the dispersion
relations in a nontrivial fashion.

\begin{figure}[hbt]
\begin{center}
\vspace*{-1.25cm}
\includegraphics[width=0.85\textwidth]{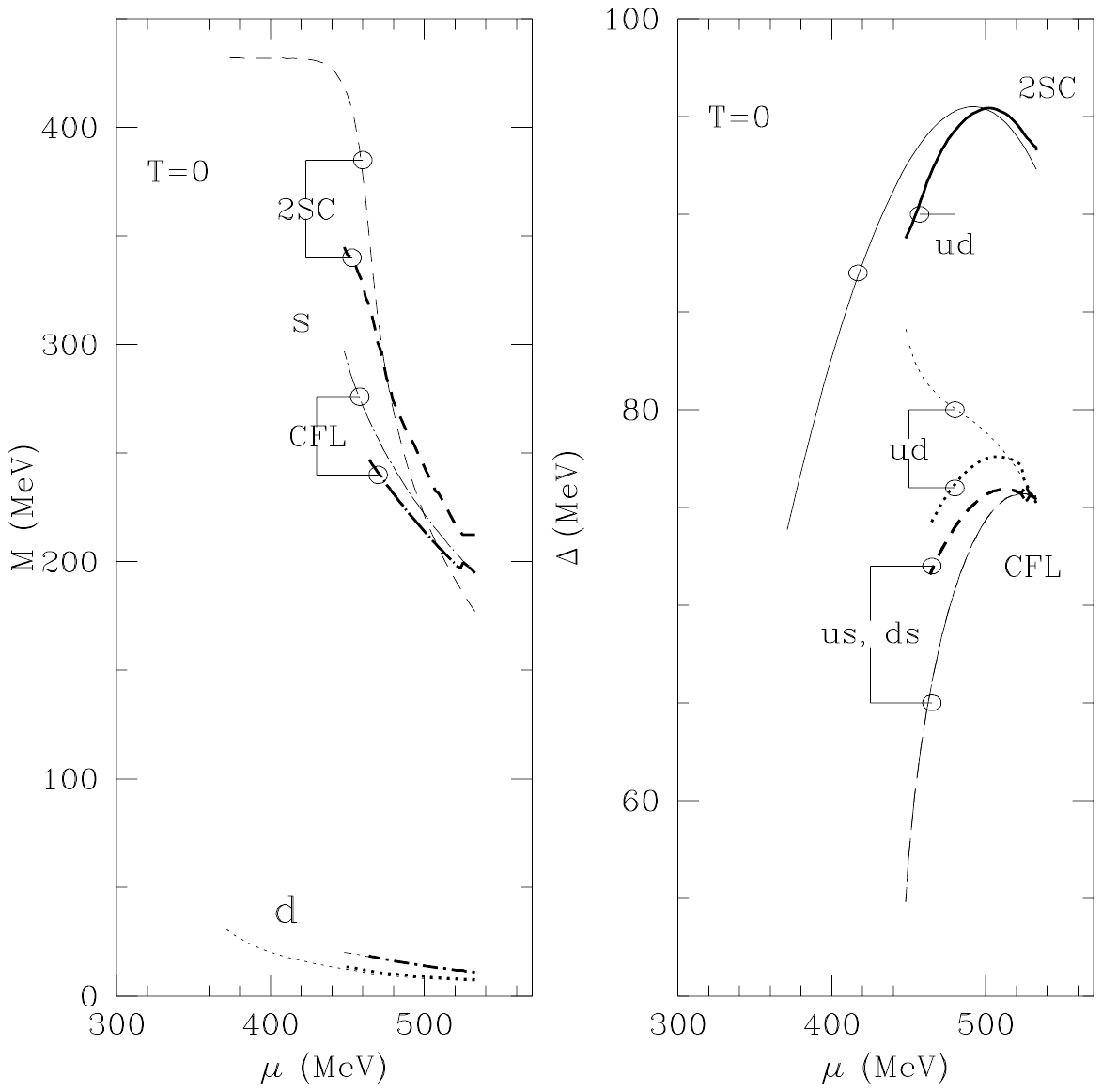}
\vspace*{-2.75cm}
\end{center}
\caption{Down and strange dynamical quark masses (left panel) and
  superconducting gaps (right panel) as a function of density from
  \cite{Steiner02}. The abscissa shows  the quark chemical potential
  $\mu$. The thin (bold) curves show values when  color
  neutrality is not (is) enforced. Different results are obtained in
  the CFL and 2SC phases as shown. 
  }
\label{Fig:csc_gaps}
\end{figure}

Results of calculations based on the above model are displayed in
\fig{Fig:csc_gaps} using \eq{eq:cfl_lag}. The dynamically
generated quark masses are larger than the current quark masses at low
density where chiral symmetry is spontaneously broken. In the CFL
phase, the superconducting gaps form among all three combinations of
unlike flavors, up-down, up-strange, and down-strange. In the 2SC
phase, the only pairing is between up and down quarks. This model
exhibits a first-order phase transition between the gapped and
ungapped phases, so the gaps do not continuously go to zero at low
densities. The decrease in the gaps as a function of increasing
density or large values of the quark chemical potential $\mu$ is an
artifact of the ultraviolet cutoff (a necessity imposed by the
nonrenormalizable Lagrangian). HDET models show that the gaps increase
with increasing $\mu$.

\subsection{The Many Phases of Quark Matter}

\begin{figure}[hbt]
\vspace*{-3.0in}
\begin{center}
\includegraphics[width=0.9\textwidth]{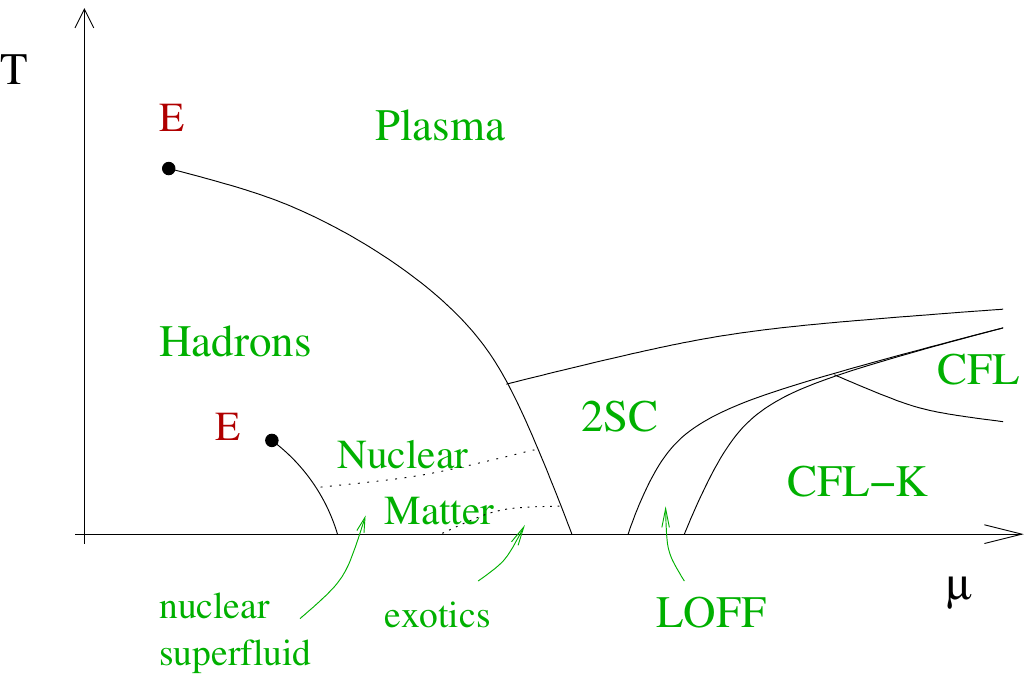}
\end{center}
\caption{A schematic model QCD phase diagram, adapted from
  \cite{Schafer03b}.}
\label{Fig:CFL}
\end{figure}

There are many different possible pairing configurations in
addition to the CFL and 2SC phases described above, including gapless
phases~\cite{Shovkovy03}, and color-spin locked pairing. Color
superfluids also admit a new set of Goldstone bosons associated with
flavor rotations of the pairing condensate which have a similar group
structure to the pseudoscalar Goldstone bosons in QCD ($\pi,K$, etc.).
These bosons can condense~\cite{Bedaque02}, forming a new phase of
superconducting quark matter. The most common is the ``CFL-K'' phase
which contains CFL quarks with a $K$ meson condensate. All of these
phases have their own associated exotic neutrino emissivities,
including their own associated quark PBF neutrino cooling processes. A
caricature phase diagram is shown in \fig{Fig:CFL}.

For densities near the deconfinement phase transition, the ground
state of the quark superfluid may be similar to the
Larkin-Ovchinnikov-Fulde-Ferrell~\cite{Fulde64,Larkin65} (LOFF)
pairing observed in condensed matter systems. LOFF pairing occurs when
two species participating in a pairing interaction have different
Fermi momenta thus creating Cooper pairs with nonzero momentum. This
pairing configuration breaks translational symmetry and encourages the
formation a crystal. This creates a novel mechanism for pulsar
glitches: the superfluid vortices pin to the crystalline part of the
quark phase and during a glitch event the vortices move outwards by
unpinning and repinning to the lattice~\cite{Alford01}. This mechanism
has not been either verified or ruled out by the data.

\subsection{Detecting Quark Matter}

The appearance of quark degrees of freedom often lowers the pressure
at high densities, yielding neutron stars with smaller radii and
smaller maximum masses compared to those in which quarks are absent.
This is not guaranteed, however, as quark-quark interactions are
sufficiently uncertain that quark matter can be nearly
indistinguishable from matter consisting entirely of neutrons,
protons, electrons and muons~\cite{Alford04}. The presence (or absence) of
quark matter will be difficult to determine from observations of neutron star
structure alone. The neutron star mass-radius relation is dependent on
the pressure of matter at a given energy density, but is insensitive
to the nature of the particular degrees of freedom which provide that
pressure. For this reason, it is natural to look to neutron star
cooling to discern the composition of a neutron star's core.  This topic
will be considered in Sec.~\ref{Sec:Quark_cool}.

\section{Neutrino Emission Processes}
\label{Sec:Neutrinos}

The thermal evolution of neutron stars with ages $\simle 10^5$ yrs is
driven by neutrino emission.  We will here briefly describe the
dominant processes; the interested reader can find a detailed
description in \cite{Yakovlev:2001dq} and an alternative point of view
in \cite{Voskresensky:2001cr}.  Table~\ref{Tab:Nu} presents a short
list of neutrino processes with estimates of their emissivities.  Most
noticeable is the clear distinction between processes involving 5
degenerate fermions with a $T^8$ dependence, which are labeled as
``slow", and those with only 3 degenerate fermions with a $T^6$
dependence, which are several orders of magnitude more efficient and
labeled as ``fast".  The difference in the $T$ dependence is important
and is simply related to phase space arguments which are outlined in
Sec.~\ref{Sec:nuT}.  The last subsection, \S~\ref{Sec:PBF}, describes
the ``PBF`''(Pair Breaking and Formation) process , which can provide
a signal of the onset of a pairing phase transition that may be
directly observable (and has likely been recently observed as
described in Sec.~\ref{Sec:CasA}).

\subsection{The Direct Urca Process}

The simplest neutrino emitting processes are \eq{Eq:DU1} and \eq{Eq:DU2} (see also  Table~\ref{Tab:Nu}),   
which collectively are generally referred to as the nucleon {\em direct Urca}
(``DU'' or ``DUrca") cycle.  By the condition of $\beta$-equilibrium, both
reactions naturally satisfy energy conservation, but momentum
conservation is more delicate.  Due to the high degree of degeneracy,
all participating particles have momenta $p(i)$ equal (within a small
$T \ll T_F$ correction) to their Fermi momenta $p_F(i)$.  As $p_F(i)
\propto n_i^{1/3}$ and $n_p \sim n_e \ll n_n$ in neutron star matter,
momentum conservation is not {\em a priori} guaranteed.  In the
absence of muons and hence with $n_p = n_e$, the ``triangle rule'' for
momentum conservation requires that the proton concentration $x_p >
1/9 \simeq 11$\%, whereas at $\rho \sim \rhonuc$ we have $x_p \simeq
4$\%.  In the presence of muons, which appear just above $\rhonuc$,
the condition is stronger and one needs $x_p$ larger than about 15\%
\cite{Lattimer:1991kx}.  The proton fraction $x_p$ grows with density
(see \eq{betaeq}), its growth being directly determined by the
growth of the nuclear symmetry energy, so that the critical proton
fraction for the DU process is likely reached at some supra-nuclear
density \cite{Lattimer:1991kx}.  For the APR EOS \cite{Akmal:1998fk}
that we will frequently use, the corresponding critical neutron star
mass for the onset of the nucleon DU process is $1.97 \Msol$, but other
EOSs can predict smaller critical densities and masses.

It should also be noted that the direct Urca process, and for that
matter, all the processes discussed in this section, can involve other
leptons.  Thus, for $\rho\simge \rhonuc$, where $\mu^-$ appear, one
also has
\be\label{muonurca} 
n\longrightarrow p+\mu^-+\bar\nu_\mu,\qquad p+\mu^-\longrightarrow n+\nu_\mu.
\ee
\begin{table}
\begin{tabular}{llrc} 
\hline 
     Name           &\hspace*{1.5cm}Process&Emissivity\,\,\,\,& \hspace*{-.2cm}Efficiency\\ 
&&\hspace*{-.0cm}erg cm$^{-3}$ s$^{-1}$&\\
\hline 
\vspace{3pt}
\parbox[c]{2.6cm}{Modified Urca\\ (neutron branch)} &
$\hspace*{-.21cm}\begin{cases} n+n^\prime \rightarrow p+n^\prime+e^-+\bar\nu_e \\ p+n^\prime+e^- \rightarrow n+n^\prime+\nu_e \end{cases}$  & 
$\hspace*{-.4cm}\sim 2\!\times\! 10^{21}  R  T_9^8$ \rule[-0.2cm]{0.0cm}{0.85cm} & Slow 
\\
\vspace{3pt}
\parbox[c]{2.6cm}{Modified Urca\\ (proton branch)}  &
$\hspace*{-.21cm}\begin{cases} n+p^\prime \rightarrow p+p^\prime+e^-+\bar\nu_e \\ p+p^\prime+e^- \rightarrow n+p^\prime+\nu_e \end{cases}$  & 
$\hspace*{-.4cm}\sim 10^{21}  R  T_9^8$ & Slow \\
\vspace{3pt}
Bremsstrahlung          &
$\hspace*{-.21cm}
\begin{cases} 
n+n^\prime \rightarrow n+n^\prime+\nu+\bar\nu \\ 
n+p \rightarrow n+p+\nu+\bar\nu  \\
p+p^\prime \rightarrow p+p^\prime+\nu+\bar\nu\\
\end{cases}
\hspace*{-1cm}$                                     & 
$\hspace*{-.4cm}\sim 10^{19}  R  T_9^8$  & Slow \\
\vspace{3pt}
\parbox[c]{2.6cm}{Cooper pair}          &
$\hspace*{-.21cm}
\begin{cases}    
n+n \rightarrow [nn] +\nu+\bar\nu \\ 
p+p \rightarrow [pp] +\nu+\bar\nu 
\end{cases}$  & 
$\begin{array}{l}
\hspace*{-1cm}\sim 5\!\times\! 10^{21}  R  T_9^7 \\
\hspace*{-1cm} \sim 5\!\times\! 10^{19}  R  T_9^7 \rule[0pt]{0pt}{11pt}
\end{array}$ & Medium\\
\vspace{3pt}
\parbox[c]{2.6cm}{Direct Urca\\ (nucleons)} &
$\hspace*{-.21cm}\begin{cases} n \rightarrow p+e^-+\bar\nu_e \\ p+e^- \rightarrow n+\nu_e \end{cases}$              & 
$\hspace*{-.4cm}\sim 10^{27}  R  T_9^6$ & Fast \\
\vspace{3pt}
\parbox[c]{2.6cm}{Direct Urca\\ ($\Lambda$ hyperons)} &
$\hspace*{-.21cm}\begin{cases} \Lambda \rightarrow p+e^-+\bar\nu_e \\ p+e^- \rightarrow \Lambda+\nu_e \end{cases}$              & 
$\hspace*{-.4cm}\sim 10^{27}  R  T_9^6$ & Fast \\
\vspace{3pt}
\parbox[c]{2.6cm}{Direct Urca\\ ($\Sigma^-$ hyperons)} &
$\hspace*{-.21cm}\begin{cases} \Sigma^- \rightarrow n+e^-+\bar\nu_e \\ n+e^- \rightarrow \Sigma^-+\nu_e \end{cases}$              & 
$\hspace*{-.4cm}\sim 10^{27}  R  T_9^6$ & Fast \\
\vspace{3pt}
\parbox[c]{2.6cm}{Direct Urca\\ (no-nucleon)} &
$\hspace*{-.21cm}\begin{cases} \Lambda +e^-\rightarrow \Sigma^-+\nu_e \\ \Sigma^- \rightarrow \Lambda+e^-+\bar\nu_e \end{cases}$              & 
$\hspace*{-1cm}\sim2\!\times\! 10^{27}  R  T_9^6$ & Fast \\
\vspace{3pt}
\parbox[c]{2.6cm}{Direct Urca\\ ($\pi^-$ condensate)} &
$\hspace*{-.21cm}\begin{cases} n+<\pi^-> \rightarrow n+e^-+\bar\nu_e \\ n+e^- \rightarrow n+<\pi^->+\nu_e \end{cases}$ &
$\hspace*{-.4cm}\sim 10^{26}  R  T_9^6$  & Fast \\
\vspace{3pt}
\parbox[c]{2.6cm}{Direct Urca\\ ($K^-$ condensate)}   &
$\hspace*{-.21cm}\begin{cases}n+<K^-> \rightarrow n+e^-+\bar\nu_e \\ n+e^-\rightarrow n+<K^->+\nu_e \end{cases}$  &
$\hspace*{-.4cm}\sim 10^{25}  R  T_9^6$  & Fast \\
\hline 
\vspace{3pt}
\parbox[c]{2.6cm}{Direct Urca cycle\\ ($u-d$ quarks)} &
$\hspace*{-.21cm}\begin{cases} d \rightarrow u+e^-+\bar\nu_e \\u+e^- \rightarrow d+\nu_e \end{cases}$              & 
$\hspace*{-.4cm}\sim 10^{27}  R  T_9^6$ \rule[-0.2cm]{0.0cm}{0.85cm} & Fast \\
\parbox[c]{2.6cm}{Direct Urca cycle\\ ($u-s$ quarks)} &
$\hspace*{-.21cm}\begin{cases} s \rightarrow u+e^-+\bar\nu_e \\ u+e^- \rightarrow s+\nu_e \end{cases}$              & 
$\hspace*{-.4cm}\sim 10^{27}  R  T_9^6$ & Fast \\
\hline 
\end{tabular}
\caption{
A sample of neutrino emission processes. 
$T_9$ is temperature $T$ in units of $10^9$ K and the $R$'s are control factors
to include the suppressing effects of pairing (see Sec.~\protect\ref{Sec:Pairing_Effects}).
}
\label{Tab:Nu}
\end{table}

\subsection{The Modified Urca Process}

At densities below the threshold density for the nucleon DU process, where the DU process is forbidden at low temperatures, a variant of this process, 
the {\em modified Urca} (``MU" or ``MUrca") process (see Table \ref{Tab:Nu})
can operate, as advantage is taken of a neighboring nucleon in the medium \cite{Friman:1979bh} to conserve momentum.
As it involves the participation of five degenerate particles, the MU process is much less efficient than the DU process.
Unlike the nucleon DU process, which requires sufficient amount of protons, 
both branches of the MU process operate at any density when neutrons and protons are present.
\subsection{Bremsstrahlung}

Related to the MU processes is another class of processes, bremsstrahlung, made possible through neutral currents 
\cite{Flowers:1975qf}.  These differ from MU processes in that each reaction
results in the production of a $\nu\bar\nu$ pair, and the pair can have any neutrino flavor.   
Bremsstrahlung reactions
are less efficient, by about 2 orders of magnitude, than the MU processes, but may
make important contributions in the case that the MU process is suppressed by pairing of neutrons or protons. 
Bremsstrahlung involving electron-ion scattering is also an important source of neutrino emission in neutron star crusts:
\be\label{nucbrehms}
e^- + (A,Z) \longrightarrow e^- + (A,Z) +  \nu + \overline{\nu} \, ,
\ee
where (A,Z) designates the participating ion.

\subsection{Exotic Matter: Hyperons, Deconfined Quarks, Meson Condensates}

In the presence of hyperons, DU processes which are obvious generalizations of the nucleon-only
process, can also occur \cite{Prakash:1992vn} and several are displayed in Table~\ref{Tab:Nu}.
When they appear, the $\Lambda$'s initially have a density much smaller than that of the neutron
and hence
a smaller Fermi momentum. Consequently, momentum conservation in the $\Lambda$ DU cycle
is easily satisfied, requiring a $\Lambda$ concentration $x_\Lambda\sim 3$\%.
Notice that if the nucleon DU process is kinematically forbidden, the $\Sigma^-$ DU process
is also kinematically forbidden, whereas no-nucleon DU processes, of which
one example in shown in Table \ref{Tab:Nu},
{\it are} possible.  This particular no-nucleon DU process requires relatively low $\Lambda$ and $\Sigma^-$ threshold concentrations.  Other examples involving $\Sigma^-, \Sigma^0, \Sigma^+, \Xi^-$ and $\Xi^0$ hyperons are given in \cite{Prakash:1992vn}.

In deconfined quark matter, DU processes involving all three flavors are possible, as indicated in Table \ref{Tab:Nu}.  
Rates for these processes have been calculated by Iwamoto \cite{Iwamoto82}.

Although not shown in Table~\ref{Tab:Nu}, hyperons or quarks could
also be involved in MU-like and bremsstrahlung processes (a quark MU process would
involve an additional quark in the entrance and the exit channels, for example), but
with greatly reduced rates compared to their corresponding DU
processes and a $T^8$ dependence.
These processes are usually neglected since the DU processes are almost invariably
allowed in the presence of hyperons or quarks.

In the presence of a meson condensate, copious neutrino emission in the processes listed in Table \ref{Tab:Nu}
occurs \cite{Maxwell:1977zr,Brown:1988ly}.  As the meson condensate is
a macroscopic object, there is no restriction arising from momentum
conservation in these processes.

\subsection{Temperature Dependence of Neutrino Emission}
\label{Sec:nuT}

We turn now to briefly describe how the specific temperature
dependence of the neutrino processes described above emerges.
Consider first the simple case of the neutron $\beta$-decay.  The weak
interaction is described by the Hamiltonian ${\cal H}_I =
(G_F/\sqrt{2}) B_\mu L^\mu$, where $G_F$ is Fermi's constant, and
$L^\mu = \overline{\psi}_e \gamma^\mu (1-\gamma_5) \psi_\nu$ and
$B_\mu = \overline{\psi}_p \gamma_\mu (C_V 1-C_A \gamma_5) \psi_n$ are
the lepton and baryon weak currents, respectively.  In the non relativistic
approximation, one has $B^0 = \cos \theta_c \Psi^\dagger_p \Psi_n$ and
$B^i = -\cos \theta_c \, g_A \; \Psi^\dagger_p \, \sigma^i \, \Psi_n$
where $\theta_c$ is the Cabibbo angle and $g_A$ the axial-vector
coupling.  Fermi's Golden rule gives us for the neutron decay rate
\be
W_{i \rightarrow f} =
\int \!\!\!\! \int \!\!\!\! \int
\frac{d^3 p_\nu}{(2 \pi)^3} \frac{d^3 p_e}{(2 \pi)^3}
\frac{d^3 p_p   }{(2 \pi)^3}
 (2 \pi)^4 \delta^4(P_f-P_i) \cdot |M_{fi}|^2 \, ,
\ee
i.e., a sum of $(2 \pi)^4 \delta^4(P_f-P_i) \cdot |M_{fi}|^2$ over the phase space of all final states 
$f=(\vec{p}_{\overline{\nu}},\vec{p}_e,\vec{p}_p)$.
The integration gives the well known result
$W_\beta =G_F^2 \cos^2 \theta_c (1+3 g_A^2) m_e^5 c^4 w_\beta/( 2 \pi^3)$,
where $w_\beta \sim 1$ takes into account small Coulomb corrections.
This gives the neutron mean life, $\tau_n \simeq 15$ minutes.  Alternatively, a measurement of $\tau_n$ determines
$G_F$ (modulo $\cos \theta_c$ and $w_\beta$).

\begin{figure}[hbt]
\begin{center}
\includegraphics[width=.9\textwidth]{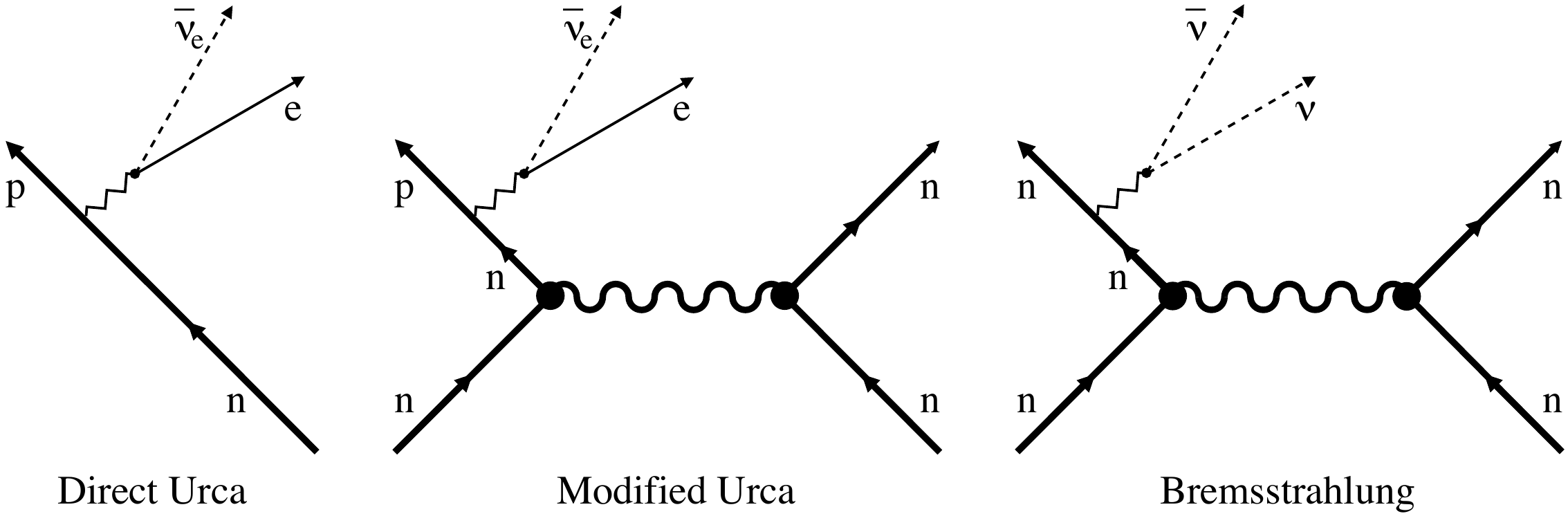}
\end{center}
\caption{Feynman diagrams for the indicated neutrino emitting processes.}
\label{Fig:Feynman}
\end{figure}

The emissivity $\epsilon^\mathrm{DU}$ of the DU process 
(the Feynman diagram for this process is shown in \fig{Fig:Feynman})
can be obtained by the same method, leading to
\be
\epsilon^\mathrm{DU} =
\int \!\!\!\! \int \!\!\!\! \int \!\!\!\! \int 
\frac{d^3 p_{\overline{\nu}}}{(2 \pi)^3} \frac{d^3 p_e}{(2 \pi)^3}
\frac{d^3 p_p   }{(2 \pi)^3} \frac{d^3 p_n}{(2 \pi)^3}
(1-f_e) (1-f_p) f_n  (2 \pi)^4 \delta^4(P_f-P_i) |M_{fi}|^2  E_\nu
\label{Eq:E_DU}
\ee
with an extra factor $E_\nu$ for the neutrino energy and the phase space sum now includes the initial $n$.
The $f_i$ terms, $f_i$ being the Fermi-Dirac distribution for particle $i$ at temperature $T$,
take into account:
(1) the probability to have a $n$ in the initial state, $f_n$, and
(2) the probabilities to have available states for the final $e$ and $p$, denoted by $(1-f_e)$ and $(1-f_p)$, respectively.
We do not introduce a Pauli blocking factor $(1-f_{\overline{\nu}})$ for the anti-neutrino as it is assumed
to be able to freely leave the star (i.e., $f_{\overline{\nu}}=0$).
When performing the phase space integrals, each degenerate fermion gives a factor $T$, as particles are
restricted to be within a shell of thickness $k_BT$ of their respective Fermi surfaces.  The  anti-neutrino phase space
gives a factor $T^3$. The factors $E_\nu$ is $ \sim T$ and the delta function $\delta^4(P_f-P_i)$ gives a factor
$T^{-1}$ from $\delta(E_f-E_i)$.
Altogether, we find that
\be
\epsilon^\mathrm{DU} \propto
T^3 \cdot T  \cdot T  \cdot T \cdot \frac{1}{T} \cdot (1)^2 \cdot T = T^6 \,,
\label{Eq:DU_T}
\ee
where the $(1)^2$ factor emphasizes that the squared matrix element $|M_{fi}|^2$ is T-independent. 
An explicit expression for the neutrino emissivity for the DU process can be found in \cite{Lattimer:1991kx}. 

Figure~\ref{Fig:Feynman} shows 
a Feynman diagram for the MU process.  There are two more such diagrams
in which the weak interaction vertex is attached to one of the two incoming legs.
In this case, the $T$-power counting gives
\be
\epsilon^\mathrm{MU} \propto
T^3 \cdot T  \cdot T  \cdot T \cdot T \cdot T \cdot \frac{1}{T} \cdot (1)^2 \cdot T = T^8 \,.
\label{Eq:MU_T}
\ee
In this case, the $|M_{fi}|^2$ 
involves two strong interaction vertices, connected by the wavy line in \fig{Fig:Feynman},
which is momentum independent and hence 
$T$-independent. 
The numerical difference $\sim 10^{-6} T_9^2$ between the MU and the DU rates, see Table~\ref{Tab:Nu},
comes mostly from the extra phase space limitation $\propto T^2$ from the two extra nucleons:
as a dimensionless factor it is $(k_BT/E_F)^2 \simeq 10^{-6} T_9^2$ with $E_F \simeq $ 100 MeV 
and $k_B 10^9$ K $\simeq 0.1$ MeV.
Notice that in the MU case, the internal neutron is off-shell by an amount $\simeq \mu_e$ which does not
introduce any extra $T$-dependence as we are working in the case 
$E_F(e) \simge 100$ MeV $\gg T$.
Reference \cite{Friman:1979bh} contains the expression from which neutrino emissivity from the MU process can be calculated.

Turning to the  $n-n$ bremsstrahlung process, 
one diagram is shown in \fig{Fig:Feynman} and there are three more diagrams with the weak interaction vertices attached 
to the other three external lines. The $T$-power counting now gives
\be
\epsilon^\mathrm{Br} \propto
T^3 \cdot T^3  \cdot T  \cdot T \cdot T \cdot T \cdot \frac{1}{T} \cdot \left(\frac{1}{T}\right)^2 \cdot T = T^8
\ee
with two $T^3$ factors for the neutrino pair. 
The factor  
$(T^{-1})^2$ arises from the matrix element as
the intermediate neutron is almost on-shell, with an energy deficit $\sim T$, and its propagator gives us
a $T^{-1}$ dependence for $M_{fi}$. 
A working expression for the bremsstrahlung process can be found in \cite{Friman:1979bh}.

\subsection{The Cooper Pair Neutrino Process}
\label{Sec:PBF}

The formation of the fermonic pair condensate also triggers a new
neutrino emission process \cite{Flowers:1976vn,Voskresensky:1987uq,Senatorov:1987ve}
which has been termed the ``pair breaking
and formation", or PBF, process \cite{Schaab:1997qf}. 
Whenever any two fermions form a Cooper pair, the
binding energy can be emitted as a $\nu-\overline{\nu}$ pair.
Under the right conditions, this PBF process can be the dominant cooling
agent in the evolution of a neutron star~\cite{Page:1998bh}. Such
efficiency is due to the fact that the pairing phase transition is
second order in nature. During the cooling of the star, the phase
transition starts when the temperature $T$ reaches $T_c$ when pairs
begin to form, but thermal agitation will constantly induce the
breaking of pairs with subsequent re-formation and possible neutrino
pair emission.

\begin{figure}
\begin{center}
\includegraphics[width=.80\textwidth]{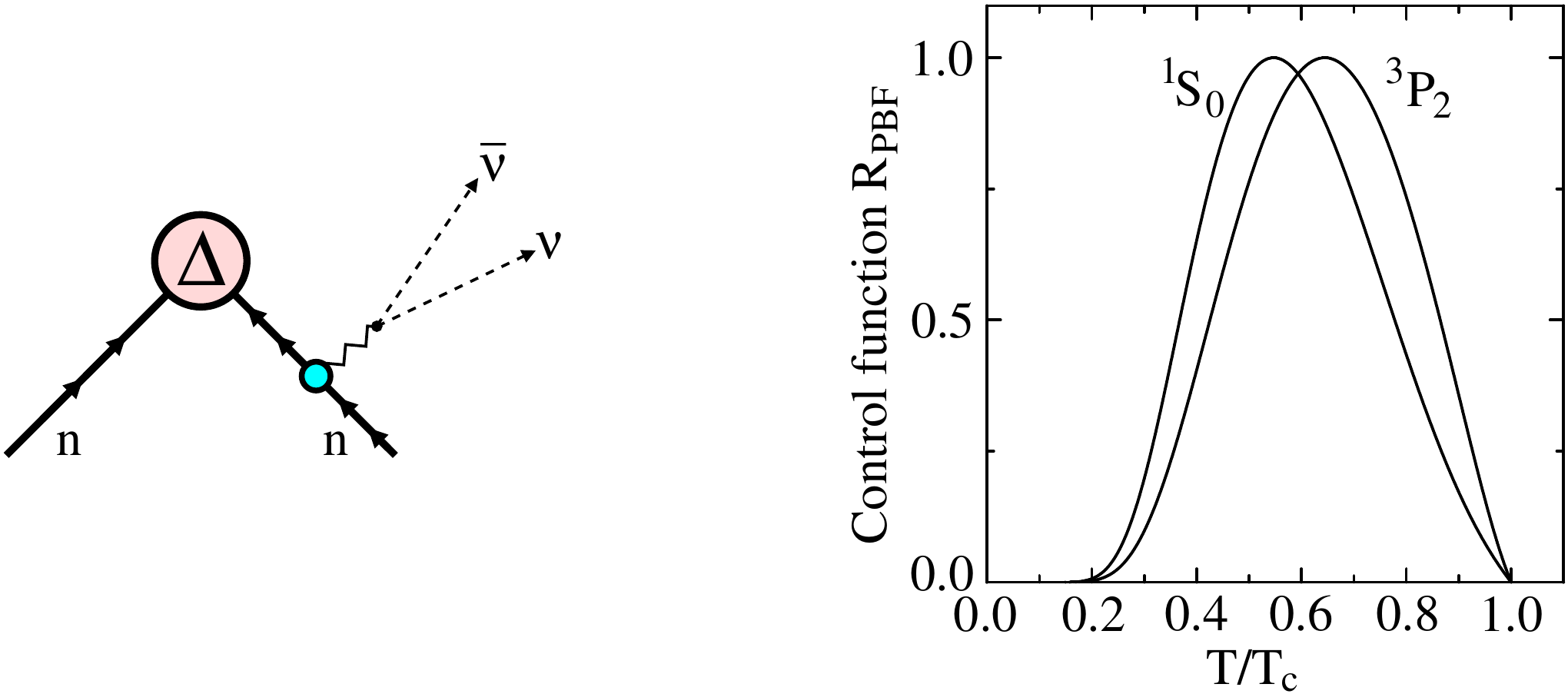}
\end{center}
\caption{
Left panel: Feynman diagram for $\nu \overline{\nu}$ emission from the 
formation of a $nn$ Cooper pair (pair breaking and formation, PBF, process). 
Right panel: control functions $R_\mathrm{PBF}$ for the PBF process.
}
\label{Fig:PBF}
\end{figure}

The emissivity of the PBF process (see the left panel of \fig{Fig:PBF} for a Feynman diagram) can be written as
\be
\epsilon^\mathrm{PBF} =
\int \!\!\!\! \int \!\!\!\! \int \!\!\!\! \int 
\frac{d^3 p_{{\nu}}}{(2 \pi)^3} 
\frac{d^3 p_{\overline{\nu}}}{(2 \pi)^3} 
\frac{d^3 p }{(2 \pi)^3} 
\frac{d^3 p^\prime}{(2 \pi)^3}
f(E_p)f(E_{p^\prime}) 
\cdot  (2 \pi)^4 \delta^4(P_f-P_i) |M_{fi}|^2 \cdot E_\nu
\label{Eq:E_GPBF}
\ee
Under degenerate conditions, the expression above can be reduced to read as
\ba
\epsilon^\mathrm{PBF} = \frac{12 G_F^2 m_{f}^{*} p_{F,f}}{15 \pi^5 \hbar^{10} c^6}
\left(k_B T\right)^{7}  a_{f,j} R_j\left[\Delta_j(T)/T\right] 
\nonumber
\\
= 3.51\times 10^{21}~
\frac{\mathrm{erg}}{\mathrm{cm}^3~\mathrm{s}} \times
\tilde{m}_f \, \tilde{p}_{F,f}
\, T_9^7 \, a_{f,j} \;R_j\left[\Delta_j(T)/T\right]
\label{Eq:Q_PBF}
\ea
for a fermion $f$ in a pairing state $j = \Singlet$ or $\triplet$. The
coefficients $a_{f,j}$ depend on the type of fermion and on the vector and
axial couplings $C_V$ and $C_A$ (see, e.g., \cite{Page:2009qf}). The
control functions $R_j$ are plotted in the right panel of
\fig{Fig:PBF}. These functions encapsulate the effect that the
PBF process turns on when $T$ reaches $T_c$ and practically turns off
at $T \simle 0.2 \, T_c$ when there is not enough thermal energy to
break pairs. 

The PBF process has had 
an interesting history. It was first
discovered by Flowers, Ruderman, and Sutherland \cite{Flowers:1976vn}
and, independently, by Voskresensky and Senatorov
\cite{Voskresensky:1987uq}. It was, however, 
overlooked for 20 years, until  implemented in a cooling
calculation in \cite{Schaab:1997qf} and its importance emphasized in
\cite{Page:1998bh}. Then in 2006, Leinson and Perez~\cite{Leinson:2006fk}
showed that the previous computations of the PBF emissivity
were incompatible with
vector current (e.g. baryon number) conservation. Neutrino
pair-production is mediated by the weak interaction, which can be
decomposed in the traditional manner into vector and axial-vector
parts. In pure neutron matter, the vector part of the PBF emissivity
is suppressed because of vector current conservation by a factor of order 
{\bf $(p_F/m^*c)^4 = (v_F/c)^4$}. 
This is equivalent to the simple observation
that a one-component system of charges does not have a time-varying
dipole moment~\cite{Steiner:2009uq}. The axial part of the PBF process is,
however, unimpeded and dominates the emissivity. \\

\subsubsection*{Temperature dependence of the PBF neutrino emissivity}

The temperature dependence of the PBF process (left panel of \fig{Fig:PBF}) can be ascertained from \eq{Eq:E_GPBF} according to the following $T$-power counting: 
\be
\epsilon^\mathrm{PBF} \propto
T^3 \cdot T^3  \cdot T \cdot 1  \cdot \frac{1}{T}  \cdot R (\Delta/T)  \cdot T= T^7~R(\Delta/T) \,,
\ee
where the two $T^3$ and the first $T$ factors arise from the phase space integrations of the neutrino pair and
the first participating nucleon, respectively.
The factor $1$ 
results from the phase space integration of the second nucleon. 
As there are only two degenerate fermions in this process (in 
contrast to the Urca and bremsstrahlung processes that
involve 3, 4, or 5 degenerate fermions), the momenta of the neutrino pair and the first nucleon are chosen the
momentum of the second nucleon is 
fixed by the three-momentum conserving delta function.
Thus, this second nucleon does not introduce any $T$ dependence.
The  $T^{-1}$ dependence arises from the energy conserving delta function.
The last $T$ factor is from the neutrino pair's energy, whereas
the $T$ and $\Delta$ dependence of the matrix element of the reaction are included in the function $R(\Delta/T)$,
which is shown in the right panel of \fig{Fig:PBF}.

An alternative way of looking at the PBF process is simply as an
interband transition of a nucleon \cite{Yakovlev:1999cr}. Considering
the particle spectrum in a paired state (the right panel of
\fig{Fig:SP_Spectrum}), the lower branch (with $\epsilon <
\epsilon_F - \Delta$) corresponds to paired particles whereas the
upper branch to excited ones, i.e., 
the ``broken pair" 
leaves a hole in the lower branch. A transition of a particle from
the upper branch to a hole in the lower branch corresponds to the
formation of a Cooper pair.

\subsubsection*{Dominance of triplet-pairing}

In the non-relativistic limit for nucleons, the leading contribution from the axial-vector part is proportional to $(v_F/c)^2$. To this order, the control function $R(z=\Delta/T)$ receives a contribution from the axial-vector part which can be expressed as \cite{Yakovlev:1999cr}
\be
R(z) = \frac {c_A^2}{8\pi} \int d\Omega \int_0^\infty \frac {z^6~dx}{(e^z+1)^2}~I \,,
\ee
where $x=v_F(p-p_F)/T$ and the quantity $I=I_{xx}+I_{yy}+I_{zz}$ with 
\be
I_{ik} = \sum_{\eta\eta^\prime} 
\langle B|\hat \Psi^\dagger\sigma_i\hat\Psi|A\rangle
\langle B|\hat \Psi^\dagger\sigma_k\hat\Psi|A\rangle^*\,,
\ee    
where $\hat \Psi$ is the second-quantized non-relativistic spinor wave function of the nucleons in superfluid matter (see \cite{Yakovlev:1999cr} for its detailed structure in terms of the Bogoliubov transformation matrix elements $U_{\sigma\eta}({\bf p})$  and  $V_{\sigma\eta}({\bf p}))$, 
$|A\rangle$ is the initial state of the system and $|B\rangle$ its final state. 
The total spin states of the pair $\eta$, $\eta^\prime$ and $\sigma$ each take on values $\pm 1$ and  $\sigma_i$'s are the Pauli spin matrices.   The energy of a paring quasiparticle is given by 
$E = {\sqrt { \epsilon^2 + \Delta_{\bf p}^2}}$, where $\epsilon = v_F(p-p_F)$. 
For singlet-state pairing, the momentum-dependent gap $\Delta_{\bf p}$ is independent of ${\bf p}$, so that the occupation probabilities $u_p$ and $v_p$ associated with the matrix elements 
 $U_{\sigma\eta}({\bf p})$  and  $V_{\sigma\eta}({\bf p})$ depend only on $p = |{\bf p}|$. In the case of singlet pairing, the Bogoliubov matrix elements satisfy the symmetry properties $V_{\alpha\beta}(-{\bf p}) = V_{\alpha\beta}({\bf p})$ and $V_{\alpha\beta}({\bf p}) = -V_{\beta\alpha}({\bf p})$, so that 
the diagonal elements of this  $2\times 2$ matrix are zero, and the non-diagonal elements are $v_p$ and $-v_p$, respectively. These symmetry properties, together with the traceless property of the Pauli matrices, ensure that the quantity $I=0$. Thus, to order $(v_F/c)^2$, the axial-vector part does not contribute in the spin-singlet channel rendering the triplet pairing channel, which does not vanish, to be the sole contribution to the PBF process.

\subsubsection*{Time history of the PBF process}

The Cooper pair neutrino process operates at different times in a
neutron star's cooling history according to the time during which the
local temperature is nearly equal to the critical temperature of any
superfluid gap. In neutron stars consisting of neutrons,
protons, and electrons, there are three relevant superfluid gaps:
singlet neutron superfluidity, singlet proton superfluidity, and
triplet neutron superfluidity at high densities when the singlet
channel of the neutron-neutron interaction becomes repulsive above the
saturation density. In neutron stars which contain exotic matter in
their interiors, each additional superfluid fermion potentially opens
up new Cooper pair cooling processes. If neutron stars contain
deconfined quark matter in their cores, then pairing between quark
flavors creates new Cooper pair neutrino processes which involve
pairing between unlike fermions~\cite{Jaikumar:2001zr}.


\section{Cooling of Neutron Stars}
\label{Sec:Cooling}

The study of neutron star cooling is a Sherlock Holmes investigation,
following the tracks of energy.  At its birth, some 300 B (1 Bethe =
$10^{51}$ ergs) of gravitational energy are converted largely into
thermal energy.  About 98\% of it is emitted in neutrinos during the
first minute, the {\em proto-neutron star} phase, 1\% is transferred
to the supernova ejecta (with 1\% of this 1\% powering the light show),
and the remainder is left in thermal energy of the {\em
  new-born neutron star}, i.e., the star produced during the
proto-neutron star phase.  Following the tracks of energy, the
subsequent evolution of the neutron star can, in a simplified way, be
described by an energy balance equation
\be
\frac{dE_{th}}{dt} = C_V \frac{dT}{dt}
                   = -L_\nu - L_\gamma + H \,,
\label{Eq:cooling}
\ee
where $E_{th}$ is the star's total thermal energy, $C_V$ its specific
heat, and $L_\gamma$ and $L_\nu$ its photon and neutrino luminosities,
respectively.  The term $H$, for ``heating'', represents possible
dissipative processes, such as friction from differential rotation or
magnetic field decay.  In this simplified equation it is assumed that
the star's interior is isothermal with temperature $T$, a state
reached within a few decades after birth in the core-collapse
supernova (see Sec.~\ref{Sec:Numerical}).  A more detailed study would
include general relativistic effects and consider a local energy
balance equation for each layer in the star, instead of the global one
of \eq{Eq:cooling}, complemented by a heat transport equation, in
order to follow the evolution of the temperature profile in the
stellar interior (see, e.g., \cite{Page:2004zr} and references
therein).

After the proto-neutron star phase, matter is highly degenerate within
most of the star, except the outermost, lowest density, layers.  As a
consequence, the gross structure of the star does not evolve with time
and is determined, once and for all, by solving the
Tolman-Oppenheimer-Volkoff equations \cite{Oppenheimer:1939uq} of hydrostatic equilibrium.  An
equation of state is required to not only solve these
equations, which determine 
the mass and radius of the star, but also to evaluate the internal chemical
composition of each species of nucleus and particle
as well as their effective masses, chemical potentials, specific heats, etc.
A complete 
cooling model requires, moreover, inclusion of neutrino and surface
photon emissions as well as a description of the pairing
properties of matter, i.e., the pairing gaps for each fermonic
species, together with their respective density dependences.

Within the isothermal approximation of \eq{Eq:cooling}, the three
major ingredients needed for the study are $C_V$, $L_\gamma$, and
$L_\nu$.  Neutrino emission processes were described in
Sec.~\ref{Sec:Neutrinos} and the specific heat and photon emission are
briefly described below.  We continue this section by describing simple
analytical solutions of \eq{Eq:cooling} and displaying the results of
representative numerical simulations of the complete set of general
relativistic evolutionary equations.

\subsection{Specific Heat}
\label{Sec:Cv}

\begin{figure}
\begin{center}
\includegraphics[width=.5\textwidth]{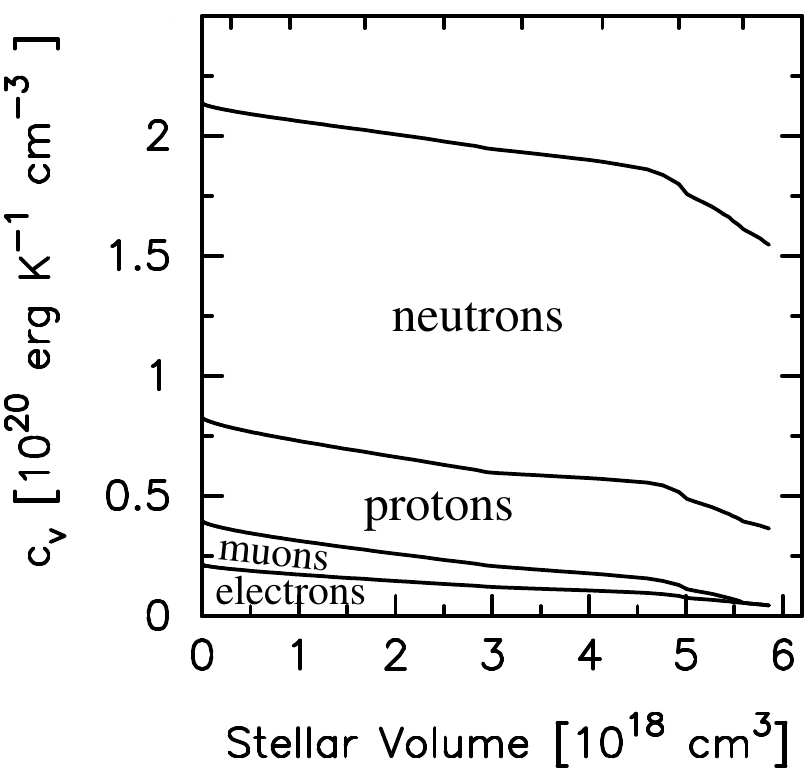}
\end{center}
\caption{Cumulative specific heats of e, $\mu$, p, and n as a function
  of stellar volume within the core of a 1.4 $M_\odot$ star built
  using the APR EOS at $T=10^9$ K.  Nucleons are assumed to be
  unpaired which implies $c_\mathrm{v} \propto T$.  No hyperons or quarks are
  permitted by the EOS.  This figure is adapted from
  \cite{Page:2004zr}.}
\label{Fig:Cv}
\end{figure}

The dominant contributions to the specific heat $C_\mathrm{v}$ come
from the core, which makes up more than 90\% of the total volume and
98\% of the mass.  Its constituents are quantum liquids of leptons,
baryons, mesons, and, possibly, deconfined quarks at the highest
densities.  Hence, one has
\be
C_\mathrm{V} = \sum_i C_{\mathrm{V}, i}
\;\;\;\;\;\;
\mathrm{with}
\;\;\;\;\;\;
 C_{\mathrm{V}, i} = \int \!\!\!\! \int \!\!\!\! \int c_{\mathrm{v}, i} \, d\mathrm{v} \,,
\label{Eq:CvTot}
\ee
where $c_{\mathrm{v}, i}$ is the specific heat per unit volume of constituent $i$
($i$ = e, $\mu$, n, p, hyperons, quarks), but those of meson condensates is usually neglected.
For normal (i.e., unpaired) degenerate fermions, the standard Fermi liquid
result \cite{Baym:2004nx}
\be
c_{\mathrm{v} \; i} = N(0) \frac{\pi^2}{3} k_B^2 T
\;\;\;\;\;\; \mathrm{with} \;\;\;\;\;\;
N(0) = \frac{m_i^* p_{F \, i}}{\pi^2 \hbar^3}
\label{Eq:Cv}
\ee
can be used, where $m^*$ is the fermion's effective mass.
In \fig{Fig:Cv}, the various contributions to $C_V$ are illustrated.

When baryons, and quarks, become paired, as briefly described in
Sec.~\ref{Sec:Pairing_Effects}, their contribution to $C_\mathrm{V}$ is
strongly suppressed at temperatures $T \ll T_c$
($T_c$ being the corresponding critical temperature).
Extensive baryon, and quark, pairing can thus significantly reduce
$C_\mathrm{V}$, but by no more than a factor of order ten because
leptons do not pair.  The crustal specific heat is, in principle,
dominated by neutrons in the inner crust but, as these are certainly
extensively paired, only the nuclear lattice and electrons
contribute in practice.

\subsection{Photon Thermal Luminosity and the Envelope}
\label{Sec:envelope}

The photon thermal luminosity $L_\gamma$ is commonly expressed through
the {\em effective temperature} $T_e$ defined by
\be
L_\gamma = 4 \pi R^2 \sigma_\mathrm{SB} T_e^4 \, ,
\label{Eq:Lgamma1}
\ee
where $\sigma_\mathrm{SB}$ is the Stefan-Boltzmann constant.  Thermal
photons from the neutron star surface are effectively emitted at the
{\em photosphere}, which is usually in an atmosphere, but could be
located on the solid surface if a very strong magnetic field exists \cite{Lai:2001uq}.
The atmosphere, which is only a few centimeters thick, contains a
temperature gradient; $T_e$ gives an estimate of its average
temperature.  The opacity in the atmosphere receives a strong
contribution from free-free scattering that has a strong ($\sim E^{-3}$) energy
dependence.  As a result photons of increasing energy
escape from deeper and hotter layers and the emitted thermal spectrum
shows an excess of emission in its Wien's tail compared to a blackbody
of the same temperature $T=T_e$ \cite{Romani:1987kx}.  
In the presence of heavy elements,
``metals'' in astronomical parlance, which may be not fully ionized,
absorption lines increase the bound-free opacity contributions, and
push the Wien's tail of the observable spectrum closer to the
blackbody one.  The presence of a strong magnetic field also alters
the opacity in such way as to mimic a blackbody with the same chemical
composition and $T_e$.  There were great expectations that, with the
improved spectral capabilities of {\em Chandra} and {\em Newton} observatories, many
absorption lines would be observed and allow the determination of the
gravitational redshifts and chemical composition of isolated neutron
star surfaces.  This expectation has, unfortunately, not been
fulfilled; only in a very few cases have lines been detected, and their
interpretation is controversial.

Observationally, $L_\gamma$ and $T_e$ are red-shifted and
\eq{Eq:Lgamma1} is rewritten as
\be
L^\infty_\gamma= 4 \pi R_\infty^2 \sigma_\mathrm{SB} (T^\infty_e)^4 \,,
\label{Eq:Lgamma2}
\ee
where $L^\infty_\gamma = \mathrm{e}^{2\phi} L_\gamma$, $T^\infty_e = \mathrm{e}^{\phi} T_e$,
and $R_\infty = \mathrm{e}^{-\phi} R$.
Here $\mathrm{e}^{-\phi} = 1+z$, with $z$ being the redshift, and $\mathrm{e}^{2\phi}$ is the $g_{00}$
coefficient of the Schwarzschild metric, i.e.,
\be
\mathrm{e}^\phi \equiv \sqrt{1-\frac{2GM}{Rc^2}} \,.
\label{Eq:redshift}
\ee Notice that $R_\infty$ has the physical interpretation of being
the star's radius corresponding to its circumference divided by
$2\pi$, and would be the radius one would measure trigonometrically,
if that were possible \cite{Page:1995vn}.

$L^\infty_\gamma$ and $T^\infty_e$ are the observational quantitities that are
compared with theoretical cooling models.  In principle, both are
independently observable: $T^\infty_e$ is deduced from a fit of the observed
spectrum while $L^\infty_\gamma$ is deduced from the observed total flux 
\footnote{The flux must, however, be corrected for interstellar
  absorption.}, and knowledge of the distance $D$, via $L_\gamma = 4
\pi D^2 F$.  The star's distance can be deduced either from the radio
signal dispersion measure, if it is a radio pulsar, or from the
distance of an associated supernova remnant, if any.  Then
\eq{Eq:Lgamma1} or (\ref{Eq:Lgamma2}) provides a consistency check:
the inferred radius $R$ should be of the order of 10 - 15 km.  Given
the lack of determination of the atmospheric composition from spectral
lines, this consistency check is generally the only criterion to
decide on the reliability of a $T_e$ measurement from an atmosphere
model spectral fit (besides the obvious requirement that the model must
give a good fit to the data, i.e. a $\chi^2 \simeq 1$).

In a detailed cooling calculation, the time evolution of the
temperature profile in the star is followed.  However, the uppermost
layers have a thermal time-scale much shorter than the interior of the
star and are practically always in a steady state.  It is, hence,
common to treat these layers separately as an {\em envelope}.
Encompassing a density range from $\rho_b$ at its bottom (typically
$\rho_b = 10^{10}$ g cm$^{-3}$) up to $\rho_e$ at the photosphere
($\rho_e\simle1$ g cm$^{-3}$), and a temperature range from $T_b$ to
$T_e$, the envelope is about one hundred meters thick.  Due to the
high thermal conductivity of degenerate matter, stars older than a few
decades have an almost uniform internal temperature, except within the
envelope which acts as a thermal blanket insulating the hot interior
from the colder surface.  A simple relationship between $T_b$ and
$T_e$ can be formulated \cite{Gudmundsson:1983kx}:
\be
T_e \simeq 10^6 \, \mathrm{K} \times \left(\frac{T_b}{10^8 \, \mathrm{K}}\right)^{0.5+\alpha}
\label{Eq:TbTe}
\ee
with $\alpha \ll 1$.  
The precise $T_e - T_b$ relationship depends on the chemical composition of the envelope.  
The presence of light elements like H, He, C, or O, which have large thermal conductivities,
leads to a larger $T_e$ for the same $T_b$ relative to the case of a heavy element, such as iron, envelope.  
Light elements are not expected to survive densities larger than $\sim 10^{10}$ g cm$^{-3}$
due to pycnonuclear reactions.
Thus, the maximum possible 
mass in light elements amounts to
$\Delta M_\mathrm{Light} \simeq 10^{-6} M_\odot$,
which is enough to raise $T_e$ by a factor of two.
Magnetic fields also alter the
$T_e - T_b$ relationship, 
but 
 to a lesser extent (see, e.g.,
\cite{Page:2006ly} for more details) unless they are super-strong as
in the case of magnetars, i.e. $B_s\sim 10^{15}$ G.

\subsection{Analytical Solutions}
\label{Sec:Analytical}

As the essential ingredients entering \eq{Eq:cooling} can all be
approximated by power-law functions, one can obtain simple and
illustrative analytical solutions (see also \cite{Lattimer:1994fk}).
We adopt the notation
\be
C_\mathrm{V} = C_9  T_9 \, ,
\;\;\;\;\;\;
L_\nu = N_9  T_9^8 \, ,
\;\;\;\;\;\; {\rm and} \;\;\;\;\;\;
L_\gamma = S_9  T_9^{2+4\alpha} \, ,
\label{Eq:PL1}
\ee
where $T_9 \equiv T/(10^9 \; \mathrm{K})$ refers to the isothermal
temperature $T_b$ in the star's interior. As written, $L_\nu$
considers slow neutrino emission involving five degenerate fermions
from the modified Urca and the similar bremsstrahlung processes,
summarized in Table~\ref{Tab:Nu}. The photon luminosity $L_\gamma$ is
obtained from \eq{Eq:Lgamma1} using the simple expression in
\eq{Eq:TbTe}.  We will ignore redshift.  Typical parameter
values are $C_9 \simeq 10^{39}$ erg K$^{-1}$, $N_9 \simeq 10^{40}$ erg
s$^{-1}$ and $S_9 \simeq 10^{33}$ erg s$^{-1}$ (see Table 3 in
\cite{Page:2006ly} for more details).  In young stars, neutrinos
dominate the energy losses (in the so-called {\em neutrino cooling era}),
and photons take over after about $10^5$ years (the {\em photon
  cooling era}).

\noindent
{\bf Neutrino cooling era:} In this case
$L_\gamma$ can be neglected in \eq{Eq:cooling}, so that
\be
t = \frac{10^9 C_9}{6N_9} \left( \frac{1}{T_9^6} - \frac{1}{T_{0, 9}^6} \right)
\;\;
\rightarrow
\;\;
T_9 = \left({\tau_\mathrm{MU}\over t}\right)^{1/6}
\;\; (\mathrm{when} \;T \ll T_0) \!
\label{Eq:Simple_nu}
\ee
with a MU cooling timescale $\tau_\mathrm{MU} = 19^9 C_9/6N_9 \sim 1$
year when the star reaches the asymptotic stage 
($T \ll T_0$,
$T_0$ being the initial temperature at time $t= 0$).
The observed slope of the {\it cooling track}  during the asymptotic stage is
\be\label{slope}
\left|{d\ln T_e\over d\ln t}\right|\simeq{1\over12}+{\alpha\over6}\,,
\ee
using the core-envelope relation \eq{Eq:TbTe}.  This slope is not sensitive to the core-envelope relation.

\noindent
{\bf Photon cooling era:} In this era,  
$L_\nu$ is negligible compared to $L_\gamma$.  Since $|\alpha| \ll 1$, one finds
\be
\label{Eq:Simple_gamma}
t=t_1+\tau_\gamma\left[\ln T_9 - \ln T_{1,9}\right]\quad\rightarrow\quad
T_9=T_{1,9}e^{(t_1-t)/\tau_\gamma} \,,
\ee
where $T_1$ is $T$ at time $t=t_1$ and
$\tau_\gamma=10^9C_9/S_9\sim3\times10^7$ years, the photon cooling
timescale.  The observed slope of the cooling track is, when $T \ll T_1$,
\be
\label{slope1} 
\left|{d\ln T_e\over d\ln t}\right|\simeq{t\over2\tau_\gamma}.  
\ee
This slope becomes steeper with time.

\subsection{Numerical Simulations and the Effects of Pairing}
\label{Sec:Numerical}

Numerical simulations of a cooling neutron star use an evolution code
in which the energy balance and energy transport equations, in their
fully general relativistic forms, are solved, usually assuming
spherical symmetry and with a numerical radial grid of several hundred
zones \footnote{Such a code, \texttt{NSCool}, is available at:
  \texttt{http://www.astroscu.unam.mx/neutrones/NSCool/}}.  Compared
to the analytic solutions discussed in Sec.~\ref{Sec:Analytical},
detailed conductivities, neutrino emissivities, and possible pairing
are taken into account.

\begin{figure*}[bt]
\begin{center}
\includegraphics[width=.60\textwidth]{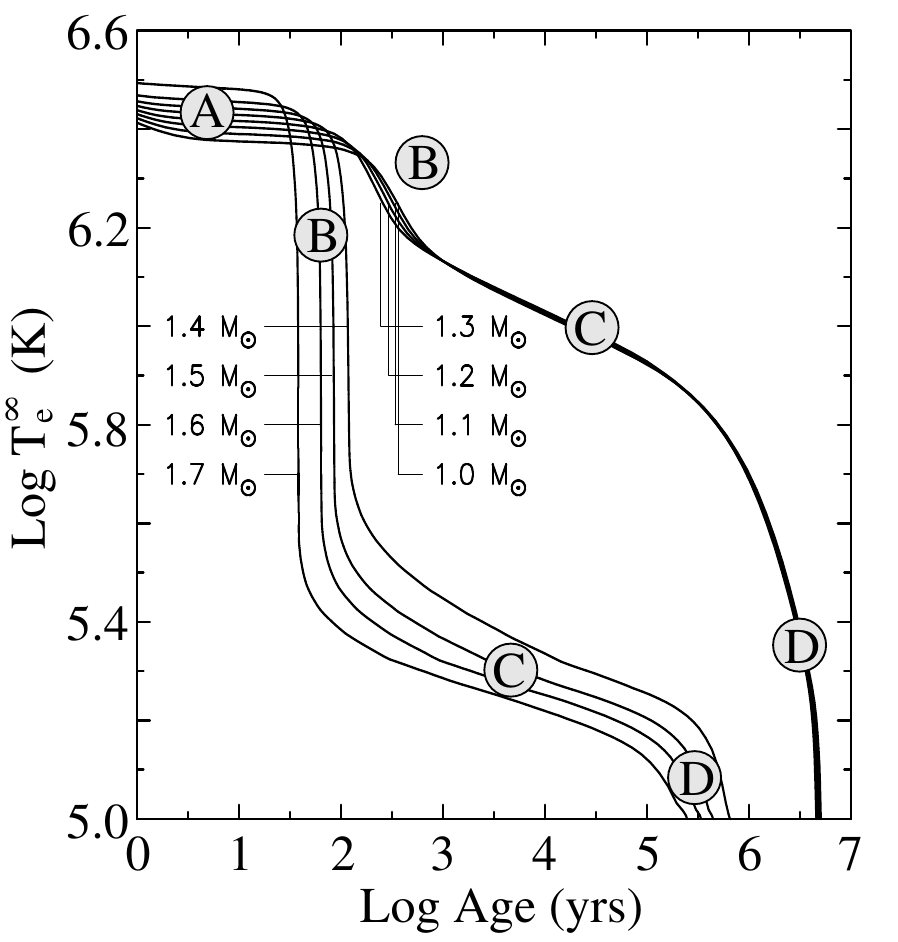}
\end{center}
\caption{Cooling curves illustrating the difference between slow cooling driven by the
modified Urca process, for masses below $1.35 M_\odot$, and fast cooling from the 
direct Urca process in more massive stars
  \cite{Page:1992nx}.}
\label{Fig:Cooling_PAL}
\end{figure*}

A set of cooling curves that illustrate the difference between cooling driven by the modified Urca
and the direct Urca processes is presented in \fig{Fig:Cooling_PAL}.
Cooling curves of eight different stars of increasing mass are shown, using an equation of state model
from \cite{Prakash:1988oq}, which allows the DU process at densities above 
$1.35\times 10^{15}$ g cm$^{-3}$, i.e., above a critical neutron star mass of 1.35 $M_\odot$.
Notice that the equation of state used has parameters  
{\em specifically adjusted} to obtain a critical mass of 1.35 $M_\odot$ which falls within
the expected range of isolated neutron star masses; other equations of state can
result in very different critical masses.
The difference arising from  {\em slow} and {\em fast} neutrino processes is clear.
The various cooling phases, A to D are discussed below.

\begin{figure}
\begin{center}
\includegraphics[width=.99\textwidth]{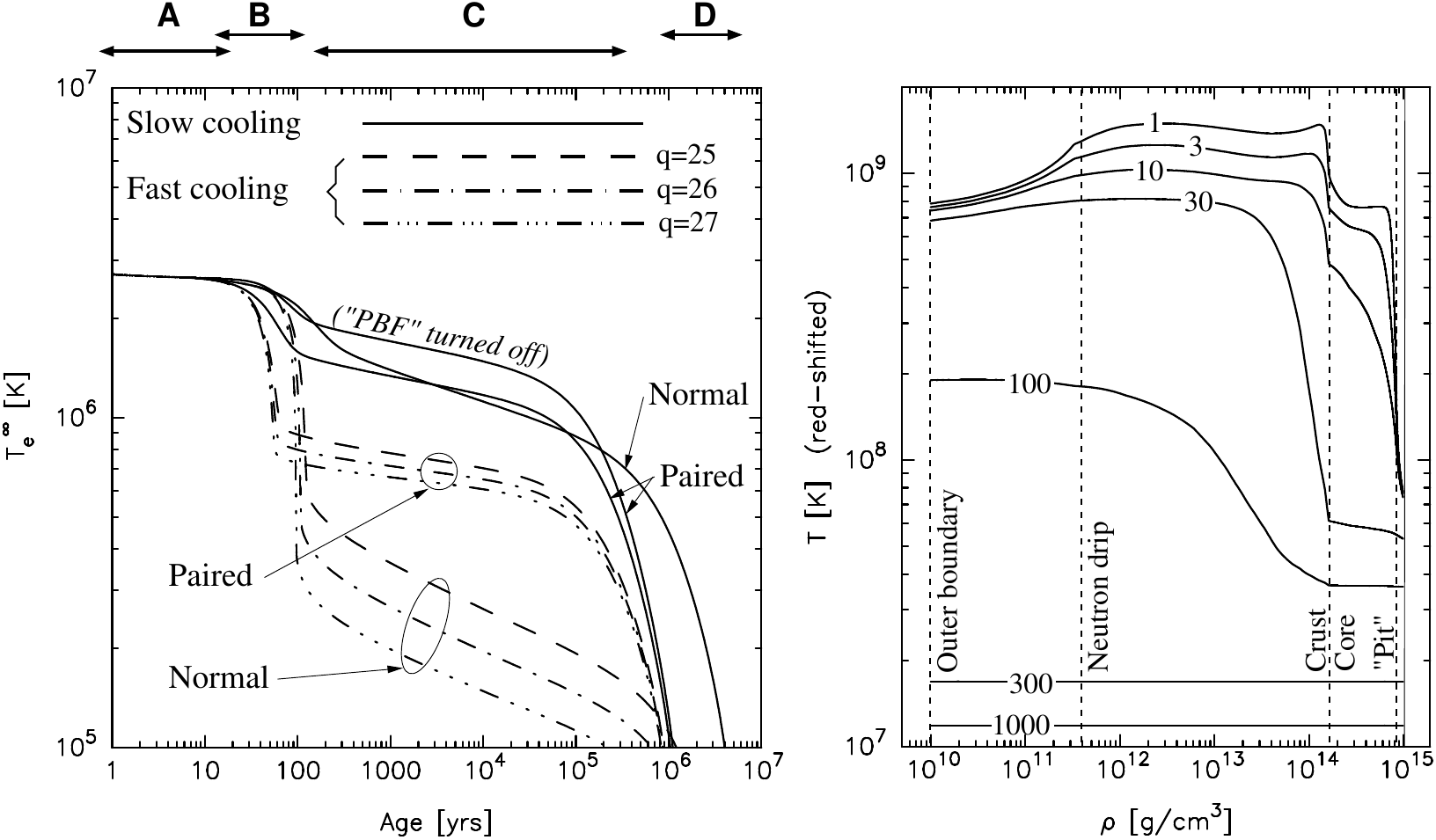}
\end{center}
\caption{ Left panel: cooling curves for various illustrative cooling
  scenarios.  Right panel: temperature profile evolution for the fast
  cooling model with q = 26.  The numbers on the curves give the age
  of the star, in years. See text for description.  Figure taken from
  \cite{Page:2012vn}.}
\label{Fig:Cooling2}
\end{figure}

To further illustrate some of the possible cooling behaviors of neutron stars,
and the effects of pairing,
we show in \fig{Fig:Cooling2} simulations based on a $1.4 M_\odot$
star built with the APR EOS \cite{Akmal:1998fk} and a heavy-element
envelope.  The ``slow cooling'' models include, in the core, the slow
neutrino processes of Table~\ref{Tab:Nu} and the PBF process only.
For the ``fast cooling'' models, a fast process with emissivity
$\epsilon_{Fast}^{(q)} = 10^q \cdot T_9^6$ erg cm$^{-3}$ s$^{-1}$,
with $q=25$, $26$, and $27$, was added at $\rho > 3\rhonuc$.
These $q$ values simulate neutrino emission from a
kaon condensate, a pion condensate, or a direct Urca, respectively.
These models, being based on the same EOS, are not self-consistent,
but they have the advantage that the only differences among them is
the presence or absence of the fast cooling process with
$\epsilon_{Fast}^{(q)}$ and the presence or absence of nucleon
pairing.  The models with pairing include the neutron $\Singlet$ gap
``SFB'' of \fig{Fig:Tc_n1S0} in the inner crust, the $^1\! S_0$ proton
gap model ``T'' of \fig{Fig:Tc_p}, and the phenomenological neutron
$\Triplet$ gap ``b'' of the right panel of \fig{Fig:Tc_n}.

The distinctive phases of evolution are labelled ``A'',
``B'', ``C'', and ``D'' on the cooling curves in \fig{Fig:Cooling_PAL} and above the cooling curves in the left panel of
\fig{Fig:Cooling2}.  Phases A and B are determined by the evolution of
the crust while C and D reflect the evolution of the core.
We describe these four phases in more detail according to \fig{Fig:Cooling2}:
\\\\
{\bf Phase A:} The effective surface temperature $T_e$ here is determined
by the evolution of the outer crust only.  At such early stages, the
temperature profile in the outer crust is independent of the rest of
the star and, as a result, all models have the same $T_e$.
\\\\
{\bf Phase B:} The age of the star during this phase is similar to the thermal
relaxation timescale of the crust, 
the heat flow in which controls the
evolution of $T_e$.  The evolution of the temperature profile for the fast
cooling model with $q=26$ in the absence of pairing (marked as
``Normal'' in the left panel of \fig{Fig:Cooling2}) is depicted in the
right panel of this figure \ref{Fig:Cooling2}.  Very early in the evolution, a cold
``pit'' develops in the core where fast neutrino emission is
occurring.  During the first 30 years, heat flows from the outer core
into this pit until the core becomes isothermal.  Afterwards, heat
from the crust rapidly flows into the cold core and the surface
temperature $T_e$ drops rapidly.  
Well before 300 years, during phase C, the
stellar interior becomes isothermal and it is only within
the shallow envelope, not 
shown in this figure to preserve clarity, that a
temperature gradient is still present.

Notice that models of with pairing have shorter crust relaxation times
due to the strong reduction of the neutron specific heat in the inner
crust by $\Singlet$ neutron superfluidity there.
\\\\
{\bf Phase C:} This is the ``neutrino cooling phase'' in which the
star's evolution is driven by neutrino emission from the core: 
$L_\nu \gg L_\gamma$.  The difference between ``slow'' and ``fast''
neutrino emission, with or without core superfluidity, is clearly seen.

Quite noticeable is the effect of pairing-induced suppression of the neutrino
emissivity in the fast cooling models.  Once $T$ drops below $T_c$,
which happens only a few seconds or minutes after the star's birth,
neutrino cooling is quenched.  Which fast cooling process occurs is
much less important in the presence of pairing than in its absence.  It
takes half a minute, if $q=27$, or half an hour, if $q=25$, for the
``pit's'' temperature to fall below $T_c$, which does not matter when
looking at the star thousands of years later.  The evolution is more
dependent on $T_c$ than on $q$.

In the case of the slow cooling models, the effects of pairing are
more subtle than that in the fast cooling models, if one ignores
the artificial case with the PBF processed turned off.  
The burst of neutrino emission occurring when $T \simeq T_c$, from the
PBF process, induces an additional, if temporary, rapid cooling
episode.  The impact of the PBF process, however, depends on the size
of the neutron $\Triplet$ gap.  If the gap is large enough, the PBF
cooling occurs during stage B and is hidden; if the gap is small enough,
the PBF cooling occurs during stage D and is again hidden.  
{\it Only intermediate size gaps reveal the presence of the PBF process.}
The effect of this gap on the evolution is considered in more detail in
Sec.~\ref{Sec:Minimal}.   The effect of the proton $\Singlet$ gap is more subtle still, 
and its effects are considered in detail in Sec.~\ref{Sec:CasA}.
\\\\
{\bf Phase D:} At late times, $L_\nu$ has decreased significantly due
to its strong $T$ dependence and photon emission, $L_\gamma$, which is
less $T$-dependent, now drives the evolution.  This is reflected by
the larger slopes of the cooling curves.  During this ``photon cooling
era'', models with pairing cool faster due to the 
reduction of the specific heat from superfluidity.

\subsubsection*{The effect of a very strong magnetic field}

In this contribution, we have neglected effects of the presence of a magnetic field.
A strong magnetic field can alter the cooling of a neutron star in two ways.
In the envelope and the crust, heat is transported by electrons.
A surface magnetic field of strength $\sim 10^{12}$ G, a typical value for   
the majority of pulsars, is sufficient to induce anisotropy in the thermal conductivity of the envelope resulting in a non-uniform surface temperature \cite{Page:1995vn}
that manifests itself as a modulation of the thermal flux with the pulsar's rotational period.  As mentioned in Sec.~\ref{Sec:envelope}, this effect only 
slightly alters the $T_b - T_e$ relationship of \eq{Eq:TbTe} \cite{Potekhin:2001ve}.
However, a strong magnetic field  deep in the crust \cite{Geppert:2004ly} will have larger effects. For example, a $\simge 10^{14} G$ toroidal field within the crust can act as an efficient insulator, rendering most of the star's surface very cold, but having two hot spots on the symmetry axis of the torus \cite{Perez-Azorin:2006ve,Geppert:2006fk}.  This results in peculiar cooling trajectories \cite{Page:2007bh}.
The second effect of a magnetic field is that a slowly decaying field can act as a source of energy (i.e., the ``H" term in \eq{Eq:cooling}) which can keep old neutron stars warm.  If the magnetic energy reservoir is large enough, and small scale magnetic structures exist that can decay rapidly\footnote{The Ohmic decay time, e.g., is $\propto l^2$, $l$ being the length-scale of the field structure.}, 
aided by the Hall drift, the thermal evolution is significantly altered 
\cite{Aguilera:2008qf}.

\subsection{Observations of Cooling Neutron Stars}
\label{Sec:Cool_Data}

\begin{figure*}
\begin{center}
\includegraphics[width=.80\textwidth]{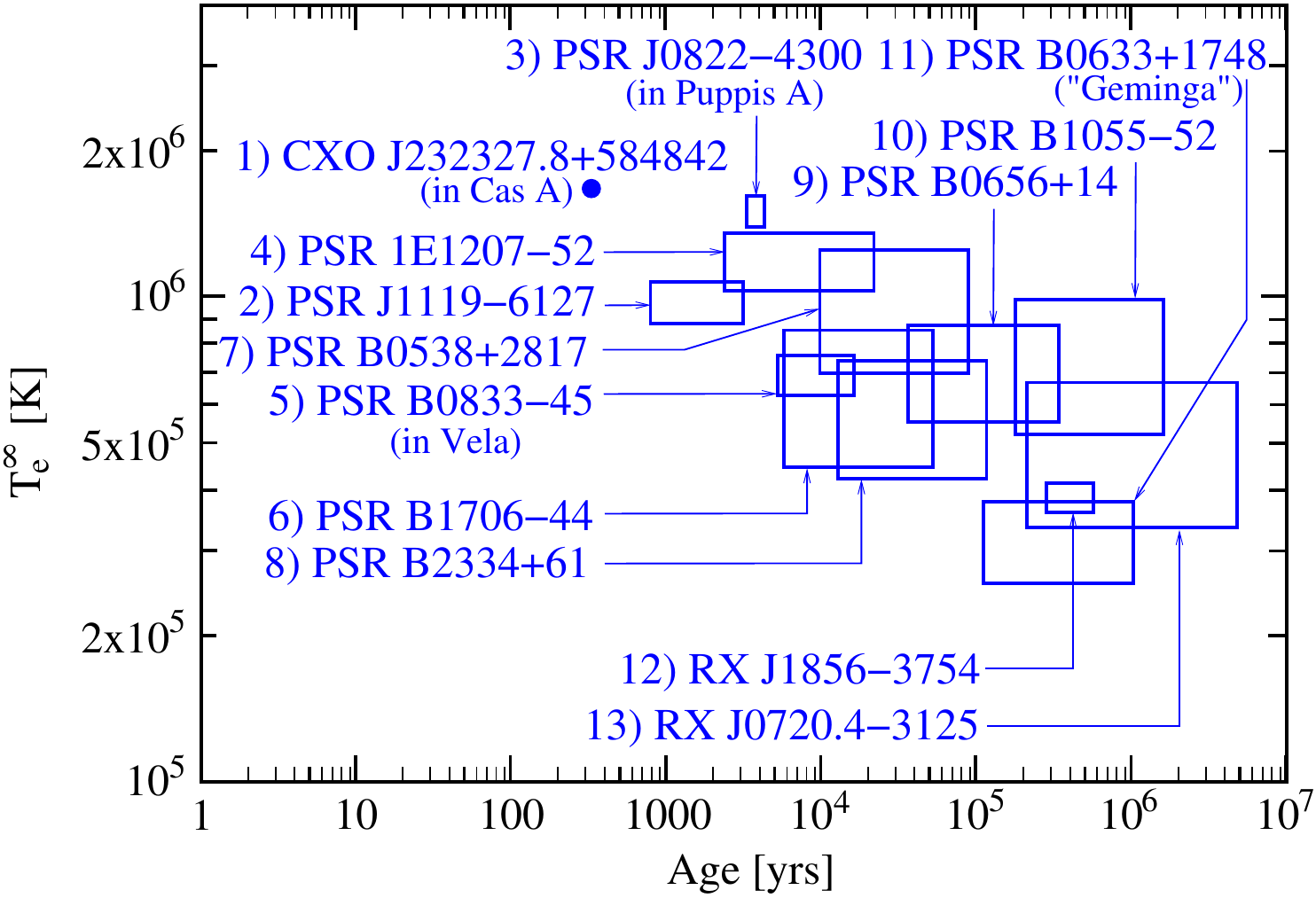}
\\
\vspace{10pt}
\includegraphics[width=.80\textwidth]{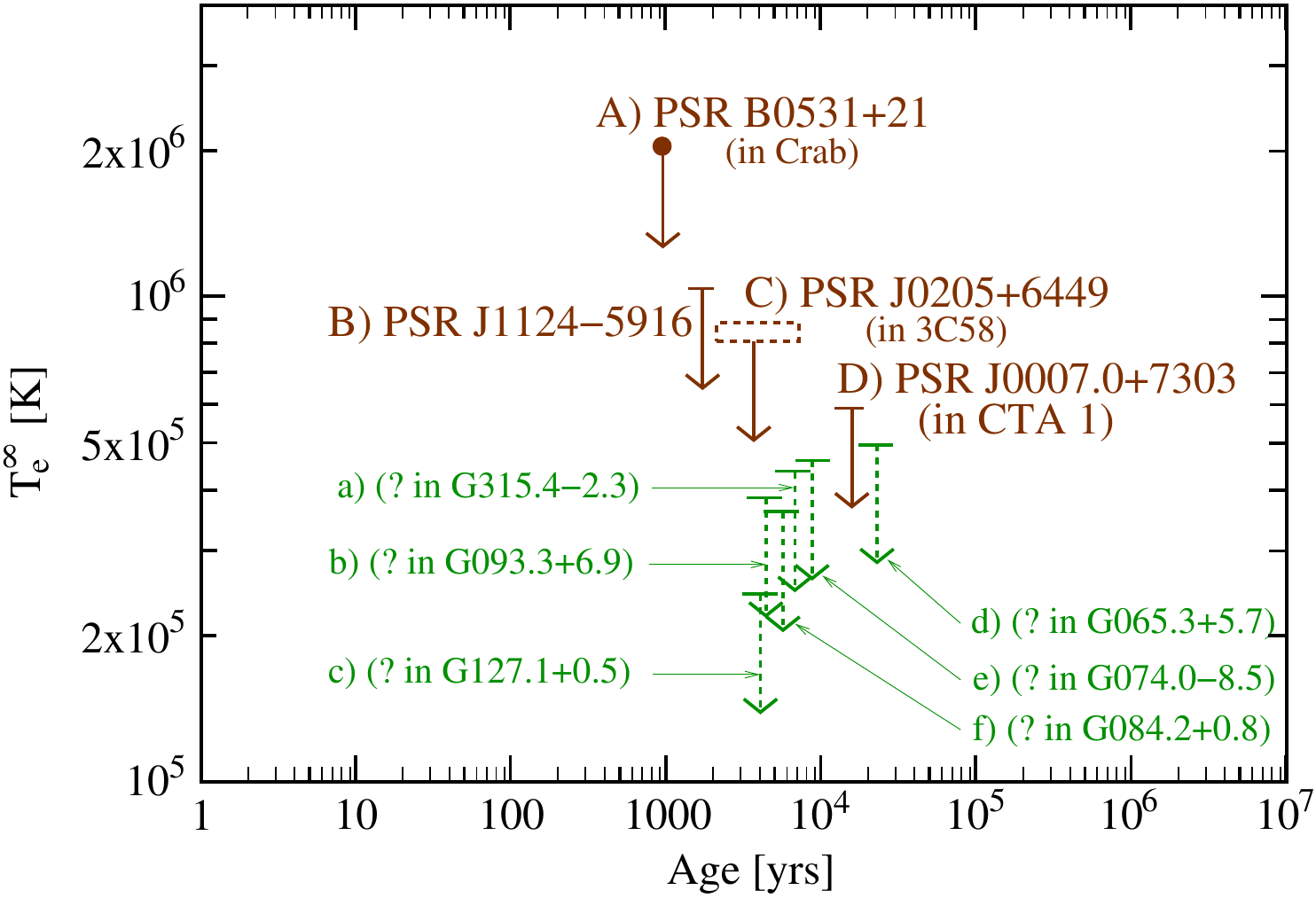}
\end{center}
\caption{The present data set of cooling neutron stars. See text for presentation.}
\label{Fig:Cool_Data}
\end{figure*}

It is useful at this juncture to compare these general behaviors with observations.
We summarize the observational information relevant to neutron star
cooling in \fig{Fig:Cool_Data}. 
The data is separated into three subsets of stars.
In the upper panel of \fig{Fig:Cool_Data} are presented 
13 stars
for which a thermal spectrum, in the soft X-ray band (0.1 - 3.0 keV)
is clearly detected. 
In the lower panel, we show 
data from 
4 pulsars, labeled (A) through (D), that are
seen in the X-ray band with a power-law spectrum, but whose detected emission is 
of magnetospheric origin.
Since the surface thermal emission 
from these 
4 stars is undetected, being 
covered by the magnetospheric emission, only 
upper limits on their effective temperatures 
could be inferred.
Finally, 
the lower panel of this figure contains 
6 upper limits resulting  
from the absence of detection of any emission from 
compact objects in 6 gravitational collapse
supernova remnants, labeled (a) through (f), from a search by 
Kaplan and collaborators \cite{Kaplan:2004nx}.
Since no compact object has been detected in these supernova remnants, 
some of them may contain an isolated black hole, 
but that is unlikely to be the case for 
all 6 of them.
The observations and estimates of ages and temperatures, or upper limits,
are detailed in \cite{Page:2004zr} and updated in \cite{Page:2009qf}.  

Estimates of surface
temperatures (redshifted to infinity) require atmospheric modeling in
which three factors are involved: (1) the composition of the
atmosphere (i.e., H, He or heavy element); (2) the column density of
X-ray absorbing material (mostly H) between the star and the Earth;
and (3) the surface gravitational redshift.  The column density is
important because the bulk of the emitted flux is absorbed before
reaching the Earth.  The redshift does not affect blackbody models, but can
influence heavy-element atmosphere models.  In most cases, this
quantity is not optimized, but set to the canonical value for $M=1.4
M_\odot$ and $R=10$ km stars.

Narrow spectral lines are not observed in any of these sources, so their
atmospheric compositions are unknown.  However, some information can be
deduced from the shape of the spectral distribution as heavy-element
atmospheres 
closely resemble blackbodies.  Fitting the flux and
temperature of a source to a model yields the neutron star radius, if
the distance is known.  In some cases, 
clothing a star with a light-element atmosphere results in a predicted
radius much larger than the canonical value, and one can infer the
presence of a heavy-element atmosphere.  Chang \& Bildsten
\cite{Chang:2004bh} have noted from such radius arguments that there
exists a trend for stars younger than $10^5$ years to be better fit by
light-element atmospheres and stars older than $10^5$ years to be
better fit by heavy-element atmospheres.  The possible evolution of
stars leading to this trend is discussed further in
\cite{Page:2004zr}.

Before embarking on the presentation of detailed cooling scenarios and comparing them
with data, it is worthwhile to discuss one general feature here.
The four oldest stars, numbered 10 to 13 in \fig{Fig:Cool_Data}, will appear to be much warmer than
most theoretical predictions.
These stars are prime candidates for the occurence of some late ``heating", i.e. the ``$H$" term in
\eq{Eq:cooling}.
Models can be easily adapted to incorporate such heating (which has nothing to do with neutrino emission or pairing) that may become important
at these times \cite{Page:2006ly}.
We will therefore concentrate on younger objects in the comparison with data.

\subsection{An Example: Fast Cooling with Pairing}
\label{Sec:Fast_Pair}

The thermal evolution of a neutron star undergoing fast neutrino
emission was illustrated in \fig{Fig:Cooling2} which showed that such
a star, after the initial crust relaxation phase, has a very low $T_e$
unless the neutrino emission is suppressed by pairing.  In the
presence of a gap that quenches fast neutrino emission, the resulting
$T_e$ is more sensitive to the size of the quenching gap than to the
specific fast neutrino process in action.

\begin{figure}
\begin{center}
\includegraphics[width=.80\textwidth]{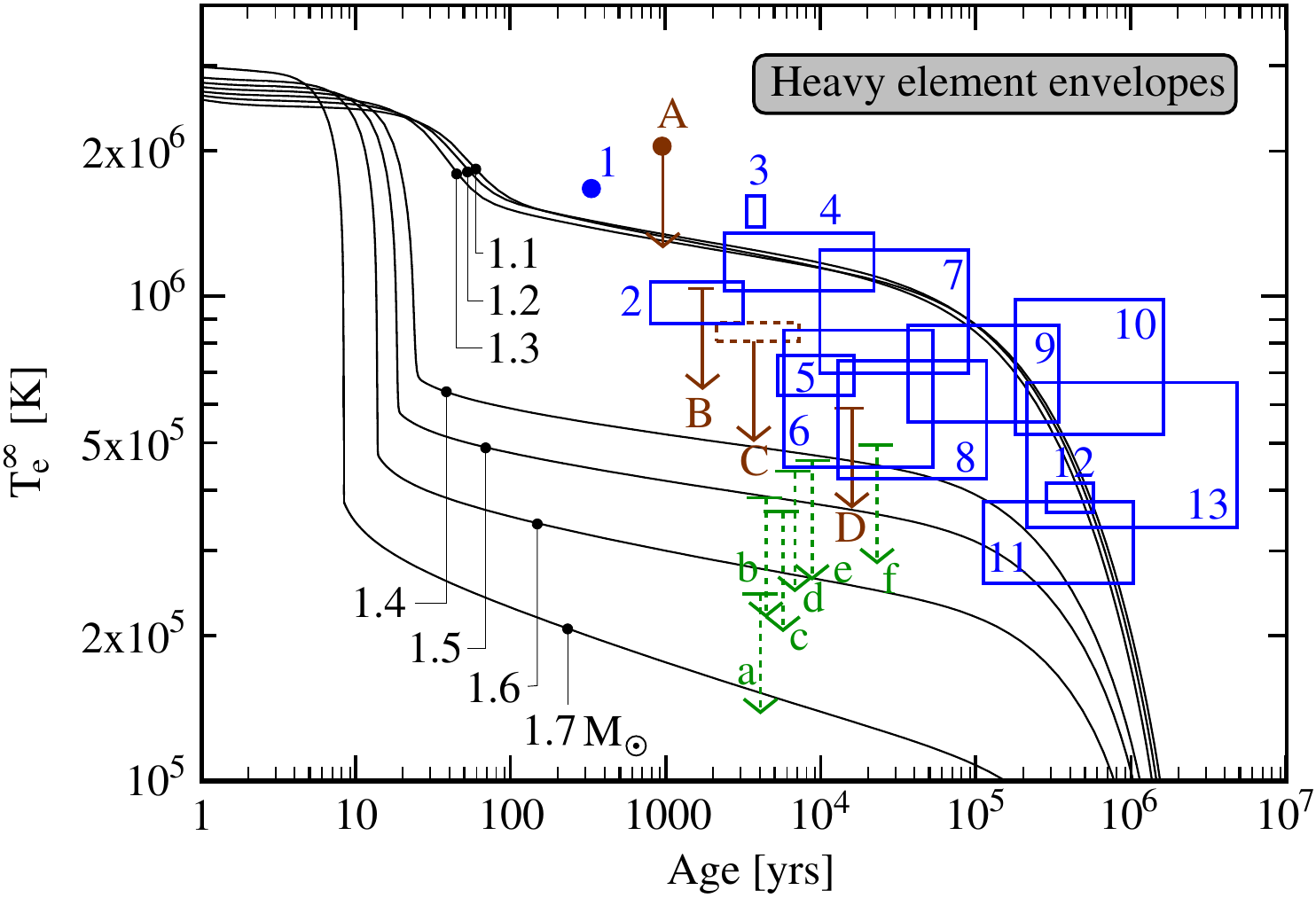}
\\
\vspace{10pt}
\includegraphics[width=.80\textwidth]{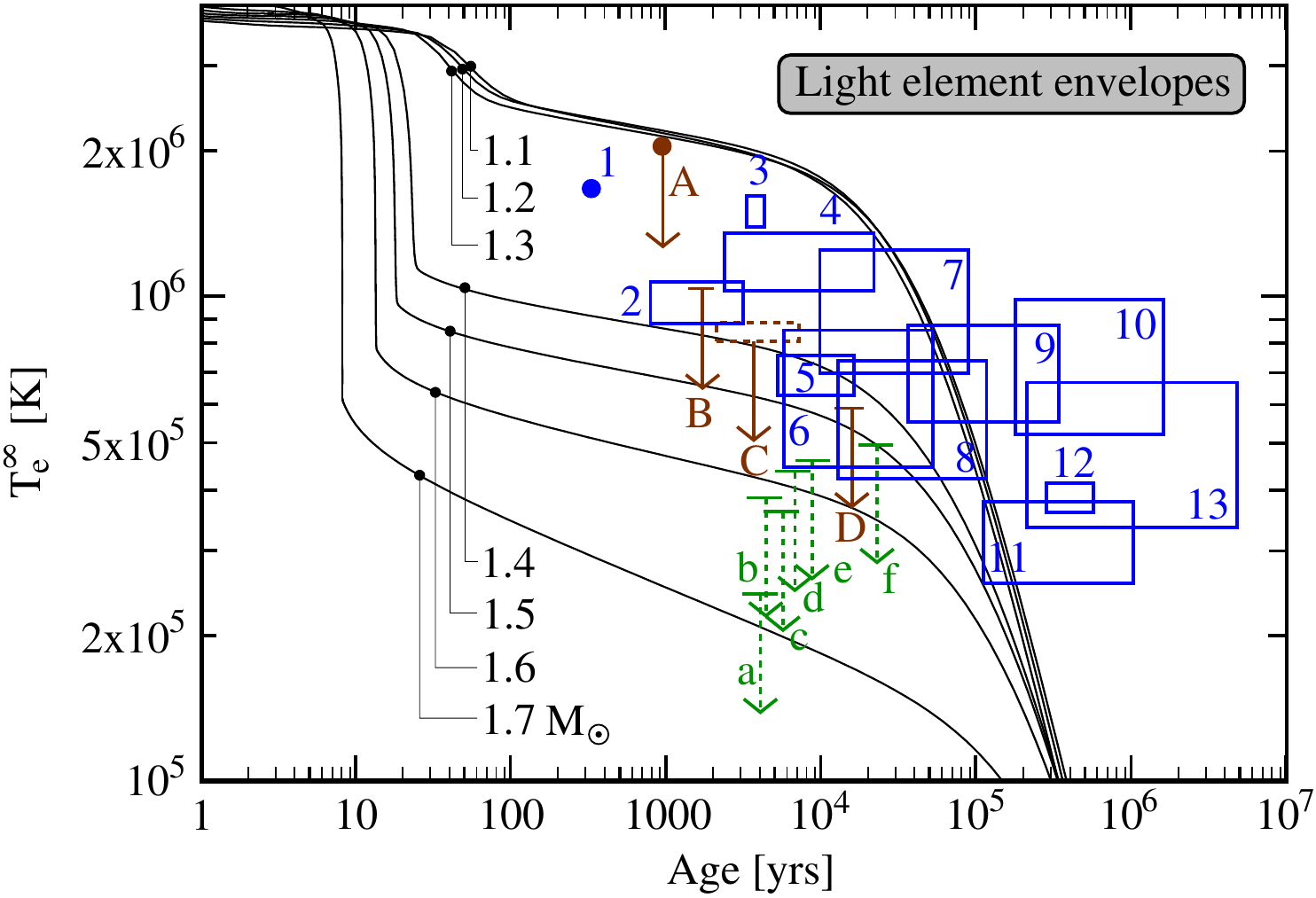}
\end{center}
\caption{Comparision of a cooling scenario with data.  The dense
  matter EOS used to build the stars is a PAL \cite{Prakash:1988oq}
  version \cite{Page:1992nx} that has only nucleons and leptons, and
  allows for the direct Urca process to occur in stars of mass larger
  than $M_\mathrm{cr}= 1.35 M_\odot$.  Cooling trajectories are shown
  for seven different masses and two different envelope chemical
  compositions.  The neutron $\Triplet$ gap from model ``b" of
  \fig{Fig:Tc_n} is implemented while the neutron and proton $\Singlet$
  gaps are from models ``SFB" of \fig{Fig:Tc_n1S0} and ``T" from
  \fig{Fig:Tc_p}, respectively.}
\label{Fig:Cool_PAL}
\end{figure}

In \fig{Fig:Cool_PAL}, we compare one specific scenario with the
observational data described in the previous Sec.~\ref{Sec:Cool_Data}.  
This scenario
employs the same EOS as in the models of \fig{Fig:Cooling_PAL} and is more realistic than what was described in
Sec.~\ref{Sec:Numerical} 
insofar as it takes into account uncertainties concerning the chemical
composition of the
envelope.
If the neutron star has an envelope containing 
the maximum possible amount of light elements,
($\sim 10^{-6} M_\odot$), its $T_e$ is raised by a factor of two compared to the case when its envelope 
contains only heavy elements.
Results for these two extreme scenarios of envelope composition are separately shown in the two panels of \fig{Fig:Cooling2}.
In each panel, the first noticeable feature is the mass dependence of results when the mass exceeds $1.35 M_\odot$
and the direct Urca process is activated.
This is partially due to the increasing size of the fast neutrino emitting ``pit" as the star's mass increases,
but more dominantly due to the decrease in
the neutron $\Triplet$ gap with increasing density. 
More massive neutron stars, at least in this model, will have the direct Urca process acting in their inner cores
for a longer time until their central temperatures drop below the corresponding central values of $T_c$.
Notice that 
in a model where the $\Triplet$ gap keeps growing with density as, in the ``3bf" model of \fig{Fig:Tc_n},
this mass dependence 
essentially disappears.

When comparing the  
cooling trajectories of \fig{Fig:Cool_PAL} with the displayed data we
find that, with the exception of some of the oldest stars, all
observed temperatures can be reasonably well fit.
This result implies that different neutron stars have different masses,
and some stars require
an envelope containing light elements. 
This conclusion is common to 
 all fast cooling scenarios that involve either 
the simple direct Urca process with nucleons or
any direct Urca process with hyperons or quarks, as well as 
scenarios with charged mesons condensates.
The successful {\it recipe} consists of having some fast neutrino process allowed in massive enough neutron stars.
Then, in order to prevent intermediate mass stars from cooling too much, it is necessary to have a gap, or several gaps, 
large enough so that all fast neutrino processes are quenched at some point.

\subsection{Can Quark Matter Be Detected from Cooling Observations?}
\label{Sec:Quark_cool}

The previous subsection presented one specific fast cooling scenario, 
while Sec. \ref{Sec:Numerical} only described very general trends of
fast cooling scenario with pairing.
It is interesting, however, to briefly consider
if observations could determine not only whether or not fast cooling
has occurred but also which fast cooling mechanism operates.  
One of the first studies
of hybrid stars with superconducting
quarks~\cite{Page:2000kl} showed, in fact, that it will be difficult
to establish that neutron stars contain deconfined quark matter based
solely on the cooling data of isolated neutron stars.  When quark
superconductivity is present, for example in the 2SC phase, neutrino
emission is suppressed by a factor of $e^{-\Delta/T}$ for each flavor
and color quark which is paired. Cooling in the CFL phase is special
because of the Goldstone bosons, which provide new cooling processes,
and other pairing configurations also have their own cooling
behaviors.  Quark gaps that are typically larger than baryonic gaps
therefore imply that hybrid star cooling is driven by other degrees of
freedom present at lower densities.  On the other hand, small quark
gaps, or the absence of quark matter entirely, imply similar cooling
behaviors that are easily reproduced by models with varying masses.

\begin{figure}[hbt]
\begin{center}
\includegraphics[width=0.48\textwidth]{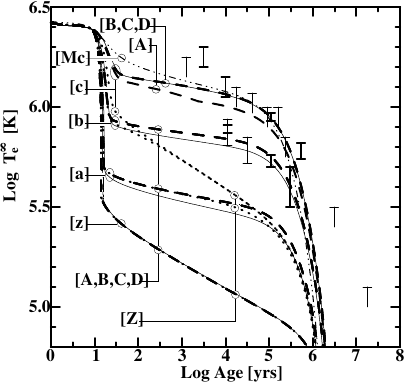}
\includegraphics[width=0.48\textwidth]{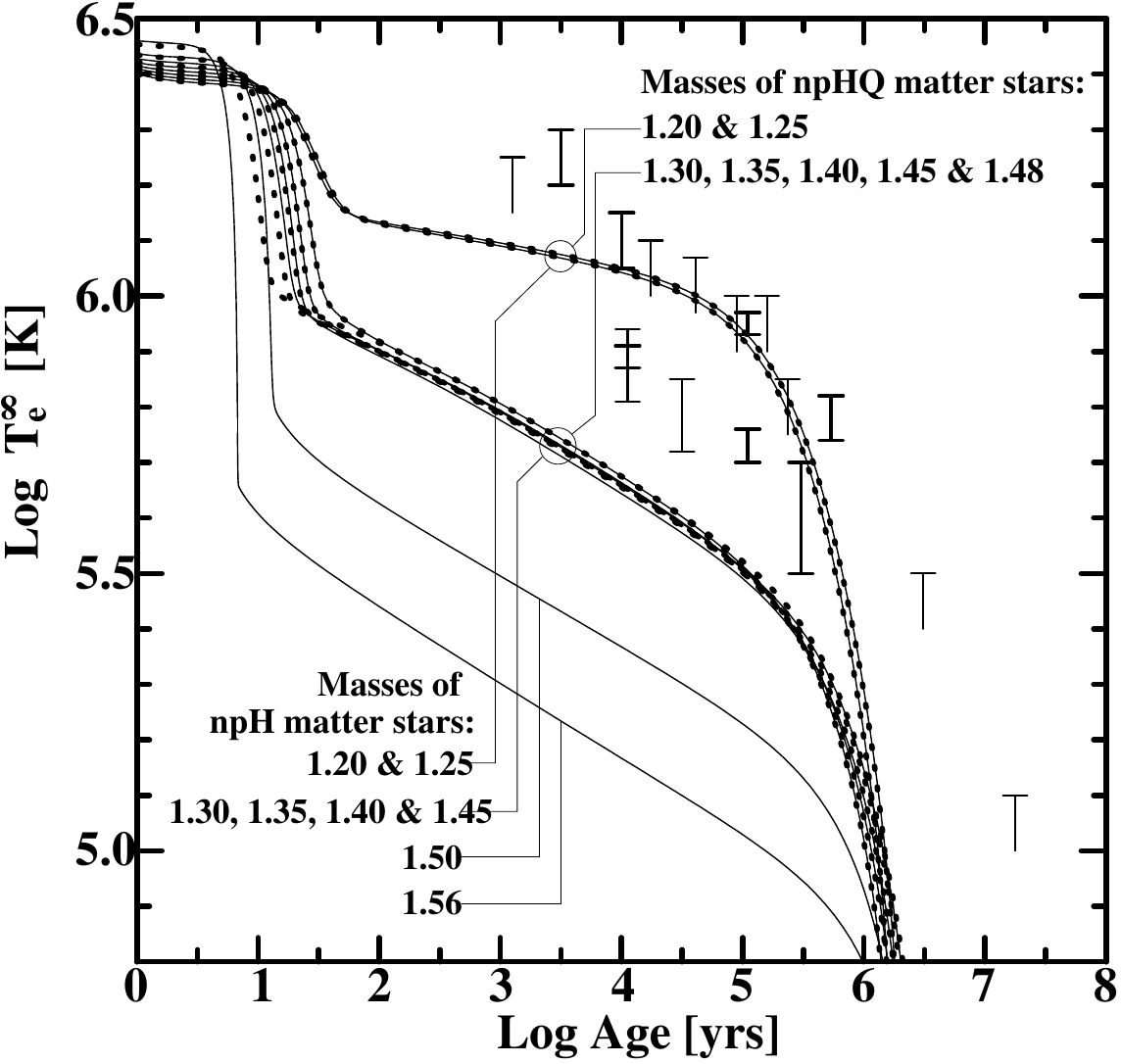}
\end{center}
\caption{
Left panel: cooling of a $1.4 M_\odot$ neutron star composed only of nucleons (continuous lines)
or nucleons and quark matter (dashed and dotted curves), with various assumptions on the pairing gaps.
Right panel: mass dependent cooling histories for two dense matter scenarios: ``npH" matter contains nucleons and
hyperons; ``npHQ" matter also includes deconfined quarks. In these two scenarios, pairing gaps are assumed fixed.
Figures from \cite{Page:2000kl}.}
\label{Fig:ppls_fig3}
\end{figure}

The above conclusions are demonstrated in \fig{Fig:ppls_fig3}, in different ways in the two panels,
using two pairs of EOSs.
In the left panel, cooling curves with a fixed neutron star mass and 
employing two EOSs, one containing only nucleons (np) and the 
other with nucleons with quark matter (npQ), are shown.
These simulations utilize a range of nucleon and quark
pairing gaps.
In the right panel, cooling curves of neutron stars of various masses 
and employing two EOSs, one with nucleons and hyperons (npH) and 
the other with quark matter added (npHQ), are compared.
This second set of simulations utilizes fixed nucleon/hyperon/quark 
pairing gaps.

We refer the reader to \cite{Page:2000kl} for details, but the figure illustrates that, given our poor knowledge
of the size and density extent of the various pairing gaps, nearly indistinguishable cooling curves
can be obtained with very different dense matter models, i.e with or without hyperons and/or with or without quark matter.
Complementary studies, as, e.g.,  in \cite{Blaschke:2000kl,Grigorian:2005kl,Hess:2011kl,Negreiros:2012kl},
confirm this conclusion.

\subsection{Can Superfluid Gaps Be Measured, or Predicted?}
\label{Sec:Good&Bad_Conclusion}

We are led, by the discussions in Sec.~\ref{Sec:Fast_Pair} and Sec.~\ref{Sec:Quark_cool}, to an intriguing  result:
{\it Any equation of state that permits  
fast neutrino emission is compatible with the presently available cooling
data on isolated neutron stars IF pairing is present with gap(s) of the appropriate size(s)}.\footnote{The ultimate
theory of neutron star matter will yield the equation of state and the superfluid gaps. However, at present, this
is far from being the case since the gaps, as described in Sec.~\ref{Sec:Pairing}, are very sensitive to many intricate 
Fermi surface processes  to which the equation of state, which is a bulk property, is quite insensitive.}
This result, unfortunately, is ambiguous: it means that theoretical
uncertainties in the description of dense matter make it difficult to
determine the composition of neutron star cores from isolated neutron
star cooling observations alone.

One could, however, take an optimistic approach and reverse the argument.
As the cooling of neutron stars is very 
sensitive to the presence, and the sizes, of  pairing gaps, we can 
{\it apply this fact to attempt to measure the gap(s)} \cite{Page:1992nx}.
This is in line with the idea of using neutron stars as extra-terrestrial laboratories to study dense matter.
However, as 
massive neutron stars may contain hyperons, meson condensates or deconfined quarks, 
the question arises:
{\it which gap would we be measuring} \cite{Page:2000kl}{\it ?}
  This situation, as demonstrated in the {\it maximally} complicated models of \fig{Fig:ppls_fig3},  
may appear to render further studies of neutron star cooling a waste of time.

However, our discussion so far has assumed the presence of ``fast" cooling.
Is fast cooling, in fact, actually necessary?  This question motivates
the {\it minimal} cooling scenario, to which we turn in Sec.~\ref{Sec:Minimal}, which
considers only models in which enhanced cooling processes do not occur.


\section{Minimal Cooling and Superfluidity}
\label{Sec:Minimal}

The comparison in the previous section of theoretical cooling models with available data suggests the presence of extensive pairing in the cores of neutron stars. 
However, the plethora of possible scenarios makes it difficult to go beyond this 
generic conclusion.
And, in spite of the fact that many theoretical models of dense matter predict
the presence
of some form of  ``exotica", one should still ask the question 
``do we need them?" and, if yes, how strong is the evidence for them?
To address these questions, the {\it minimal cooling paradigm} was developed 
in \cite{Page:2004zr,Page:2009qf}. In this scenario, all possible fast neutrino emission processes (from direct Urca processes involving baryons or quarks, and from any form of exotica)  are excluded {\it a priori}.  
Superfluid effects along with the PBF process are, however, included in the slow neutrino emission processes, such as the modified Urca processes involving neutrons and protons. 
This is a very restrictive scenario that minimizes the number of degrees of freedom, 
but fully incorporates 
uncertainties associated with the equation of state, envelope composition and its mass, etc.
A detailed presentation of this scenario can be found in \cite{Page:2004zr,Page:2009qf}
and a variant of it was developed by the Ioffe group in \cite{Gusakov:2004ys,Kaminker:2006hc}.
Extensive studies of these two groups have pinned down 
two major sources of uncertainties: (i)  
our present lack of knowledge on the chemical composition of the envelope, and (ii) 
the size and extent of the neutron $\Triplet$ gap.
It turns out that, in the absence of any ``fast 
process'', neutrino emissivity resulting from the PBF process involving 
$\Triplet$ Cooper pairs  
is a major factor in the thermal evolution of a neutron star and its effect is strongly dependent on the size of the neutron $\Triplet$ gap.
We describe in some detail effects of the PBF process in the next subsection 
before turning to a comparison of  
the predictions of the minimal cooling paradigm with available data.

\subsection{Effects of the PBF Induced Neutrino Process}

The PBF process is distinctive 
in the sense that, in any layer in the star's interior,
it turns on when the temperature $T$ reaches the corresponding $T_c$ of 
the ambient density. Then the pairing phase transition is 
triggered and while $T$ is not too much lower
than $T_c$, there is constant formation and breaking of Cooper pairs induced by thermal excitation. 
However, when $T$ falls
below $\sim 0.2 \, T_c$, there is not 
enough thermal energy
to break pairs and the process 
shuts off (see its control function displayed
in the right panel of \fig{Fig:PBF}).
As described in Sec.~\ref{Sec:PBF} the neutrino emissivity of the PBF process is significant only
in the case of anisotropic pairing, as in the case of a $\Triplet$ gap.

\begin{figure*}[ht]
\begin{center}
\hspace*{-.1cm}
\includegraphics[width=.70\textwidth]{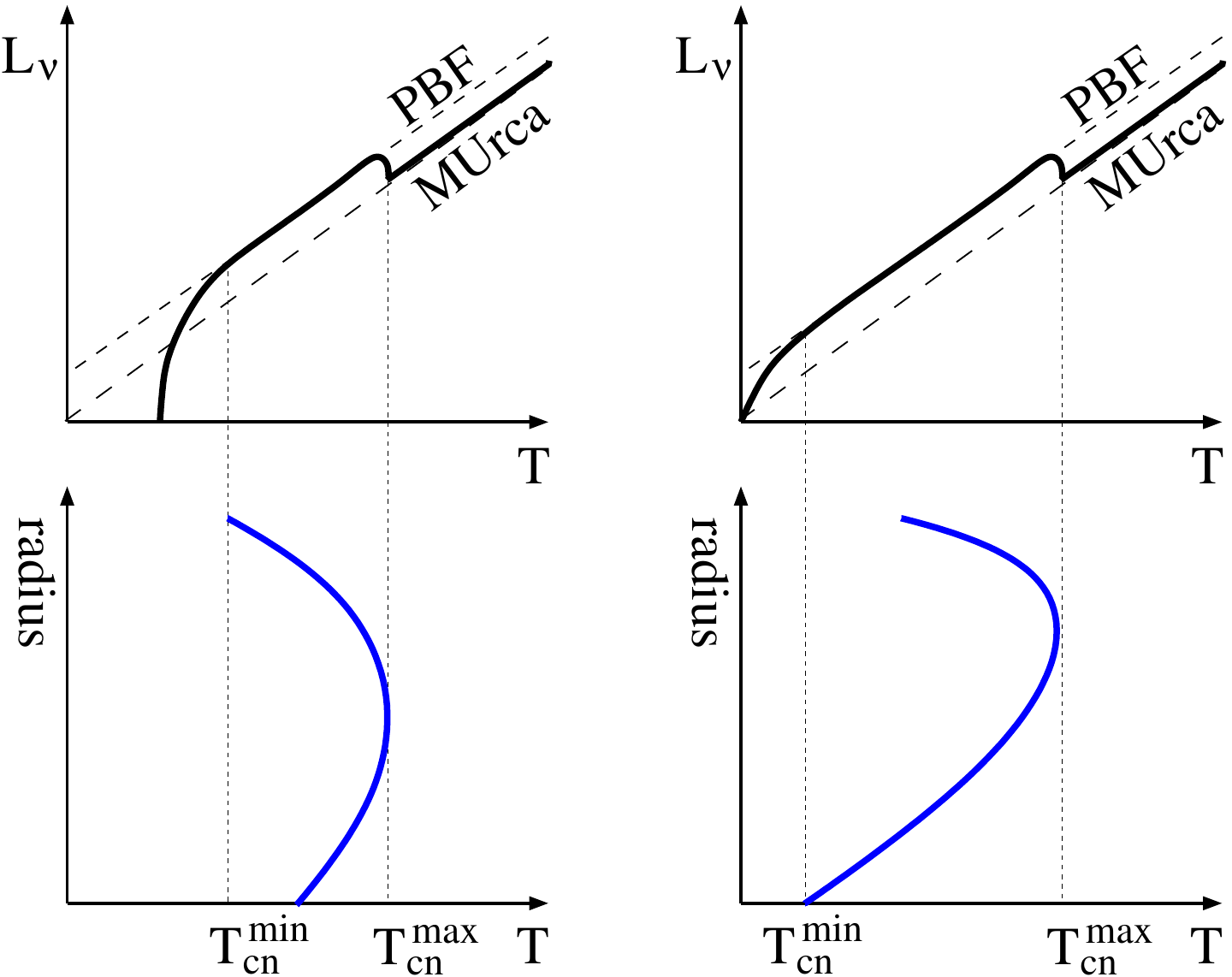}
\end{center}
\caption{Schematic illustration of the PBF process induced neutrino luminosity as controlled by the  shape of the $T_c$ curve.
See text for a detailed description.}
\label{Fig:Coop-Scheme}
\end{figure*}

The schematic \fig{Fig:Coop-Scheme} illustrates the effect of the pairing phase transition
on the  total neutrino luminosity of the star, $L_\nu$, 
depicted in the upper part of each 
panel as a function of the core temperature $T$.
The long-dashed curves labelled ``MUrca" show the modified Urca luminosity, $L_\nu^\mathrm{MU} \propto T^8$
in the absence of neutron $\Triplet$ pairing
while the short-dashed lines labelled ``PBF" show the maximum possible PBF luminosity, 
$L_\nu^\mathrm{PBF} \propto T^7$, i.e. from \eq{Eq:Q_PBF} with the control function $R_j$ set to 1.
The lower part of each 
panel shows a neutron $\Triplet$ $T_c$ versus radius curve 
(displayed with its $T$ axis horizontal to coincide with the upper part's $T$ axis).
Assuming the temperature $T$ to be nearly uniform in space in the entire core,
the core $T$ profile is just a vertical straight line that 
moves from the right to the left as the star cools down.
Each panel considers a different type of $T_c$ curve and 
in its upper part, the thick continuous line shows the actual evolution of the total $L_\nu$ in the presence of pairing.
As the star cools, and while $T>T_{cn}^\mathrm{max}$, the total $L_\nu$ is dominated by the MUrca process 
with small contributions from other much less efficient processes. 
When $T$ reaches $T_{cn}^\mathrm{max}$, there is a thick layer in which the phase transition starts and
where neutrino emission from the PBF process is triggered which  increases $L_\nu$.
As $T$ decreases, but is still larger then $T_{cn}^\mathrm{min}$, there will be one or two layers in the star 
where $T$ is only slightly lower 
than the corresponding value of $T_c$ in that layer(s) and where the PBF process is efficiently acting.
(At these times, neutrino emission is strongly suppressed
in the layers where $T$ is much smaller than $T_c$.)
Finally, when $T$ drops
well below $T_{cn}^\mathrm{min}$, neutrino emission by the dominant processes 
is suppressed everywhere in the core and $L_\nu$ drops rapidly 
to reflect the much less efficient neutrino emissivity from the crust.
The higher the value of $T_{cn}^\mathrm{max}$ the earlier in the star's history will the enhanced cooling from the PBF
process be triggered, 
while the lower the value of $T_{cn}^\mathrm{min}$ the longer this enhanced cooling will last,
as illustrated by the difference between the left and right panels of  \fig{Fig:Coop-Scheme}.
This description will be pursued with a simple analytical model in Sec.~\ref{Sec:CasA_analytical} and \ref{Sec:CasA_proton}.

\begin{figure*}[h]
\begin{center}
\hspace*{-.2cm}\includegraphics[width=12.5cm]{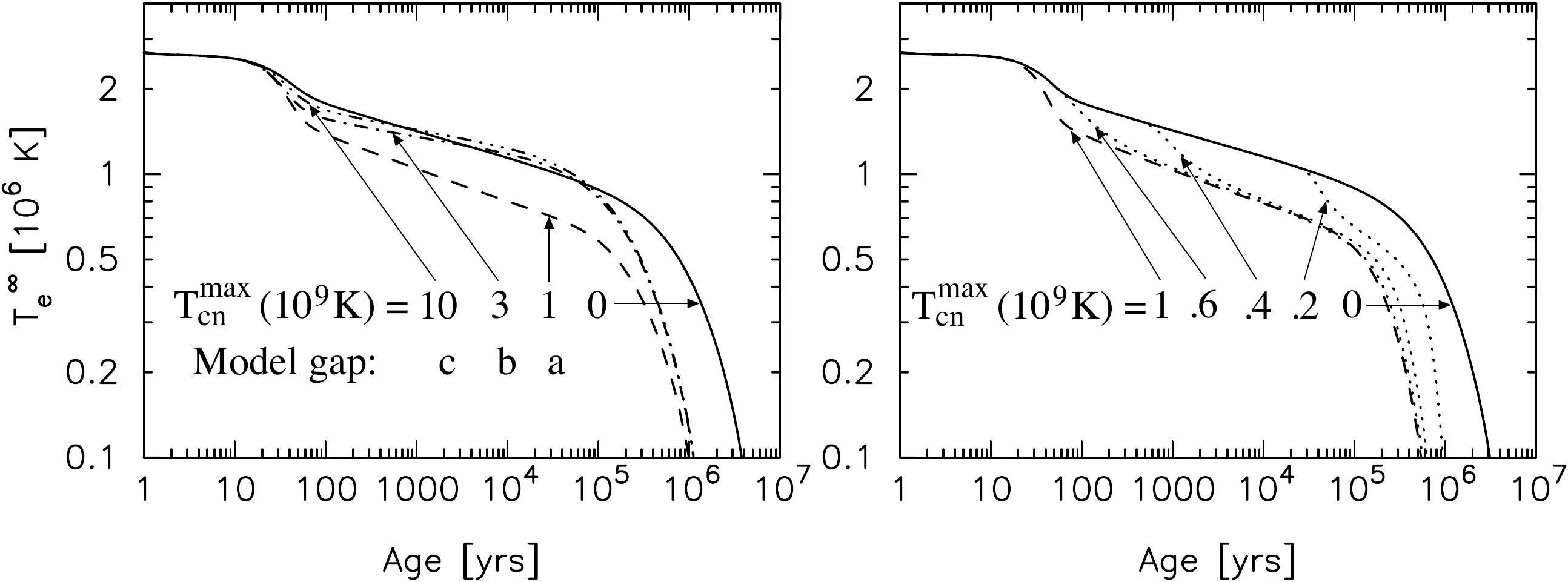}
\end{center}
\caption{Neutron star cooling trajectories varying the magnitude of
  the neutron $\Triplet$ gap.
  See text for description.}
\label{Fig:Cooling_SF}
\end{figure*}

From the above discussion,  
we learn 
that the effect of the PBF process depends 
strongly on 
the size and density dependence of the neutron $\Triplet$ gap. 
To explore 
the possible impact of these two physical ingredients, 
we show in \fig{Fig:Cooling_SF} two sets of cooling trajectories resulting from 
different assumptions. 
For contrast, 
both panels show a curve with no pairing, i.e., the same as the slow cooling model marked ``Normal"
in the left panel of \fig{Fig:Cooling2}.
The various cooling curves are labelled by 
$T_{cn}^\mathrm{max}$, 
the maximum value of $T_c$ reached in the stellar core.
In the left panel, large values of  $T_{cn}^\mathrm{max}$ corresponding to 
the three phenomenological gaps ``a", ``b", and ``c"
of the right panel of \fig{Fig:Tc_n}, are employed.
All three gaps lead to the same result 
during the photon cooling era ($t>10^5$ yrs), but case ``a" strongly differs from the other two
cases  during the neutrino cooling era.
All three models have large enough gaps and $T_{cn}^\mathrm{max}$ that the enhanced cooling from the PBF
process started well before the end of the crust relaxation phase. 
The gaps and  $T_{cn}^\mathrm{min}$ of models ``b" and ``c" are so large that 
they induce a very strong suppression of $L_\nu$ at early times and hence high values of $T_e$
during the neutrino cooling phase.
In contrast, the gap ``a" has a moderate $T_{cn}^\mathrm{max}$ and also a very small $T_{cn}^\mathrm{min}$
which guarantees that at ages $10^2 - 10^5$ yrs there is alway a significant layer going through the pairing phase transition
in which $L_\nu^\mathrm{PBF}$ is large which results in a colder star.

As   small gaps have the strongest effect in inducing some enhanced cooling of the neutron star, we explore in the
right panel of \fig{Fig:Cooling_SF} the effect of increasingly small neutron $\Triplet$ gaps by 
scaling down the model ``a'' gap 
by a factor 0.6, 0.4, and 0.2, respectively.
A new feature now emerges:
with decreasing values of $T_{cn}^\mathrm{max}$, the pairing phase transition is initiated at 
progressively  later stages. 
This feature is signaled by cooling trajectories (dotted curves) departing from the $T_{cn}^\mathrm{max} =0$ trajectory.
This late onset of pairing then manifests itself as a sudden rapid cooling of the star
(due to the sudden increase of $L_\nu$ from $L_\nu^\mathrm{MU}$ to $L_\nu^\mathrm{PBF}$ at the moment when $T$
reaches $T_{cn}^\mathrm{max}$)
and interestingly within the age range for which we have many observations available.

Armed with these considerations, we turn now
to compare the predictions of the minimal cooling paradigm with data.

\subsection{Comparison of Minimal Cooling With Observations}
\label{Sec:Comparison}
\begin{figure}
\begin{center}
\includegraphics[width=0.80\textwidth]{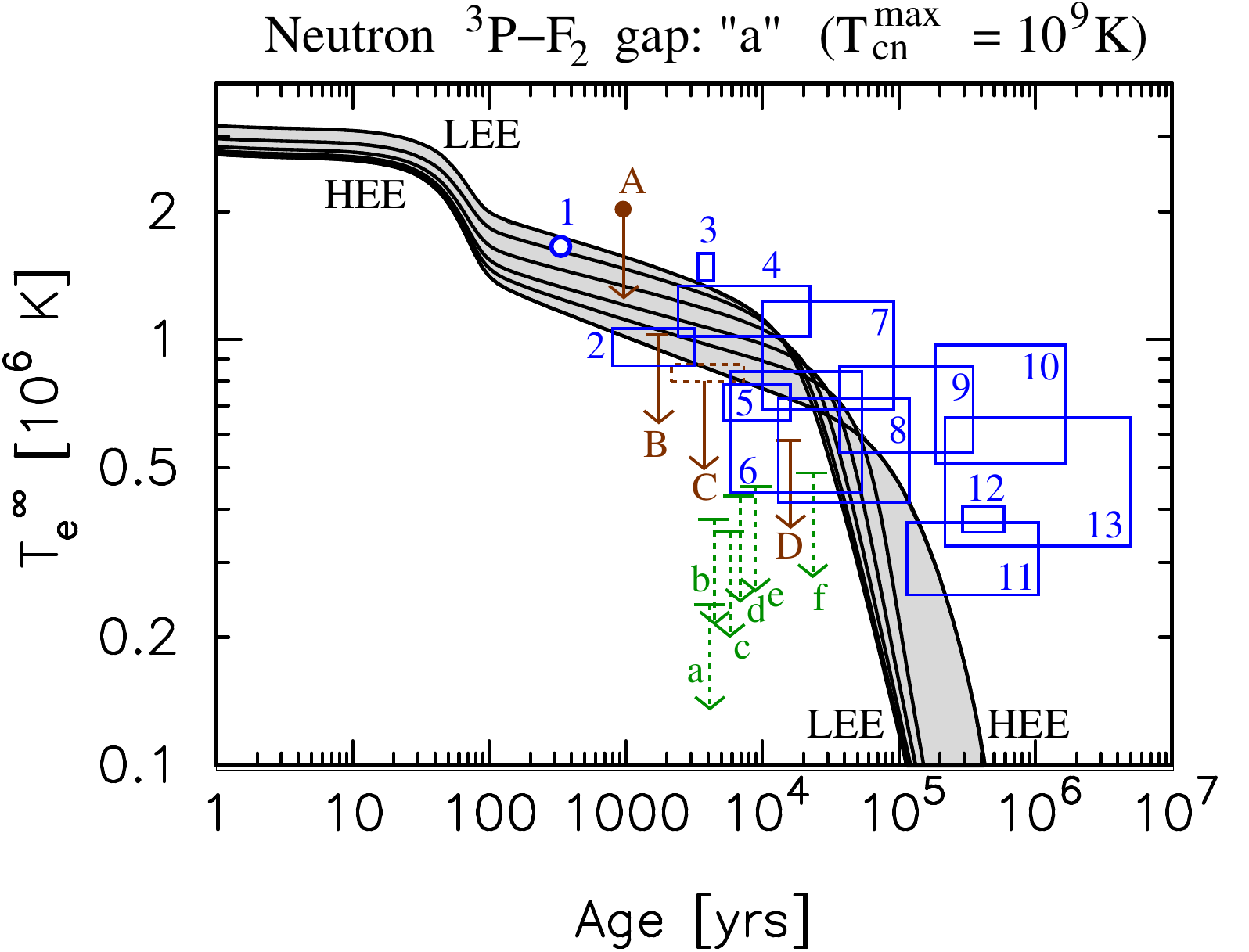}
\\
\vspace{15pt}
\includegraphics[width=0.80\textwidth]{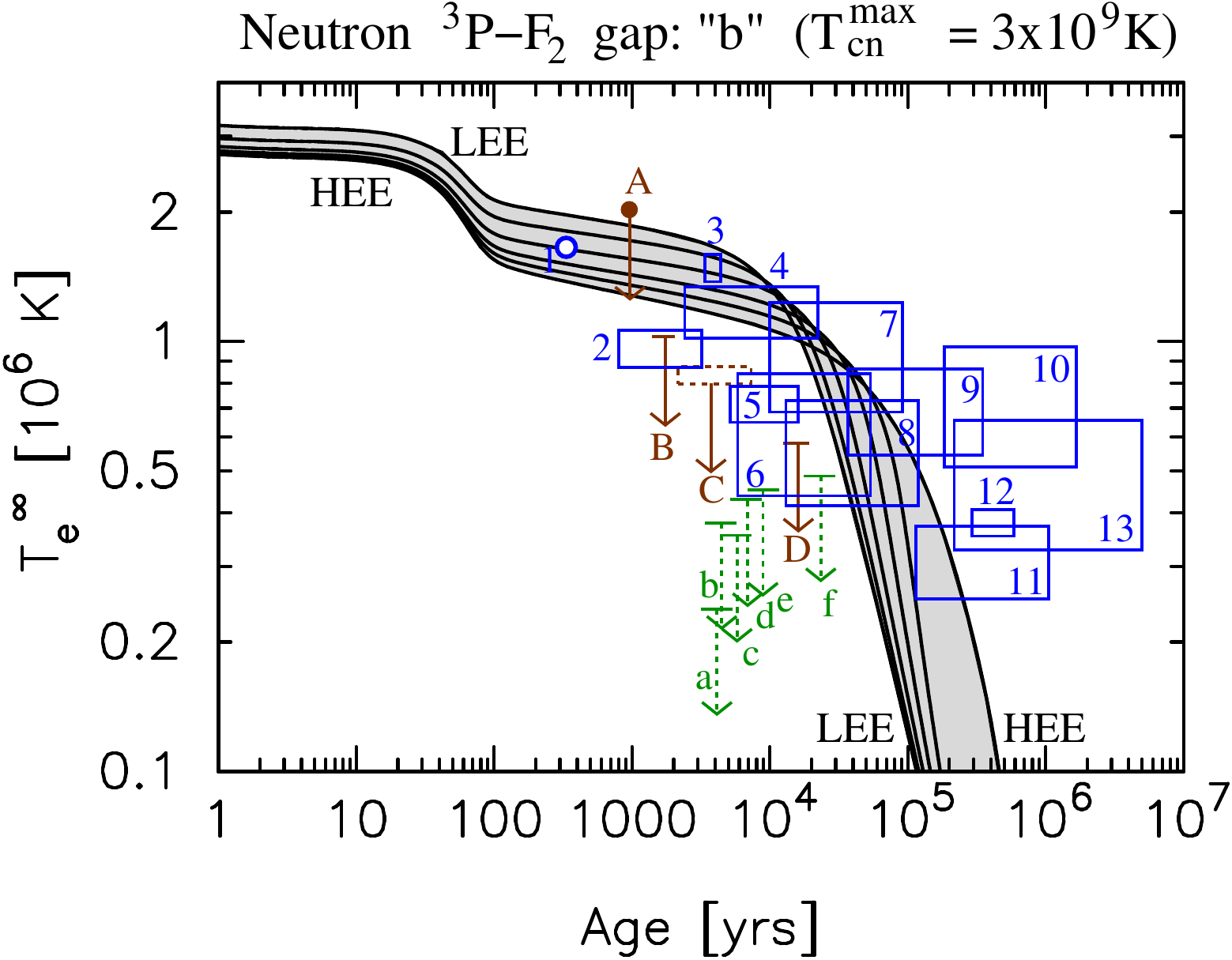}
\end{center}
\caption{Comparison of two minimal cooling scenarios with observational data.
               The neutron $\Triplet$ gaps employed are shown in \fig{Fig:Tc_n}.
               See text for description.  Adapted from the results of  \cite{Page:2009qf}}
\label{Fig:Cool_Minimal_a-b}
\end{figure}

\fig{Fig:Cool_Minimal_a-b} shows ranges of predicted thermal
evolutions of a canonical neutron star of $1.4 M_\odot$ built with the EOS of
Akmal \& Pandharipande \cite{Akmal:1998fk}.  
The density dependence of the symmetry energy in this EOS 
precludes the occurrence of the direct Urca process in a $1.4 M_\odot$ 
neutron star. 
Furthermore, this EOS does not permit
hyperons, kaon condensates, pions or deconfined quark matter, all of which could
lead to enhanced cooling. It was shown in \cite{Page:2004zr} that all EOS's compatible with
the restrictions imposed by the minimal cooling scenario 
yield almost identical predictions.
Moreover, within the range of neutron $\Triplet$ pairing gaps explored, for which $T_{cn}^\mathrm{max}$ is reached
at a density that is smaller than the central density of any neutron star (see right panel of \fig{Fig:Tc_n}),
the neutron star mass of any model also has very little effect \footnote{
The Ioffe group, in their version of minimal cooling \cite{Gusakov:2004ys,Kaminker:2006hc},
assumed neutron $\Triplet$ gaps that are small at $\rhonuc$ and grow rapidly at high densities.
Consequently, $T_{cn}^\mathrm{max}$ is reached in the center of the star and stars of increasing mass
have increasing values of $T_{cn}^\mathrm{max}$ resulting in a mass dependence of the cooling evolution.
}.

The various possible assumptions about the neutron $\Singlet$ gap have only a small effect on the early
crust thermal relaxation phase, phase (B) of \fig{Fig:Cooling2}.
Changes in the proton $\Singlet$ gap produce significantly differing  
effects, but 
dominant effects are due to changes in the 
neutron $\Triplet$ gaps.
The full range of possibilities is (almost) covered by 
considering the two neutron $\Triplet$ gaps in models 
``a" and ``b" of \fig{Fig:Tc_n} and is depicted in the two 
panels of \fig{Fig:Cool_Minimal_a-b}.
The various curves in the grey shaded areas show the uncertainty 
in the predictions due to lack of knowledge of the envelope 
compositions, which can range from pure heavy elements 
(``HEE") to pure light elements (``LEE").
In agreement with the results of \fig{Fig:Cooling_SF} and the discussion of the previous subsection, 
the large gap ``b" results in warmer stars than the smaller gap ``a"
during the neutrino cooling era.

Comparing model tracks with observations, we can conclude, following
\cite{Page:2004zr} and \cite{Page:2009qf}, that if the neutron $\Triplet$ gap 
is of small size ($T_{cn}^\mathrm{max} \sim10^9$ K) as that of model ``a", all 
neutron stars with detected thermal emission are compatible with the minimal
cooling scenario with the exception of the oldest 
objects (labelled 10, 12, and 13 in \fig{Fig:Cool_Minimal_a-b}, see the discussion at the end of  Sec.~\ref{Sec:Cool_Data}).
These older stars are candidates for the presence of internal heating,
i.e., the ``H" term in \eq{Eq:cooling} \cite{Page:2006ly}, a possibility that can easily be incorporated
within the minimal cooling scenario.
In addition, it is
found that young neutron stars must have heterogeneous envelope
compositions: some must have light-element compositions and some must
have heavy-element compositions, as noted above.
The updated comparison by \cite{Page:2009qf}
more precisely quantifies the required size of this gap.
In the notations used in \fig{Fig:Coop-Scheme}, 
maximal compatibility of 
the minimal scenario requires the neutron $\Triplet$ gap to satisfy
$T_{cn}^\mathrm{max} \simge 5\times 10^8$ K and 
$T_{cn}^\mathrm{min} \simle 2\times 10^8$ K.
The constraint on $T_{cn}^\mathrm{max}$, which determines when the PBF process will be triggered,
derives from the necessity of having a PBF enhanced $L_\nu$ already acting in the 
youngest observed stars at ages $\simle 10^3$ yrs.
On the other side, the constraint on the low $T_{cn}^\mathrm{min}$, which determines when 
core neutrino emission will be strongly suppressed, including the PBF process, 
is obtained by the requirement that $L_\nu$ should not be strongly suppressed
before the star reaches an age of a few times $10^4$ yrs.
This constraint assures
compatibility with the low $T_e$ of objects 5 and 6.
Considering the smooth $k_F$, and thus density, dependence of gaps, the constraint on $T_{cn}^\mathrm{min}$
likely prevents $T_{cn}^\mathrm{max}$ from being too large:
gaps with $T_{cn}^\mathrm{max} \gg 10^9$ K usually also have $T_{cn}^\mathrm{min} \gg 10^8$ K.

In the case the neutron $\Triplet$ gap is as large
as our model ``b'',
one reaches the opposite conclusion that minimal cooling cannot explain
about half of the 
young isolated stars, implying 
the occurence of some fast neutrino process.
The same conclusion is also reached in the case this gap is very small.
As seen in the left panel of \fig{Fig:Cooling_SF}, a vanishing gap leads to cooling trajectories very close to those obtained using 
large gaps during the neutrino cooling era.

Only one object, the pulsar  in the supernova remnant CTA 1 (object ``D" in the figure)
for which only an upper limit on $T_e$ is available, 
 stands out as being significantly below all
predictions of minimal cooling.

Finally, irrespective of the magnitude of the 
neutron $\Triplet$ gap,
if any of the six upper limits marked ``a" to ``f" in \fig{Fig:Cool_Minimal_a-b}
are, in fact, neutron stars, they can only be explained by enhanced cooling.
As it is unlikely that all of these remnants contain
black holes, since the predicted neutron star/black hole abundance ratio
from gravitational collapse supernovae is not small, these objects
provide additional evidence in favor of enhanced cooling in some
neutron stars\footnote{
Further evidence for enhanced cooling is provided by two neutron stars, 
SAX J1808.4-3658 and 1H 1905+00,  
 in transiently accreting 
low-mass X-ray binaries.
In contrast to the the six, yet to be detected, candidates ``a" to ``f",
these two stars are known to be neutron stars from their characteristics during the accretion phases,
but their thermal emission is not detectable after accretion stops, implying extremely fast
neutrino emission occurring 
in their cores \cite{Heinke:2009dq,Jonker:2007cr}.
}.  
As the nuclear symmetry energy likely 
increases continuously with density, larger-mass stars, which have larger central
densities, likely have larger proton fractions and a greater probability of
enhanced cooling.  The dichotomy between minimal cooling and enhanced
cooling might simply be due to a critical neutron star mass above
which the direct Urca process can operate before being quenched by
superfluidity.

\subsection{Conclusions from the Minimal Cooling Paradigm}

The study of the minimal cooling scenario allows 
progress beyond 
the conclusions reached in Sec.~\ref{Sec:Good&Bad_Conclusion}, 
but we are still faced with a clear dichotomy: 

\smallskip

(1) In the case the neutron $\Triplet$ gap is 
small in size (satisfying 
$T_{cn}^\mathrm{min} \simle 2\times 10^8$ K and
$T_{cn}^\mathrm{max} \simge 5\times 10^8$ K), neutron stars undergoing 
fast neutrino cooling must be relatively rare as
among the dozen of known young isolated neutron stars,
only one candidate, the pulsar in the supernova remnant CTA1, exists. 
This candidate is augmented by possible neutron stars represented by 
one or more of the(green) upper 
limits in \fig{Fig:Cool_Minimal_a-b}.  Nevertheless, the 
total number of fast cooling candidates remains small.
\smallskip

(2) In the case the neutron $\Triplet$ gap is either larger, 
or smaller than in (1), 
a much larger fraction of neutron stars appear to undergo fast neutrino cooling.

\smallskip

\begin{figure}[h]
\begin{center}
\includegraphics[width=.70\textwidth]{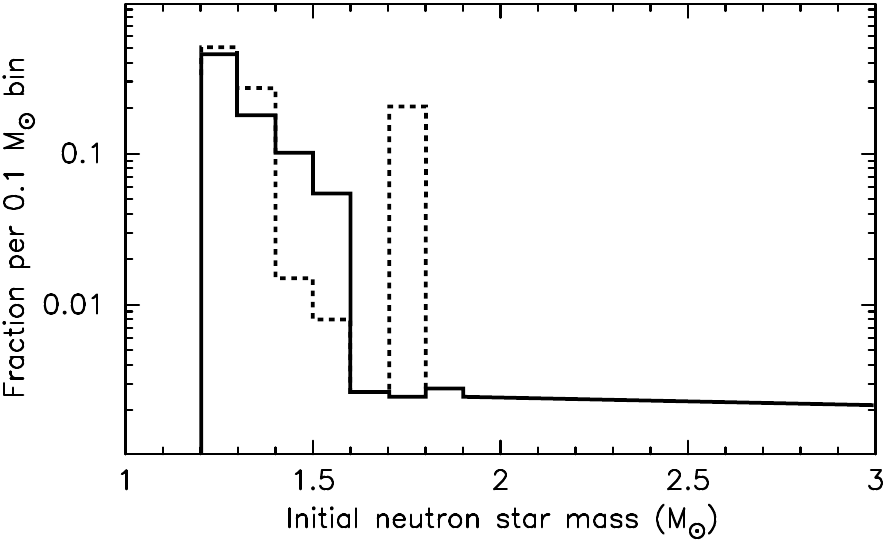}
\end{center}
\caption{The initial mass function of neutron stars as predicted by stellar evolution models.
The continuous line shows results from \cite{Fryer:2001uq} and the
dotted line is adapted from \cite{Woosley:2002cr}.
Figure from \cite{Page:2006vn}.}
\label{Fig:IMF}
\end{figure}

Both conclusions are extremely interesting, but more information is needed to choose the (hopefully) correct one.
Such information may be at hand and is presented in the next section.
These conclusions also have implications in terms of neutron star masses.
Neutron star mass measurements are only available 
for about 3 dozen pulsars in binary systems, and these may not be indicative of
the overall neutron star population.  However, we can appeal to theoretical     
predictions of the
neutron star initial mass function, which are shown in \fig{Fig:IMF}. 
The predicted distributions are actually similar to the
observed distributions from binary neutron stars \cite{Lattimer:2012}.
In the case that conclusion (1) above is correct, the critical mass $M_\mathrm{cr}$
for fast neutrino cooling would appear to be much larger than the average
mass, perhaps in the range $1.6-1.8 M_\odot$.  Alternatively,
if conclusion (2) above is correct, $M_\mathrm{cr}\sim1.4 M_\odot$, 
around the average mass.
%


\section{Cassiopeia A and its Cooling Neutron Star}
\label{Sec:CasA}

\begin{figure}[h]
\begin{center}
\includegraphics[width=.50\textwidth]{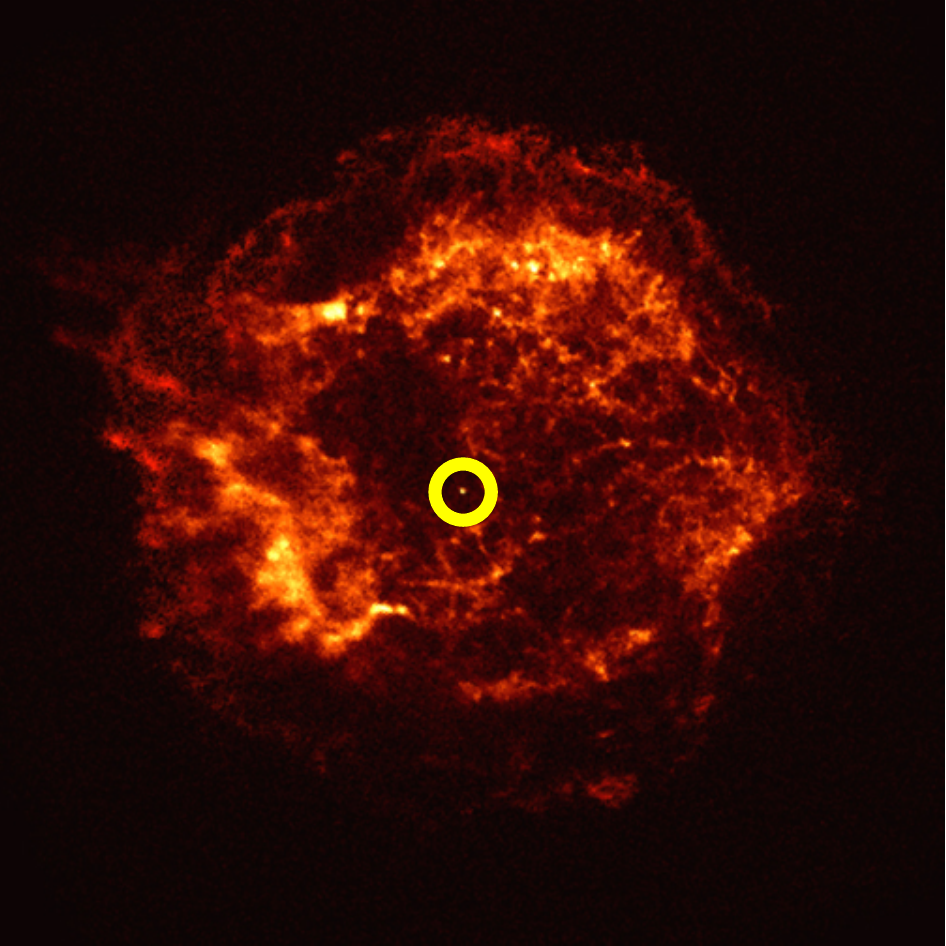}
\end{center}
\caption{The Cassiopeia supernova remnant in X-rays: first light of \texttt{Chandra}, August 1999.
(The neutron star is highlighted by the authors.) Image from \copyright NASA/CXC/SAO.}
\label{Fig:CasA}
\end{figure}

The Cassiopeia A (Cas A) supernova remnant (SNR) was discovered in radio in
1947 and is the second brightest radio source in the sky (after the
Sun).  It has since then been observed at almost all wavelengths.
Very likely, the supernova was observed by the first {\it Astronomer Royal},
John Flamsteed \cite{Ashworth:1980vn} who, on August 16, 1680,
when describing the stars in the Cassiopeia constellation, listed the star
``3 Cassiopeia" at a position almost coincident with the supernova
remnant.  This star had never been reported previously, and was never
to be seen again, until August 1999 when the first light observation of
\texttt{Chandra} found a point source in the center of the
remnant (see \fig{Fig:CasA}).

The distance to the SNR is $3.4^{+0.4}_{-0.1}$ kpc \cite{Reed:1995kl},
and the direct observation, by the \texttt{Hubble Space Telescope}, of
the remnant expansion implies a birth in the second half of the
17$^\mathrm{th}$ century \cite{Fesen:2006tg} and supports Flamsteed's
observation.  These observations give a present age of 333 yrs for the
neutron star in Cas A.  The optical spectrum of the supernova has been
observed through its light echo from scattering of the original light
by a cloud of interstellar dust and shows the supernova was of type
IIb \cite{Krause:2008hc}.  The progenitor was thus a red supergiant
that had lost most of its hydrogen envelope, with an estimated zero
age main sequence (ZAMS) mass of 16 to 20 $M_\odot$
\cite{Willingale:2003ys,Chevalier:2003ij,van-Veelen:2009fv} or even up
to 25 $M_\odot$ in the case of a binary system \cite{Young:2006bs}.
This implies a relatively massive neutron star, i.e. likely
$\simge 1.4 \; M_\odot$ \cite{Young:2006bs}.  The large amount of
circumstellar material associated with mass loss from its massive
progenitor could have diminished its visibility from Earth and could
explain why it wasn't as bright as the two 
Renaissance supernovae,
Kepler's SN 1572 and Tycho's SN 1604.

The soft X-ray spectrum of the point source in the center of the SNR
in Cas A is thermal, but its interpretation has been challenging
\cite{Pavlov:2004kl}.  With a known distance, a measurement of the
temperature implies a measurement of the star's radius, but spectral
fits with a blackbody or a H atmosphere model resulted in an estimated
radius of 0.5 and 2 km, respectively.  This suggests a hot spot, but
that should lead to spin-induced variations in the X-ray flux, which
are not observed.  It was only in 2009 that a successful model was
found: a non-magnetized C atmosphere\footnote{There is, to date, no evidence for
the presence of a significant magnetic field in the Cas A neutron star.}, 
which implies a stellar radius between 8 to 18 km
\cite{Ho:2009fk}.  Models with heavier elements, or a blackbody, produced significantly poorer fits.  With the C model, and analyzing 5 observations of
the SNR, Heinke \& Ho \cite{Heinke:2010hc} found that the inferred neutron star
surface temperature had dropped by 4\% from 2000 to 2009, from $2.12$
to $2.04 \times 10^6$ K, and the observed flux had decreased by 21\%.
The neutron star in Cas A, the youngest known neutron star, is the
first one whose cooling has been observed in real time!

\subsection{Superfluid Neutrons in the Core of Cas A}
\label{Sec:Cool_CasA}

The \texttt{Chandra} observations of Cas A, together with its
known distance, imply that the photon luminosity of the neutron star
is
\be
L_\gamma \simeq 10^{34} \; \mathrm{erg \; s}^{-1} \; .
\label{Eq:Lg_CasA}
\ee
With a measured $T^\infty_e \simeq 2 \times 10^6$ K \cite{Ho:2009fk},
we deduce an internal $T \simeq 4 \times 10^8$ K from
\eq{Eq:TbTe}.  The star's total specific heat is thus $C_V
\simeq 4 \times 10^{38}$ erg K$^{-1}$ (from \fig{Fig:Cv} or
\eq{Eq:PL1}).  The observed $\Delta T^\infty_e/T^\infty_e \simeq$ 4\%
\cite{Heinke:2010hc} gives a change of core temperature $\Delta T/T
\simeq$ 8\% over a ten years period since $T \sim (T^\infty_e)^2$.  Assuming
the observed cooling corresponds to the {\em global cooling} of the
neutron star, its thermal energy loss is
\be
\dot{E}_\mathrm{th} = C_V \dot{T} 
\simeq (4 \times 10^{38} \; \mathrm{erg \; K}^{-1}) \times (0.1 \; \mathrm{K \; s}^{-1} )
\simeq 4\times 10^{37}  \; \mathrm{erg \; s}^{-1}\,,
\label{Eq:Edot_CasA}
\ee
which is 3--4 orders of magnitude larger than 
$L_\gamma$!  For a young neutron star, neutrinos are the prime
candidates to induce such a large energy loss.

\begin{figure*}[h]
\begin{center}
\includegraphics[width=.60\textwidth]{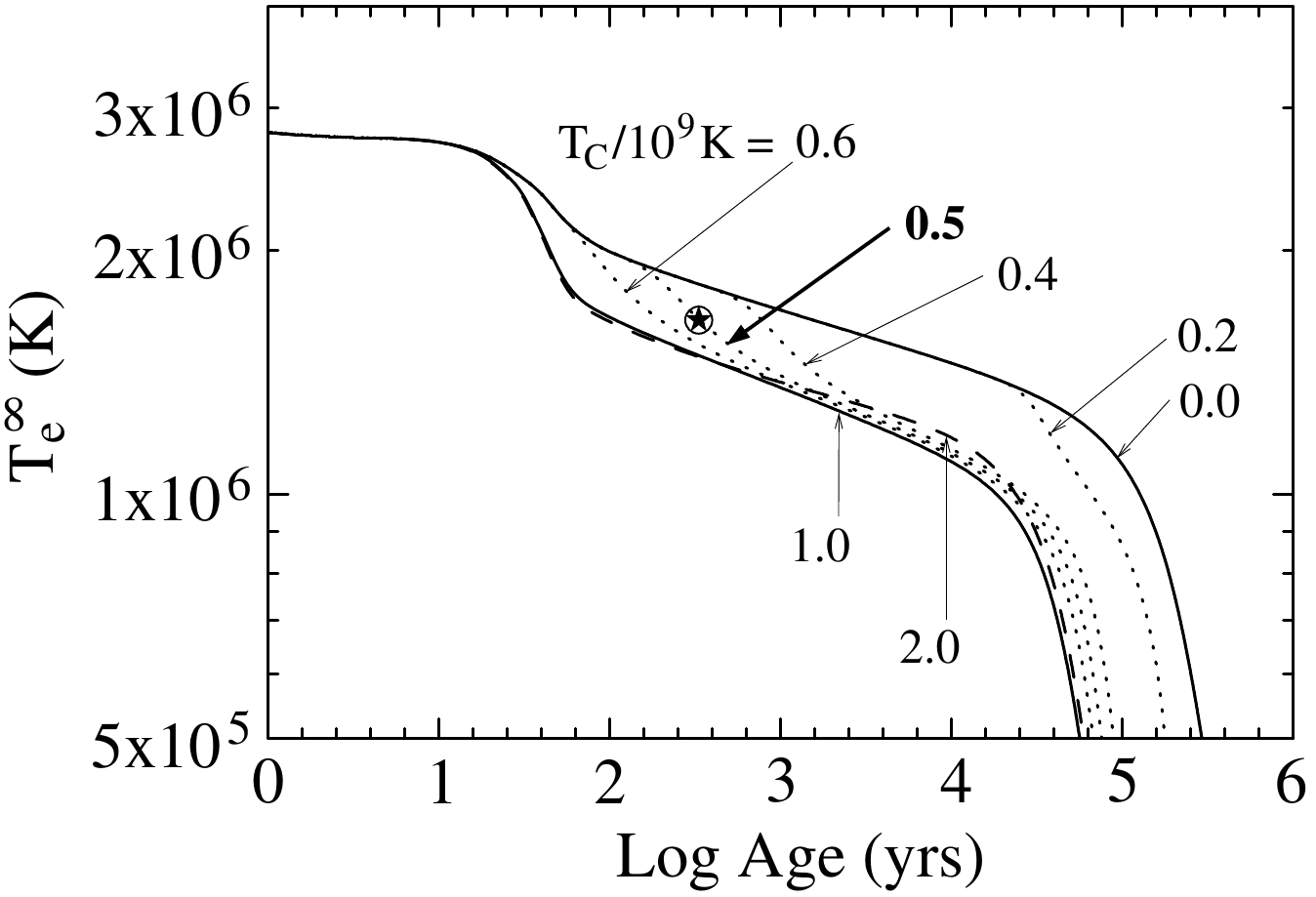}
\end{center}
\caption{Similar to the right panel of \fig{Fig:Cooling_SF}, 
 but showing the observed
  age and temperature of the Cas A neutron star (the
  star) and its consistency with $T_C\simeq5\times10^8$ K for the
  magnitude of the neutron $\Triplet$ gap.
    Taken from \cite{Page:2011ys}.
  }
\label{Fig:Cooling_CasA_1}
\end{figure*}

The cooling rate of this neutron star is so large that it must be a
transitory event, which was initiated only recently. Otherwise it
would have cooled so much it would probably now be unobservable.
Something critical occurred recently within this star!  ``Something
critical" for a cooling neutron star points toward a critical
temperature, and a phase transition is a good candidate.  The previous
section highlighted that a phase of accelerated cooling occurs when
the neutron $\Triplet$ pairing phase transition is triggered.  With
$T_{cn}^\mathrm{max} \simeq 5 \times 10^8$ K, a transitory cooling episode can occur
at an age $\simeq 300$ yrs as shown in the right panel of
\fig{Fig:Cooling_SF} and in \fig{Fig:Cooling_CasA_1}.

The interpretation that the observed rapid cooling of the neutron star
in Cas A was triggered by the recent onset of the neutron $\Triplet$ 
superfluid phase transition and consequent neutrino emission
from the formation and breaking of Cooper pairs in the neutron
superfluid was recently proposed in \cite{Page:2011ys} and,
independently, in \cite{Shternin:2011fu}.  

\subsubsection*{A simple analytical model}
\label{Sec:CasA_analytical}

\begin{figure*}[h]
\begin{center}
\includegraphics[width=0.80\textwidth]{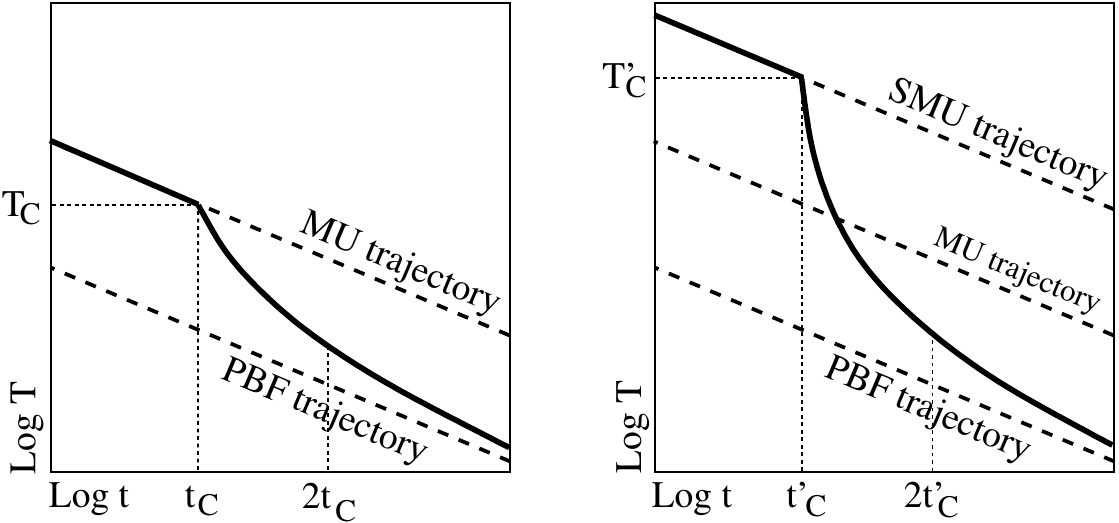}
\end{center}
\caption{ Schematic cooling trajectories (heavy curves) showing the effect 
of superconductivity.  Left panel: Without superconductivity,
                 $T$ initially follows the modified Urca (MU) trajectory,
                 $T_9 = (\tau_\mathrm{MU}/t)^{1/6}$  until 
                 $T$ reaches  the neutron critical temperature $T_{C}$ at time
                 $t_C$.  The pair breaking and formation (PBF) process turns on
                 and the neutrino luminosity $L_\nu$ abruptly increases by a factor $f$. 
                 Thereafter, $T$ rapidly transits, on a time scale
                 $\tau_{_\mathrm{TR}} = t_C/f$
                 toward the PBF trajectory, $T_9 = (\tau_\mathrm{MU}/ft)^{1/6}$.  
                 Empirically, this transition takes a time $\simle 2t_C$.
                 Right panel: When protons are superconducting, the
initial evolution follows a superconducting-suppressed modified Urca (SMU) path.
                 For the transit to start at a time $t'_C \approx t_C$
                 the trajectory requires $T'_C>T_C$.
                 The early transit has a shorter time scale,
                 $\tau_\mathrm{TR}' = t'_C/f'$ with $f'  \gg  f$,
                 and a significantly larger slope.
                 The late time evolution converges to that of the left panel. 
                 The left (right) panel corresponds to models in \fig{Fig:Cooling_CasA_1}
                 (\fig{Fig:Cool_CasA} and \ref{Fig:Cool_Minimal_CasA}).  
                 From \cite{Page:2011ys}.}
\label{Fig:Trajectories}
\end{figure*}

The simple analytical solution of \eq{Eq:Simple_nu} gives some insight into the observed behavior.  
When $T> T_C \equiv T_{cn}^\mathrm{max}$, but $\ll T_0$,
the star follows the asymptotic ``MU trajectory", $T_9 = (\tau_\mathrm{MU}/t)^{1/6}$,
and when $T$ reaches $T_C$, at time $t=t_C$, the neutrino luminosity suddenly increases
(see \fig{Fig:Coop-Scheme}).
Despite the complicated $T$ dependence of $\epsilon^\mathrm{PBF}$, \eq{Eq:Q_PBF}, 
the resulting luminosity, once integrated over the entire core (also, aided by
the bell shape of the $T_c(k_F)$ curve), is well approximated by a $T^8$ power
law in the $T$ regime in which some thick shell of neutrons is going
through the phase transition.  If we write 
\be
L_\nu^\mathrm{PBF} = f
\cdot L_\nu^\mathrm{MU} = f N_9 T_9^8,
\label{pbf1}
\ee
with $f \sim 10$, the solution
of \eq{Eq:Simple_nu}, replacing $t_0$ by $t_C$ and $T_0$ by
$T_C$, gives
\be
T = \frac{T_C}{[1+f(t-t_C)/t_C]^{1/6}} \quad \rightarrow \quad
T_9 = \left({\tau_\mathrm{MU}\over ft}\right)^{1/6}\quad(T\ll T_C).
\!\!\!\!
\label{Eq:Simple_PBF}
\ee
Thus, at late times, the asymptotic ``PBF trajectory" is a factor
$f^{-1/6}\simeq0.7$ lower than the ``MU trajectory"
while the time-scale for the transit from the ``MU trajectory"
to the asymptotic ``PBF trajectory", $T_9 = (\tau_\mathrm{MU}/ft)^{1/6}$,
is $\tau_\mathrm{TR} = t_C/f$.
This behavior is shown schematically in the left panel of
\fig{Fig:Trajectories}.  

Since the initiation of the rapid transit, when the neutron star left the MU trajectory,
must have occurred recently, i.e. $t_C \simeq (0.5-0.9) \times 333$ yrs, we obtain
\be
T_C = T_{cn}^\mathrm{max} =  10^9 \, \mathrm{K} \, (\tau_{MU}/t_C)^{1/6} \sim 5 \times 10^8 \, \mathrm{K} \, .
\label{Eq:Cas_T_C}
\ee
This is the first important result from this simple analytical model:
given the observation of rapid cooling,  the inferred value of $T_C$ depends
{\it only} on the known age, $t=333$ yrs, of this neutron star
and the value of $\tau_{MU} \sim1$ year.
This result is also in good agreement with the numerical models of \fig{Fig:Cooling_CasA_1}.
The $1/6$ exponent in \eq{Eq:Cas_T_C} confirms it to be a robust result without
much dependence on microphysical details.

\subsection{Superconducting Protons in the Core of Cas A}
\label{Sec:CasA_proton}

How ``rapid" is the observed ``rapid cooling" of the Cas A neutron star can be
quantified by the observed slope
\be
s_\mathrm{obs} = \left| \frac{d \ln T_e^\infty}{d \ln t} \right| \simeq 1.4
\label{Eq:slope_obs}
\ee
from the six data points reported in \cite{Heinke:2010hc,Shternin:2011fu}.
This slope is much larger than the one of the ``MU trajectory", 
\be
s_\mathrm{MU} = \left| \frac{d \ln T_e^\infty}{d \ln t} \right| =
  \frac{d \ln T_e^\infty}{d \ln T}  \times \left| \frac{d \ln T}{d \ln t} \right| \simeq 
\frac{1}{2} \times \frac{1}{6} \simeq 0.08 \, .
\label{Eq:slope_MU}
\ee
In contrast, the ``transit trajectory" of \eq{Eq:Simple_PBF} gives 
a much larger slope
\be
s_\mathrm{TR} =
\frac{1}{2} \times \frac{f}{6}  \, \left(\frac{T}{T_C}\right)^6 \frac{t}{t_C} \, ,
\label{Eq:slope_transit}
\ee
which has a maximum value $f/12$.  Nevertheless, to match the observed 
slope $s_\mathrm{obs}$, $f \gg 10$ is required.

\begin{figure*}[h]
\begin{center}
\includegraphics[width=0.60\textwidth]{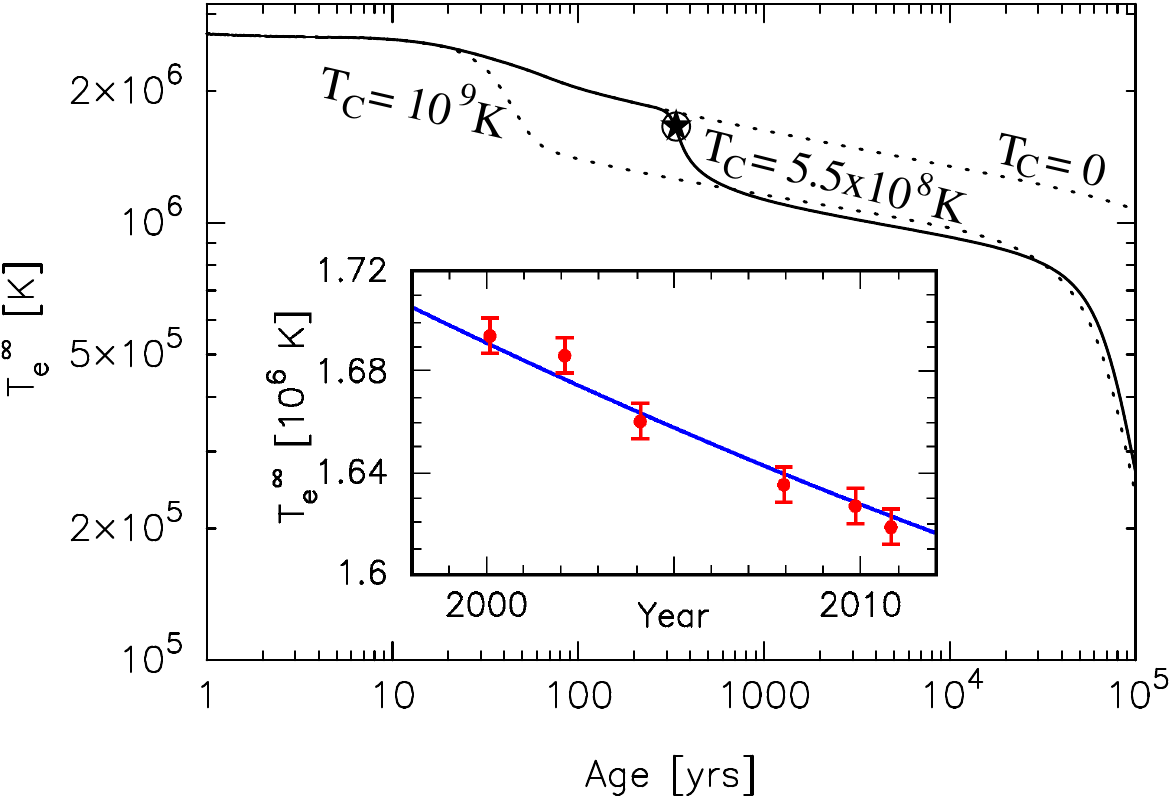}
\end{center}
\caption{ A fit to observations of the neutron star in Cas A assuming
  recent onset of neutron $\Triplet$ superfluidity and PBF cooling.  
  The $1.4 M_\odot$ model shown assumes the APR EOS \cite{Akmal:1998fk}
  with a C atmosphere \cite{Ho:2009fk}.
  With model ``CCDK"  of \fig{Fig:Tc_p} 
($T_{cp}^\mathrm{max} \gg 10^9$ K), proton are superconducting from early times.
  The neutron $\Triplet$ gap is model ``a2" of \fig{Fig:Tc_n} with
   $T_{C} = T_{cn}^\mathrm{max} = 5.5 \times 10^8$ K.
  This neutron superfluid phase transition triggered the PBF process
  that results in a sudden cooling of the neutron star.   
  Observations
  \cite{Heinke:2010hc,Shternin:2011fu} suggest $|d\ln
  T^\infty_e/d\ln t|\simeq1.4$, shown in the inset.  Two dotted
  curves with $T_C=0$ and $1\times 10^9$ K, respectively, illustrate the sensitivity to $T_C$.
  Figure adapted from \cite{Page:2011ys}.}
\label{Fig:Cool_CasA}
\end{figure*}

The value of $f$ considered above, $f = L_\nu^\mathrm{PBF}/L_\nu^\mathrm{MU} \simeq 10$,
arose from a freely acting modified Urca neutrino emission.
However, despite the many theoretical uncertainties discussed in Sec.~\ref{Sec:Pairing},
there seems little doubt that proton superconductivity exists around a few times
$\rhonuc$, as illustrated in \fig{Fig:Tc_p}.
Moreover, expected values of $T_{cp}^\mathrm{max}$ are somewhat larger than $10^9$ K,
implying that protons were likely already superconducting in some part of the core
of the Cas A neutron star when neutron anisotropic superfluidity set in.
If such was the case, the previous neutrino luminosity of this star
was due to a proton-pairing-suppressed modified Urca process with 
$L_\nu^\mathrm{SMU} < L_\nu^\mathrm{MU}$.
This implies a much higher relative efficiency of the neutron PBF process, i.e.,
$f' = L_\nu^\mathrm{PBF}/L_\nu^\mathrm{SMU} \gg f\sim10$.
The resulting transit from a ``SMU trajectory" to the ``PBF trajectory" is depicted in the
right panel of \fig{Fig:Trajectories} and exhibits a transit slope enhanced 
by a factor $f'/f$
\footnote{Presumably, at the time of the onset of proton
superconductivity, another PBF episode had occurred, but, as the
PBF process for singlet pairing is much less efficient than for triplet pairing,
and also because protons are much less abundant than neutrons, 
it did not result in a significant cooling of the star.
Moreover, this cooling occurred on
timescales much smaller than the crustal thermal timescale.}.

With the above considerations, a very good fit to the observations can be obtained,
as shown in \fig{Fig:Cool_CasA},
implying a maximum neutron $\Triplet$ pairing $T_C=5.5\times 10^8$ K
along with superconducting protons with a larger $T_{cp}^\mathrm{max}$.
Very similar results were independently obtained by \cite{Shternin:2011fu}\footnote{
These authors, however, assumed that proton superconductivity extends to very high densities,
with $T_{cp} \simge (2-3) \times 10^9$ K in the whole neutron star core.
This results in a very strong suppression of $L_\nu$ prior to the onset of
neutron $\Triplet$ superfluidity, and a neutron star much warmer than in the model of \cite{Page:2011ys},
from which a larger $T_{cn}^\mathrm{max} \simeq (6-9) \times 10^8$ K is deduced.
However, strong proton superconductivity at high densities is, at present time,
not supported by the microscopic models presented in \fig{Fig:Tc_p}.}.
This observation of the cooling of the youngest known neutron star is
unique and its interpretation potentially imposes very strong
constraints on the physics of ultra-dense matter.  

\begin{figure*}[h]
\begin{center}
\includegraphics[width=0.55\textwidth]{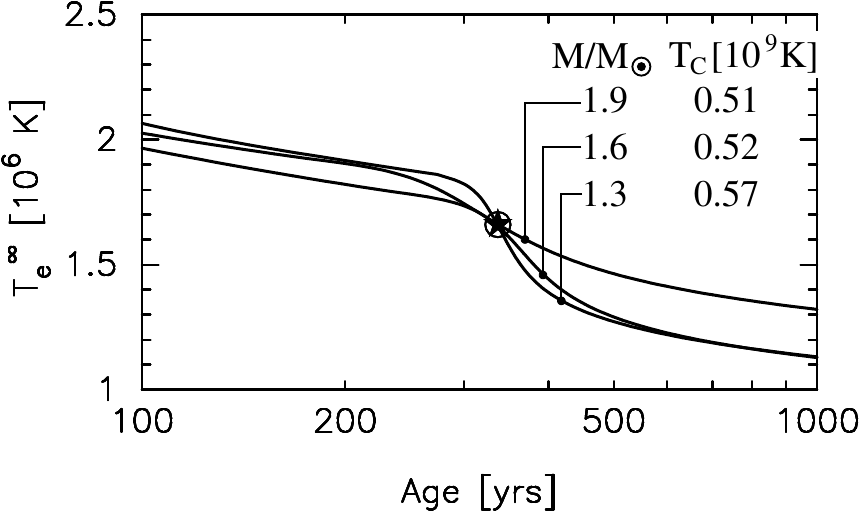}
\end{center}
\caption{Dependence of the slope $s=|d\ln T^\infty_e/d\ln t|$ 
of the cooling curve 
  on the star mass at $t=330$ years: 
$s = 1.4, 0.9$, and 0.5 for $M = 1.3, 1.6$,
  and $1.9 M_\odot$, respectively (from \cite{Page:2011ys}).}
\label{Fig:Cool_Minimal_CasA}
\end{figure*}

 The requirement that protons became superconducting before the onset
 of neutron superfluidity places a constraint not only on the proton
 $\Singlet$ pairing but also on the neutron star mass.  
If the neutron star mass is not too large, 
 models show that proton superconductivity does not extend to very
 high densities.  \fig{Fig:Cool_Minimal_CasA}
 illustrates the sensitivity to the neutron star mass.  A better
 understanding of the progenitor of Cas A and its expected neutron
 star mass will prove important in validating this
 scenario.

\begin{figure*}[h]
\begin{center}
\includegraphics[width=0.8\textwidth]{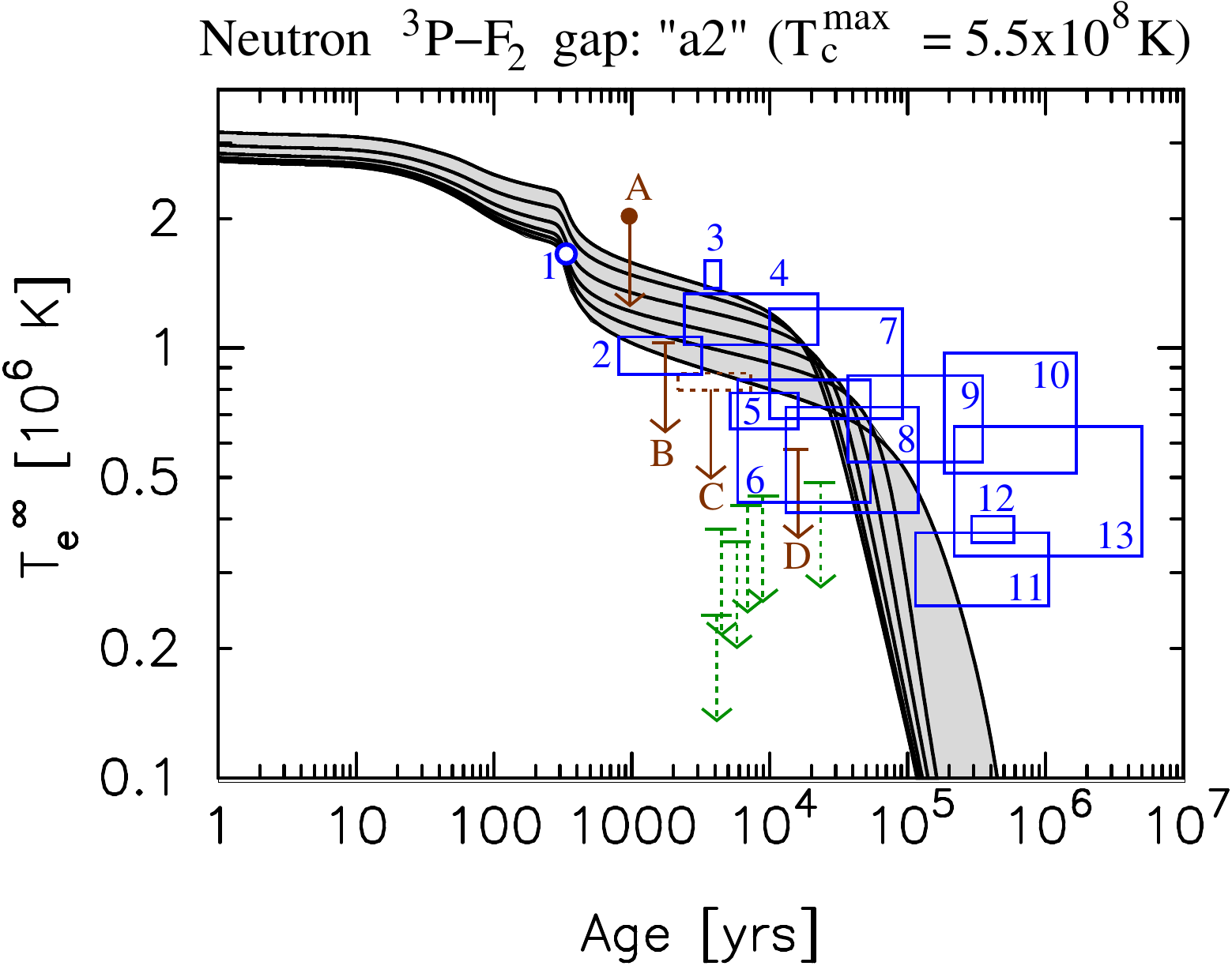}
\end{center}
\vspace*{-.5cm}
\caption{
Comparison of the cooling scenario of \fig{Fig:Cool_CasA} with data for isolated neutron stars.
As in \fig{Fig:Cool_Minimal_a-b}, the various lines show the effect of varying the amount of light elements in the envelope
 (from \cite{Page:2012vn}).}
\label{Fig:Cool_Minimal_CasA_1}
\end{figure*}

The neutron superfluid explanation for
the rapid cooling of the neutron star in Cas A fits well within the {\em minimal
  cooling} scenario \cite{Page:2004zr,Page:2009qf}.  
As described in Sec.~\ref{Sec:Comparison},
maximal compatibility of 
the minimal cooling  scenario  \cite{Page:2009qf} with data 
required 
the neutron $\Triplet$ gap to have $T_{cn}^\mathrm{max}\simge5 \times 10^8$.
This lower limit on $T_{cn}^\mathrm{max}$ was deduced for compatibility with the 
measured $T_e$ of the youngest neutron stars of age $\sim 10^3$ yrs
The  
upper limit on $T_{cn}^\mathrm{min}$ was 
deduced for compatibility with the
oldest middle-aged stars as, e.g., the Vela pulsar.
The 
compatibility of the neutron $\Triplet$
gap inferred from the cooling data of the neutron star in Cas A 
with observations of other isolated neutron stars is confimed in  
\fig{Fig:Cool_Minimal_CasA_1}.
The only marked difference between \fig{Fig:Cool_Minimal_CasA_1} and 
the left panel of \fig{Fig:Cool_Minimal_a-b} is the occurrence of the rapid cooling phase
at ages $\simeq 300$ yrs to $\sim 500$ yrs, due to the reduced
value of $T_{cn}^\mathrm{max}$ in the former,  $0.55 \times 10^9$ K
compared to $1.0 \times 10^9$ K.

\subsection{Alternative Scenarios}
\label{Sec:CasA_alternative}

Alternative explanations for the observed rapid cooling of Cas A
have been proposed.  One could consider observed cooling
of this star to be due to a significantly longer thermal
relaxation timescale in the crust or core than assumed in 
\cite{Page:2011ys,Shternin:2011fu}.  In
such a case, the estimate of \eq{Eq:Edot_CasA}, which assumes
the star to be isothermal, becomes invalid.  In all models shown in
this chapter, the crust thermal relaxation
occurs on a timescale of a few decades.  However, if the crust thermal
conductivity is, in fact, significantly smaller, it is conceivable
that the observed rapid cooling corresonds to the thermal relaxation
of the crust (see, e.g., \cite{Yakovlev:2011kl}).  
Such a low crust thermal conductivity is, however, in conflict
with the observed crust relaxation time in transiently accreting
neutron stars \cite{Shternin:2007fk,Brown:2009uq}
and is 
based on the assumption that the crust is in an amorphous
solid state instead of a crystalline one, a possibility that 
is not supported by microscopic studies \cite{Horowitz:2009kx,Daligault:2009vn}.

Similarly, the core
thermal relaxation time may be much larger than usually considered.
For example, Blaschke et al. \cite{Blaschke:2012ys} have proposed that the
inner core of the star cools rapidly and that it also takes a few
hundreds years for the star to become isothermal.  The latter time is
when the rapid decrease of $T^\infty_e$ would be observed.  This
scenario requires that the core thermal conductivity be lower than
usually considered, by a factor 4 or larger, and also requires
that neutrons {\it do not} form a superfluid until the star is much colder.
This scenario, 
based on the ``Medium-Modified Urca" neutrino emission process
\cite{Voskresenskii:1986nx,Voskresensky:2001cr}, is also compatible with the
cooling data, but only if 
 the suppression of core conductivity 
 is adjusted to fit the observed cooling of the neutron star in Cas A.
More work is required to confront
these alternative possibilities with other facets of neutron star
phenomenology.

Finally, there are important systematic uncertainties related
to the
observations which may affect our ability to interpret the
cooling of
this neutron star. Among these uncertainties are: 
the incorrect idenfication of two simultaneous
photons as a
single photon of larger energy, detector calibration issues,
and
contributions from material in the line of sight between the
neutron
star and the observing satellite. Recent analyses of these
uncertainties cannot conclusively confirm
that cooling is present, but cannot unambiguosly rule out such cooling
either~\cite{Elshamouty:2013,Posselt:2013}.



\section{Dynamical Evolution of Neutron Stars}
\label{Sec:glitch}

Pulsars are rotating neutron stars whose spin rates are generally
observed to be decreasing.  The regularity of pulsars is outstanding;
as timekeepers they rival atomic clocks.  Although the pulses are
remarkably regular, the time between pulses slowly but predictably
increases.  Their spin-downs are attributed to magnetic dipole
radiation \citep{Pacini:1968ly,Gunn:1969ve} -- the conversion of
rotational energy into electromagnetic energy -- and the pulsar wind
from ejection of the magnetospheric plasma \citep{Goldreich:1969qf}.
That the observed evolution of pulsar spins exhibits evidence for both
core and crustal superfluidity is the subject of this section.

\subsection{Magnetic Dipole Pulsar Model}
\label{Sec:dipole}
Dimensional analysis gives us a simple estimate of the pulsar energy
losses.  Energy and angular momentum are irreversibly lost when either
the magnetic field or the plasma reaches the {\em light cylinder}
where co-rotation with the pulsar implies a speed equal to $c$ and
whose radius is thus $R_{lc} =c/\Omega \simeq 50 (P/1~\mathrm{msec})$
km, $\Omega$ being the spin rate and $P=2\pi/\Omega$ the spin period.
The magnetic field at the light cylinder is $B_{lc} = (R/R_{lc})^3
B_s$ assuming a dipolar field with a strength $B_s$ at the stellar
surface.  Just writing that an energy density $B_{lc}^2/4\pi$ is lost
at the speed of light $c$ from a sphere of area $4\pi R_{lc}^2$, one
obtains
\be
\dot{E}_{PSR} \simeq \frac{1}{c^3}  (R^3 B_s)^2 \, \Omega^4 \, .
\label{Eq:E_PSR}
\ee
Energy loss by {\em in vacuum} magneto-dipolar radiation gives the
same result with just an extra factor $(2/3) \sin^2 \alpha$, $\alpha$
being the angle between the rotationnal axis and the dipolar moment.
An aligned rotator, i.e. with $\alpha = 0$, will not spin-down from
magneto-dipolar radiation but rather by plasma ejection
\cite{Goldreich:1969qf}.  Numerical and consistent calculations of the
energy loss from the ejected magnetospheric plasma have been possible
only recently (\cite{Contopoulos:1999cr} and \cite{Spitkovsky:2008dq}
for a review) and the result is that the total $\dot{E}_{PSR}$ is
given by \eq{Eq:E_PSR} with an extra factor $\simeq (1+\sin^2
\alpha)$.
Considering the rotational energy of a uniformly rotating sphere 
$E_{rot}=I\Omega^2/2$, where $I$ is the star's moment of inertia,
the pulsar's spin-down is determined by 
equating $\dot{E}_{rot}$ with $-\dot{E}_{PSR}$, giving
\be
\dot \Omega = - {K\over I} \Omega^3
\ee
with $K \simeq R^6 B_s^2/c^3$, assuming that $I$ and $B_s$ remain
constant.  This also provides an estimate of $B_s$, probably reliable
within a factor of a few,
\be
B_s \simeq\sqrt{-c^3\dot\Omega\over\Omega^3IR^6}\simeq3.2\times10^{19}(P\dot P)^{1/2}\; \mathrm{G} \simeq
1.5 \times10^{12}{P\over0.01{\rm~s}} \left({1000{\rm~yr}\over\tau_c}\right)^{1/2} \; \mathrm{G},
\label{field1}
\ee
where we assumed
$M\simeq1.4$ M$_\odot$ and $R\simeq12$ km.
It is traditional to also deduce an observable characteristic pulsar age
\be\label{tauc}
\tau_c=-{\Omega\over2\dot\Omega}
\ee
and an observable braking index
\be
n \equiv {\Omega\ddot\Omega\over\dot\Omega^2}
\label{brake}
\ee
In the case of the magnetic dipole model and assuming both $I$ and
$B_s$ are constant,  $n=3$.

The ATFN catalogue\footnote{On-line version:
  \url{http://www.atnf.csiro.au/research/pulsar/psrcat/}}
\cite{Manchester:2005jl} lists 1759 pulsars (as of December 2012) with
measured values of $\dot P$.  The measurement of a second derivative
$\ddot \Omega$, allowing the determination of $n$, requires a very
accurate long term and smooth fit of $\dot \Omega$ and is much more
difficult due in part to timing noise and to glitches.  To date, there
are only 11 published values of $n$ \cite{Espinoza:2012dq}, all with
values less than 3 and ranging from $-1.5$ to $+2.91$ (see
Table \ref{Tab:Brake}).  Anomalous values $n < 3$ may be due to a growing
magnetic field\cite{Muslimov:1995dz} and/or to a growing fraction of
the star's core becoming superfluid \cite{Ho:2012tg}.  The latter
could be due to the onset and growth of $\Singlet$ p and  $\Triplet$ n
components (further evidence for this stems from observations of the
cooling of the neutron star in the Cassiopeia A supernova remnant,
which is discussed in Sec.~\ref{Sec:CasA}).

\subsection{Anomalous Braking Indices}
\label{Sec:brake}

Consider the evolution of the spin of a pulsar if we allow changes in
the surface field $B_s$ and the moment of inertia $I$.  Assume the
star has both normal matter and superfluid matter with moments of
inertia $I_n$ and $I_s$, respectively.  Although the
onset of superfluidity is unlikely to significantly change the total
moment of inertia, the portion of the star that is superfluid may be
considered to spin frictionlessly and its spin rate $\Omega_s$ to remain
constant.  The observed spin rate of the star is that of the star's
surface, which is composed of normal matter.  We therefore have
\be\label{idiv}
I=I_n+I_s,\qquad\dot I=0,\qquad\Omega_n=\Omega,\qquad\dot\Omega_s=0.
\ee
With these two components, the rate of change of angular momentum in
the magnetic dipole model is
\begin{equation}
{d\over dt}\left[I_n\Omega_n+I_s\Omega_s\right]=
I_n\dot\Omega+\dot I_n\left(\Omega-\Omega_s\right)=-K\Omega^3.
\label{ang}
\end{equation}
Taking a time derivative of the above, one finds
\begin{equation}
n
=3+4\tau_c\left[{\dot I_n\over I_n}-{\dot B_s\over B_s}\right]
+\tau_c\left({\Omega_s\over\Omega}-1\right){\dot I_n\over
 I_n}\left[6+4\tau_c\left({\ddot I_n\over\dot I_n}-2{\dot B_s\over B_s}\right)\right].
\label{ang2}
\end{equation}
Since $\Omega_s-\Omega \ll \Omega$, we can drop the last term in
\eq{ang2}.  Thus, the growth of either the surface magnetic field and/or
the superfluid component leads to $n<3$.
In fact, the growth of a core superfluid component 
could lead to field ejection from the core and an increase in the
surface field strength.  Note that there has been a long history of
study of the opposite possibility of {\em decay} of pulsar magnetic
fields. However, statistical studies of the pulsar population
\cite{Faucher-Giguere:2006hc} no longer support this idea and the
present consensus is that magnetic field decay is significant only in
the case of super-strong fields as in magnetars or after a long phase
of accretion in a binary system.  

Nevertheless, observed values $n<3$ do not prove that superfluidity
exists.  The growth of the surface field
of a pulsar can be obtained in the case that the initial magnetic
field was partially buried, e.g., by post-supernova
hypercritical accretion
\cite{Geppert:1999ij,Bernal:2010bs,Bernal:2012fv}, into the stellar
interior and is now slowly diffusing outwards
\cite{Muslimov:1995dz}.  Detailed numerical models show that observed
values of $n$ can be reproduced \cite{Muslimov:1996fu} without
invoking superfluidity.  Alternatively, magnetospheric currents
may deform the field into a non-dipolar geometry
\cite{Blandford:1988qa,Melatos:1997kl} that results in $n < 3$ (but $n
\ll 3$ is unreachable in these models and $n<0$ is impossible).  The
truth may be a superposition of these three types of arguments:
$\dot I_n <0$, $\dot B_s > 0$, and non-dipolar fields.

\begin{table}\label{Tab:Brake}
\begin{center}
\begin{tabular}{lcccccc}   
\hline 
Pulsar (SNR) & $P$ & $\tau_c$ & $B_s$       & $n$ & ${\cal A}$ &$2\tau_c{\cal A}$\\
             & s & kyr  &\!$10^{12}$ G\!&    &\!$10^{-9}$/d\!& \% \\
\hline 
B0531+21 (Crab)        &0.0331& 1.24  & 3.78  & 2.5  &     &       \\
J0537-6910 (N157B)     &0.0161& 4.93  & 0.925 & -1.5 & 2.40 & 0.9 \\
B0540-69 (0540-693)    &0.0505& 1.67  & 4.98  & 2.1  &      &     \\
J0631+1036             &0.2878& 43.6  & 5.55  &      & 0.48 & 1.5 \\
B0833-45 (Vela)        &0.0893& 11.3  & 3.38  & 1.7  & 1.91 & 1.6 \\
J1119-6127 (G292.2-0.5)&0.4080& 1.61  & 41    & 2.9  &      &     \\
B1338-62 (G308.8-0.1)  &0.1933& 12.1  & 7.08  &      & 1.31 & 1.2 \\
B1509-58 (G320.4-1.2)  &0.1513& 1.56  & 15.4  & 2.8  &      &     \\
J1734-3333             &1.1693& 8.13  & 52.2  &  0.9 &      &     \\
B1737-30               &0.6069& 20.6  & 17    &      & 0.79 & 1.2 \\
J1747-2958             &0.0988& 25.5  & 2.49  & $<1.3$&     &     \\
B1757-24 (G5.4-1.2)    &0.1249& 15.5  & 4.04  &  -1  & 1.35 & 1.5 \\
B1758-23 (W28?)        &0.4158& 58.3  & 6.93  &      & 0.24 & 1.0 \\
B1800-21 (G8.7-0.1(?)) &0.1367& 15.8  & 4.29  &   2  & 1.57 & 1.8 \\
B1823-13               &0.1015& 21.4  & 2.8   &   2  & 0.78 & 1.2 \\
J1846-0258 (Kes75)     &0.3266& 0.728 & 48.8  & 2.6  &      &     \\
B1930+22               &0.1445& 39.8  & 2.92  &      & 0.95 & 2.8 \\
B2229+6114 (G106.6+2.9)&0.0516& 10.5  & 2.03  &      & 0.63 & 0.5 \\
\hline 
\end{tabular}
\caption{Observed pulsar properties (``SNR'' is the
  associated supernova remnant).   Values of spin period
  $P$, characteristic age $\tau_c$ [\eq{tauc}] and dipole surface
  field strength $B_s$ [\eq{field1}] rounded from
  the ATFN catalogue \cite{Manchester:2005jl}.   Values of the braking index $n$ [\eq{brake}] rounded
  from \cite{Espinoza:2012dq}.  Values of ${\cal A}$ [\eq{cala}]
  taken from \cite{Andersson:2012tw}.    }
\end{center}
\end{table}

\subsection{Glitches}
\label{Sec:glitches}

\begin{figure}
\begin{center}
\includegraphics[width=0.8\textwidth,angle=180]{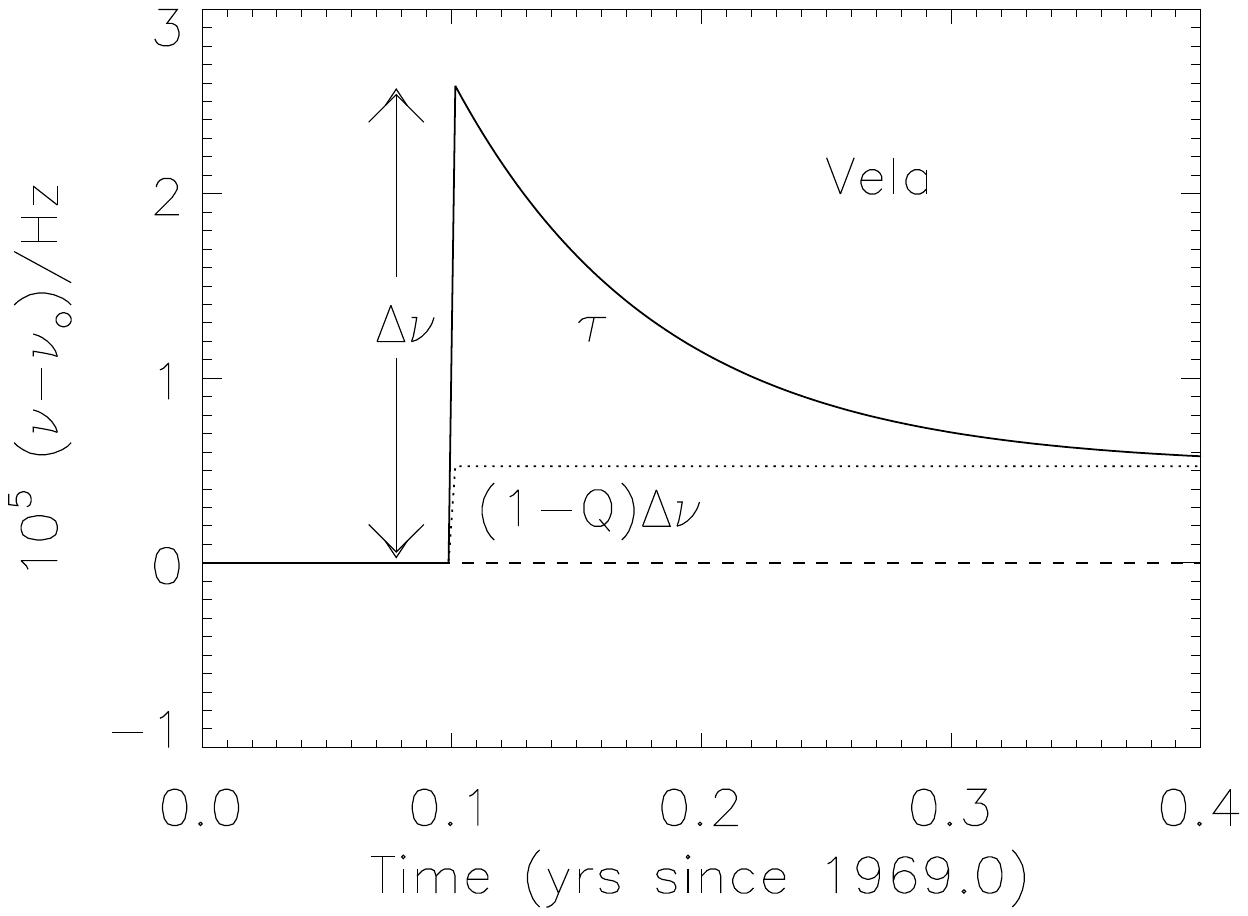}
\vspace*{-1cm}
\end{center}
\caption{Schematic illustration of a Vela pulsar glitch.  The observed
  pulsar spin frequency $\nu$ has been corrected by the average accumulated
  spindown since 1969.0 ($\nu_o=\dot\nu(t-1969.0)$) to make the glitch easier to discern. The
  glitch is observed as a relatively sudden jump $\Delta\nu$ in spin
  frequency.  The frequency afterwards relaxes roughly exponentially on a timescale
  $\tau$ to a new spindown line (dotted) which has a higher average
  frequency $(1-Q)\Delta\nu$ than the pre-glitch spindown line $\nu=\nu_0$
  (dashed).}
\label{Fig:glitch}
\end{figure}

\begin{figure}
\begin{center}
\includegraphics[width=\textwidth]{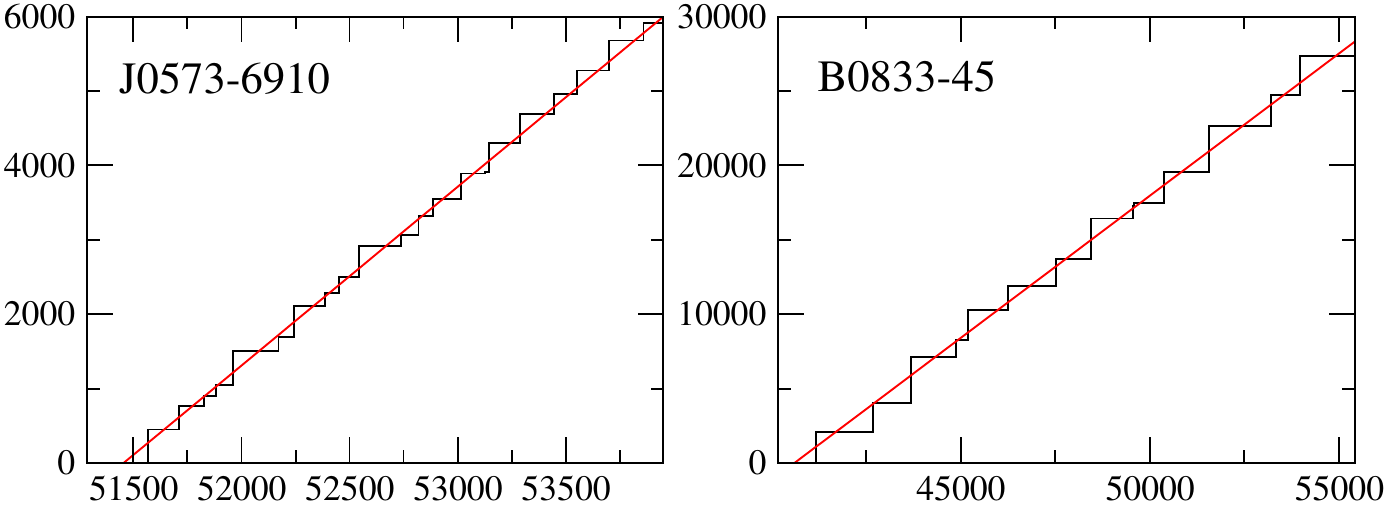}
\end{center}
\caption{The accumulated relative angular velocity
  $10^{-9}\sum_i\Delta\Omega_i/\Omega$ as a function of time in days
  for the X-ray pulsar J0537-6910 and for Vela (B0833-45).
  $\Delta\Omega_i$ is the change in angular velocity of a single
  glitch.  Linear fits are indicated, with respective slopes [
  \eq{cala}] ${\cal A}=2.4\times10^{-9}$ d$^{-1}$ and
  $1.91\times10^{-9}$ d$^{-1}$.  (From \cite{Andersson:2012tw})}
\label{Fig:glitches}
\end{figure}

The second indication of superfluidity in neutron stars stems from the
fact that many pulsars exhibit sporadic spin jumps, or glitches (see
\fig{Fig:glitch}).  The Jodrell Bank glitch
catalogue\footnote{On-line version:
  \url{http://www.jb.man.ac.uk/pulsar/glitches.html}} lists 420
glitches (as of December 2012) \cite{Espinoza:2011oq}.  These are
thought to represent angular momentum transfer between the crust and a
liquid, possibly superfluid, interior
\cite{Anderson:1975mi,Ruderman:1976pi,Pines:1985ff}.  As the star's
crust spins down under the influence of magnetic torque, differential
rotation develops between the crust (and whatever other parts of the
star are tightly coupled to it) and a different portion of the
interior containing a superfluid (recall in the above that
$\dot\Omega_s=0$), but possibly a superfluid component different from
those in the core.  The now more rapidly rotating (superfluid)
component then acts as an angular momentum reservoir which exerts a
spin-up torque on the crust as a result of an instability.  The Vela
pulsar, one of the most active glitching pulsars, glitches about every
3 years, with a fractional change in the rotation rate averaging a
part in a million \cite{McCulloch:1987fu,Cordes:1988lh}, as shown in
\fig{Fig:glitches}.

The stocastic nature of glitches implies that they represent a
self-regulating instability for which the star prepares over a waiting
interval.  The amount of angular momentum transferred is observed to
increase linearly with time, as shown in \fig{Fig:glitches} for
Vela and for the X-ray pulsar J0537-6910.  Assume that the moment of
inertia of the superfluid component responsible for glitches is $I_g$
and that of the crust and whatever parts of the star are tightly
coupled to it is $I_c$.  In an individual glitch $i$, an amount of
angular momentum $I_c\Delta\Omega_i$ is transferred to the crust,
where $\Delta\Omega_i$ is the observed jump in angular velocity.  Over
the interval $t_i$ between the last glitch and the current glitch, the
star spins down by the amount $\Delta\Omega=-\dot\Omega t_i$ and the
total angular momentum differentially stored in the superfluid
component is $I_g\Delta\Omega$.  These two angular momenta are equal,
and over a total observed time $t_{obs}=\sum_it_i$, we have
\be\label{ig} I_g\simge2\tau_c{\cal A}I_c,  \ee
where the observed quantity ${\cal A}$  is related to the accumulated jumps, 
\be\label{cala}
{\cal A}=\left(\sum_i\Delta\Omega_i/\Omega\right)t_{obs}^{-1}.
\ee
In the case of the Vela pulsar, the magnitude of ${\cal A}$ implies
that $I_g/I_c=2\tau_c{\cal A}\simge0.016$\cite{Link:1999ye}.  Minimum values
for this ratio for other pulsars are given in the last column of Table
\ref{Tab:Brake}.

Since the moment of inertia in the inner crust (with the n  $\Singlet$
type superfluid) is $I_{sc}$ and satisfies
\be\label{ig0} 
I_{sc}\simeq0.04I, 
\ee 
where $I$ is the star's total moment of inertia, glitches can be
naturally explained by the inner-crust superfluid since $I_g\simle
I_{sc}$.  Although the magnitudes of individual glitches varies
somewhat, the maximum-sized glitches have stable and limited sizes.
This would be difficult to explain if glitches originated from the
inner core's superfluid component, whose associated moment of inertia
is a much larger fraction of the star's total.

Recently it has been shown that most of the neutron mass of
the dripped neutrons in the inner crust is entrained by Bragg
scattering with the nuclear lattice, effectively increasing the
neutron mass by factors of 4--5  \cite{Chamel:2012qo,Anderson:1975mi}.
With this entrainment, \eq{ig} becomes, for Vela,
\be\label{ig1} 
I_g\simge0.08I_c, 
\ee 
precluding an inner crust superfluid explanation for $I_g$ if
$I_c\simeq I$.  Entrainment results in most of the neutron fluid
spinning down with the crust, and the unentrained conduction neutrons
cannot accumulate angular momentum at a high-enough rate to produce
the largest observed glitches \cite{Link:2013il}.

B. Link \cite{Link:2013il}, however, has maintained that typically observed glitch
behavior is almost certainly a crustal feature, or perhaps due to some
small region of the core where vortex pinning is not occurring.  In
other words, it is likely that $I_c<I$, and perhaps substantially so, as
if the outer core decouples from the inner core over timescales
ranging from weeks to years.  
In fact, this is the observed timescale $\tau$ of the post-glitch
relaxation (see \fig{Fig:glitch}), which may simply represent the
dynamical recovery of the outer core.  For older pulsars, which are
cooler, the relaxation timescale is long, of the order of years, and
this could explain the nearly step-like behavior of many pulsar
glitches.  As a consequence, for example, if $I_c\simeq I/2$,
\eq{ig1} and (\ref{ig0}) regain consistency.  The upshot
is that spin glitches could then originate in either the inner crust
or the outer core.  The observed limited maximum magnitude of
glitches retains its natural explanation in terms of the small and
regular sizes of neutron star crusts as opposed to the wider
variations in sizes of core superfluid regions.

It should be noted that the above is at odds with the results of
Sidery and Alpar \cite{Sidery:2009zt},
who obtain relaxation times of only about 2 days.  This discrepancy is
the subject of ongoing discussions.


\section{Discussion and Conclusions}
\label{Sec:Conclusions}

In this chapter, the influence of pairing, leading to neutron
superfluidity and proton (and quark) superconductivity, on key
observables of neutron stars is described.  The observables include
aspects of their thermal evolution, composed of surface temperatures,
cooling rates, and ages, and dynamical evolution, comprised of pulsar
spin-down characteristics and glitch information.  The major effects
of pairing on the thermal evolution of an isolated neutron star are
the quenching of neutrino emissions and the reduction of specific heat
of the paired fermions, be they nucleons, hyperons or deconfined
quarks, in their cores.  However, the onset of pairing also triggers
short episodes of increased neutrino emission through pair breaking
and formation (PBF) processes when the ambient temperature falls
below the superfluid critical temperature.  
The PBF emission of neutrino pairs through weak interactions 
of strongly interacting particles is unique to dense neutron-star matter,
as a similar phenomenon does not occur in nuclei for \Singlet paired fermions as it
is forbidden on the basis of symmetry.
The major effects of pairing on
dynamical evolution include the reduction of the so-called braking
index, a measure of spin deceleration, of pulsars, and the possible
triggering of glitches from weak coupling of superfluid vortices to 
neutron star crusts.

The sizes and density dependences of superfluid and superconducting
gaps play a crucial role.  Pairing is observed in nuclei as an energy
difference between even-even and odd-even nuclei with a typical
magnitude, ranging from 0.5 to 3 MeV, that decreases with atomic
number.  The basic cause of pairing in nuclei is due to the attractive
interaction between neutrons in the spin $S=0$ channel.  Gaps of
similar magnitude for nucleon pairing in neutron stars are expected.
Both spin-singlet $\Singlet$ (at lower densities) and spin-triplet
$\triplet$ (at higher densities) configurations appear possible from
scattering phase shifts arguments.  Pairing appears as a gap $\Delta$
in the single particle energy spectrum, leading to a strong
suppression ($\sim e^{-2\Delta/T}$) of both specific heat and neutrino
emissivities at low temperatures.  In the simple BCS approximation,
however, the gap depends exponentialy on the pairing potential.
Hence, uncertainties associated with in-medium effects of
strong-interactions at high density have prevented a consensus about
the sizes and density dependences of gaps from being reached. In
addition, the nature and abundance of possible candidates for pairing
(nucleons and strangeness-bearing hyperons, quarks, Bose condensates)
are also uncertain.

In the neutron star crust, the density of unbound neutrons is high
enough that $\Singlet$ pairing is expected to occur.  Most theoretical
models suggest that the associated gap disappears at neutron densities
higher than that of the core-crust interface, so it is confined to the
crust.  The latest calculations indicate a maximum gap magnitude of
about $1.5\times10^{10}$ K. 
Since proton-neutron correlations reduce the
effective mass of the proton below that of the neutron, the size of
the proton $\Singlet$ gap is smaller than that of the neutron.
``Unbound'' protons exist only at densities greater than $\rhonuc$
where nuclei disappear, so proton superconductivity in the
spin-singlet state is expected to exist from the core-crust boundary
to deep into the core once temperatures fall below a few times $10^9$
K.  The $\Singlet$ neutron gap vanishes close to the core-crust interface
and the dominant pairing for neutrons in the core occurs in the
anisotropic $\Triplet$ channel.  Uncertainties in the size and density
range for this gap are larger than for $\Singlet$ gaps, however, with maximum
magnitudes ranging from a few times $10^7$ K  to a few times $10^9$ K.

The greatest influence of pairing will be on the thermal evolution of
neutron stars.  The occurrence of pairing leads to three important
effects for neutron star cooling: alteration and eventual suppression
of nucleon specific heats, suppression of neutrino emissivities, and
triggering of PBF neutrino emission for temperatures just below the
critical temperature.  Presently, 13 isolated neutron stars with
thermal spectra in the soft X-ray band have been identified.  
In addition, there are four pulsars with detected X-rays but with 
only upper limits to thermal emission.
Finally, there are six gravitational-collapse
supernova remnants that might contain neutron stars (if not, then
black holes), but no detected thermal emission as yet.  Atmospheric
modeling of the thermal sources yields estimates of surface
temperatures, and together with age estimates, allows these stars to
be compared to theoretical cooling models. Most observed sources are
younger than a million years, during which time they cool primarily
through neutrino emission.

Neutrino cooling can be either very fast (i.e., {\it enhanced}), or
relatively slow.  Enhanced cooling occurs by way of the direct Urca
process on nucleons, hyperons, Bose condensates, or deconfined quark
matter.  It is allowed when energy and momentum can be conserved with
3 or fewer degenerate fermions involved.  If additional ``bystander''
nucleons are required to conserve momentum, neutrino emission is
suppressed by about a factor of a million.  If pairing is not present
and enhanced neutrino emission does not occur, it is found that
several observed neutron stars are too cold to match cooling models.
Therefore, an important first conclusion is that {\it either pairing
  or enhanced neutrino emission must occur in some neutron stars.}  On
the other hand, if pairing is not present and enhanced neutrino
emission does occur, {\it all} observed neutron stars are too hot to
match cooling models.  Thus, a second conclusion is that {\it in the
  presence of enhanced cooling, superfluidity and/or superconductivity
  must occur.}  However, as there are many combinations
of pairing gap sizes and extents, neutron star masses and envelope
compositions, and enhanced cooling candidates that can match
observations, it is not possible to determine either gap properties or
the specific enhanced cooling reactions involved.  Therefore, for
example, neutron star cooling cannot lead to an unambiguous detection
of hyperons, Bose condensates or deconfined quark matter in the
interior of neutron stars at this time.

A final scenario, known as the {\it minimal cooling paradigm}, assumes
enhanced neutrino emission does not occur, but allows pairing.  In
this case, {\it all} observed stars, with the possible excption of the
pulsar in the supernova remnant CTA 1, for which only an upper
temperature limit is available, and the six undetected neutron stars
in gravitational-collapse supernova remnants, none of which is certain
to be a neutron star at all, can be fit by theoretical cooling models.
However, fitting observations {\it requires} two constraints on neutron
stars:
\begin{itemize}
\item
The neutron $\Triplet$ gap must have a maximum critical temperature $T_{cn}^\mathrm{max}$
larger than $\sim 0.5\times 10^9$ K, but unlikely larger than $1.5 - 2 \times 10^9$ K; and

\item some, but not all, neutron stars must have envelopes composed of
  light elements (H/He/C) and some, but not all, must have envelopes
  composed of heavy elements.
\end{itemize}
The assumed mass of a neutron star has only a minor influence on the
results of the minimal cooling scenario.  The source CTA 1 and some of
the undetected neutron stars in gravitational-collapse supernova
remnants might then be candidates for enhanced neutrino emission via
the direct Urca process.  One obvious way in which this could occur is that,
since there is a density threshold for the onset of the direct Urca
process, only neutron stars above a critical mass could
participate.  From observed and theoretically-predicted neutron star
mass distributions, this critical mass is estimated to be in the range
$1.6-1.8 M_\odot$.  This result is itself supported by recent
experimental and observational restrictions on how fast the symmetry
energy of nuclear matter can increase with density \cite{Lattimer:2012},
which suggest that the nucleon direct Urca threshold density is much
greater than $\rhonuc$.

The recently detected rapid cooling of the neutron star in Cas A
provides an unprecedented opportunity to overcome the dilemma of
deciding whether ``exotica'' are needed, or if the minimal cooling
paradigm is sufficient, to account for observational cooling data.
The rapidity of the stellar cooling in Cas A remnant points to both
the onset of core neutron superfluidity within the last few decades and the
prior existence of core proton superconductivity with a larger critical
temperature. The PBF process, with its
time-dependent burst of neutrinos at the critical temperature, is
central to the success of model calculations fitting both Cas A
cooling observations and the entire body of data for other observed
isolated neutron stars.  Furthermore, the Cas A cooling observations
can be fit only if $T_{cn}^\mathrm{max}\simeq5-6\times10^8$ K, {\it
  precisely in the range independently found} for the minimal
cooling scenario based on other isolated neutron star cooling data.
Sherlock Holmes would not have deemed this a coincidence.  ``Enhanced''
processes appear to be ruled out in the case of Cas A, as its star
would presently be too cold.  Further observations of the cooling of
the neutron star in Cas A can confirm these conclusions.

Final indications of the existence of pairing in neutron stars is
afforded by observations of the deceleration of the spin frequency and
glitches of pulsars.  The standard paradigm to explain the pulsar
mechanism is based on the magnetic rotating dipole model, for which it
can be shown that the spin-down parameter
$n=-\Omega\ddot\Omega/\dot\Omega^2=3$.  All non-accreting pulsars for
which $n$ has been determined have $n<3$, which can be understood
either if the surface dipolar magnetic field is increasing in strength
or if there is a growing superluid or superconducting component within
the neutron star's core.  An increasing surface field could, in fact,
be due to the expulsion of magnetic flux from the core by
superconductors.  The leading model for pulsar glitches is that a
more-rapidly rotating, and essentially frictionless, superfluid in the
crust sporadically transfers angular momentum to the more
slowly-rotating crust and the parts of the star strongly coupled to
it.

In conclusion, there is abundant evidence that the ultimate
``high-tem\-per\-ature'' superfluid (or superconductor) exists in nearly
every neutron star.
Unmistakably, superfluidity and superconductivity exhibited by the constituents of neutron stars 
provide avenues by which the observed cooling behaviors of these stars can be explained. 
Theoretical predictions of pairing gaps being as yet uncertain, 
astronomical data are pointing to a path forward in their determinations.
A remarkable recent development is that, despite the nearly three-order-of-magnitude theoretical 
uncertainty in the size of the neutron $\Triplet$ gap, 
and the factor of 3 uncertainty in the size of the proton $\Singlet$ gap, 
observations of the neutron star in Cas A now appear to restrict their magnitudes to remarkably small ranges.
It will be interesting to see if these considerable restrictions translate into a more complete understanding of 
the pairing interaction between nucleons and provide insights into other aspects of condensed matter and nuclear physics.
Continued X-ray observations of this star and other isolated sources, and, hopefully, 
the discovery of additional cooling neutron stars, will further enhance these efforts.


\bigskip
\subsubsection*{Acknowledgments}

DP acknowledges support by grants from UNAM-DGAPA, \# PAPIIT-113211, and Conacyt, CB-2009-01, \#132400.
MP, JML and AWS acknowledge research support from the U.S. DOE grants DE-FG02-93ER-40746, DE-AC02-87ER40317 and DE-FG02-00ER41132, respectively, and 
by the DOE Topical Collaboration ``Neutrinos and Nucleosynthesis in Hot and Dense Matter'', contract \#DE-SC0004955.  



\begin{thebibliography}{184}
\providecommand{\natexlab}[1]{#1}
\providecommand{\url}[1]{\texttt{#1}}
\expandafter\ifx\csname urlstyle\endcsname\relax
  \providecommand{\doi}[1]{doi: #1}\else
  \providecommand{\doi}{doi: \begingroup \urlstyle{rm}\Url}\fi

\bibitem[{Cooper}(1956)]{Cooper:1956qf}
L.~N. {Cooper}.
\newblock {Bound Electron Pairs in a Degenerate Fermi Gas}.
\newblock \emph{\pr} {\bf 104}:\penalty0 1189, 1956.

\bibitem[{Bardeen} et~al.(1957){Bardeen}, {Cooper}, and
  {Schrieffer}]{Bardeen:1957dq}
J.~{Bardeen}, L.~N. {Cooper}, and J.~R. {Schrieffer}.
\newblock {Theory of Superconductivity}.
\newblock \emph{\pr} {\bf 108}:\penalty0 1175, 1957.

\bibitem[{Bohr} et~al.(1958){Bohr}, {Mottelson}, and {Pines}]{Bohr:1958fk}
A.~{Bohr}, B.~R. {Mottelson}, and D.~{Pines}.
\newblock {Possible Analogy between the Excitation Spectra of Nuclei and Those
  of the Superconducting Metallic State}.
\newblock \emph{\pr} {\bf 110}:\penalty0 936, 1958.

\bibitem[{Segr\`e}(1965)]{Segre:1965uq}
E.~{Segr\`e}.
\newblock \emph{{Nuclei and Particles}}.
\newblock W. A. Benjamin, 1965.

\bibitem[{Robledo} and {Bertsch}(2013)]{Bertsch:2013bh}
L.~M. {Robledo} and G.~F. {Bertsch}.
\newblock {Pairing in finite systems: beyond the HFB theory}.
\newblock In \emph{50 Years of Nuclear BCS: Pairing in Finite Systems},
R.~A. {Broglia} and V.~{Zelevinsky}, editors. World Scientific, 2013.
[arXiv:1205.4443].

\bibitem[{Brink} and {Broglia}(2005)]{Brink:2005kx}
D.~M. {Brink} and R.~A. {Broglia}.
\newblock \emph{Nuclear Superfluidity: pairing in finite systems}.
\newblock Cambridge University Press, 2005.

\bibitem[{Broglia} and {Zelevinsky}(2013)]{Broglia:2013vn}
R.~A. {Broglia} and V.~{Zelevinsky}.
\newblock \emph{{50 Years of Nuclear BCS: Pairing in Finite Systems}}.
\newblock World Scientific, 2013.

\bibitem[{Migdal}(1959)]{Migdal:1959bh}
A.~{Migdal}.
\newblock {Superfluidity and the moments of inertia of nuclei}.
\newblock \emph{\npa} {\bf 13}:\penalty0 655, 1959.

\bibitem[{Baade} and {Zwicky}(1934)]{Baade:1934ly}
W.~{Baade} and F.~{Zwicky}.
\newblock Cosmic rays from super-novae.
\newblock \emph{\pnas} {\bf 20}:\penalty0 259, 1934.

\bibitem[{Woosley} et~al.(2002){Woosley}, {Heger}, and
  {Weaver}]{Woosley:2002cr}
S.~E. {Woosley}, A.~{Heger}, and T.~A. {Weaver}.
\newblock {The evolution and explosion of massive stars}.
\newblock \emph{\rmp} {\bf 74}:\penalty0 1015, 2002.

\bibitem[{Hessels} et~al.(2006){Hessels}, {Ransom}, {Stairs}, {Freire},
  {Kaspi}, and {Camilo}]{Hessels:2006fk}
J.~W.~T. {Hessels}, S.~M. {Ransom}, I.~H. {Stairs}, P.~C.~C. {Freire}, V.~M.
  {Kaspi}, and F.~{Camilo}.
\newblock {A Radio Pulsar Spinning at 716 Hz}.
\newblock \emph{Science} {\bf 311}:\penalty0 1901, 2006.

\bibitem[{Lattimer} and {Prakash}(2005)]{Lattimer:2005fk}
J.~M. {Lattimer} and M.~{Prakash}.
\newblock {Ultimate Energy Density of Observable Cold Baryonic Matter}.
\newblock \emph{\prl} {\bf 94\penalty0 (11)}:\penalty0 111101, 2005.

\bibitem[{Oppenheimer} and {Volkoff}(1939)]{Oppenheimer:1939uq}
J.~R. {Oppenheimer} and G.~M. {Volkoff}.
\newblock {On Massive Neutron Cores}.
\newblock \emph{\pr} {\bf 55}:\penalty0 374, 1939.

\bibitem[{Burrows} and {Lattimer}(1986)]{Burrows:1986kx}
A.~{Burrows} and J.~M. {Lattimer}.
\newblock {The birth of neutron stars}.
\newblock \emph{\apj} {\bf 307}:\penalty0 178, 1986.

\bibitem[{Page} and {Reddy}(2006)]{Page:2006vn}
D.~{Page} and S.~{Reddy}.
\newblock {Dense Matter in Compact Stars: Theoretical Developments and
  Observational Constraints}.
\newblock \emph{\arnps} {\bf 56}:\penalty0 327, 2006.

\bibitem[{Akmal} et~al.(1998){Akmal}, {Pandharipande}, and
  {Ravenhall}]{Akmal:1998fk}
A.~{Akmal}, V.~R. {Pandharipande}, and D.~G. {Ravenhall}.
\newblock {Equation of state of nucleon matter and neutron star structure}.
\newblock \emph{\prc} {\bf 58}:\penalty0 1804, September 1998.

\bibitem[{Lai}(2001)]{Lai:2001uq}
D.~{Lai}.
\newblock {Matter in strong magnetic fields}.
\newblock \emph{\rmp} {\bf 73}:\penalty0 629, 2001.

\bibitem[{Tilley} and {Tilley}(1990)]{Tilley:1990ys}
D.~R. {Tilley} and J.~{Tilley}.
\newblock \emph{Superfluidity \& Superconductivity}.
\newblock Graduate Student Series in Physics. Institute of Physics Publishing,
  1990.

\bibitem[{Pethick} and {Potekhin}(1998)]{Pethick:1998uq}
C.~J. {Pethick} and A.~Y. {Potekhin}.
\newblock {Liquid crystals in the mantles of neutron stars}.
\newblock \emph{\plb} {\bf 427}:\penalty0 7, 1998.

\bibitem[{Bertulani} and {Piekarewicz}(2012)]{Bertulani:2012ys}
C.~A. {Bertulani} and J.~{Piekarewicz}.
\newblock \emph{Neutron Star Crust}.
\newblock Nova Publisher, 2012.

\bibitem[{Tamagaki}(1970)]{Tamagaki:1970uq}
R.~{Tamagaki}.
\newblock {Superfluid state in neutron star matter. I. Generalized Bogoliubov
  transformation and existence of $^{3}$P$_{2}$ gap at high density.}
\newblock \emph{\ptp} {\bf 44}:\penalty0 905, 1970.

\bibitem[{Gorkov} and {Melik-Barkhudarov}(1961)]{Gorkov:1961kx}
L.~P. {Gorkov} and T.~K. {Melik-Barkhudarov}.
\newblock {Contribution to the Theory of Superfluidity in an Imperfect Fermi
  Gas}.
\newblock \emph{Sov. Phys. JETP} {\bf 13}:\penalty0 1018, 1961.

\bibitem[{Lombardo} and {Schulze}(2001)]{Lombardo:2001ys}
U.~{Lombardo} and H.-J. {Schulze}.
\newblock {Superfluidity in Neutron Star Matter}.
\newblock In  \emph{Physics of Neutron Star Interiors},
{D.~Blaschke, N.~K.~Glendenning, \& A.~Sedrakian}, editors,
 vol. 578 of \emph{Lecture Notes in Physics}, Springer Verlag, page~30, 2001.
[arXiv:astro-ph/0012209].

\bibitem[{Dean} and {Hjorth-Jensen}(2003)]{Dean:2003vn}
D.~J. {Dean} and M.~{Hjorth-Jensen}.
\newblock {Pairing in nuclear systems: from neutron stars to finite nuclei}.
\newblock \emph{\rmp} {\bf 75}:\penalty0 607, 2003.

\bibitem[{Sedrakian} and {Clark}(2006)]{Sedrakian2006:gk}
A.~{Sedrakian} and J.~W. {Clark}.
\newblock {Nuclear Superconductivity in Compact Stars: BCS Theory and Beyond}.
\newblock In \emph{{Pairing in Fermionic Systems: Basic Concepts and Modern Applications}},
  A.~{Sedrakian}, J.~W. {Clark}, and M.~{Alford}, editors,
 page 135. World Scientific, 2006.
 [arXiv:nucl-th/0607028].

\bibitem[{Abrikosov} et~al.(1963){Abrikosov}, {Gorkov}, and
  E.]{Abrikosov:1963zr}
A.~A. {Abrikosov}, L.~P. {Gorkov}, and {Dzyaloshinski}~I. E.
\newblock \emph{{Methods of Quantum Field Theory in Statistical Physics}}.
\newblock {Dover Publications}, 1963.

\bibitem[{Landau} et~al.(1980){Landau}, {Lifshitz}, and
  {Pitaevskii}]{Landau:1980ly}
L.~D. {Landau}, E.~M. {Lifshitz}, and L.~P. {Pitaevskii}.
\newblock \emph{{Statistical Physics: Theory of the Condensed State. (Vol. 9 of
  Course of Theoretical Physics)}}.
\newblock {Butterworth-Heinemann}, 1980.

\bibitem[{Takatsuka}(1972{\natexlab{a}})]{Takatsuka:1972bs}
T.~{Takatsuka}.
\newblock {Superfluid state in neutron star matter. III. Tensor coupling effect
  in $^{3}$P$_{2}$ energy gap.}
\newblock \emph{\ptp} {\bf 47}:\penalty0 1062, 1972{\natexlab{a}}.

\bibitem[{Amundsen} and {{\O}stgaard}(1985{\natexlab{a}})]{Amundsen:1985nx}
L.~{Amundsen} and E.~{{\O}stgaard}.
\newblock {Superfluidity of neutron matter (II). Triplet pairing}.
\newblock \emph{\npa} {\bf 442}:\penalty0 163, 1985{\natexlab{a}}.

\bibitem[{Baldo} et~al.(1992){Baldo}, {Cugnon}, {Lejeune}, and
  {Lombardo}]{Baldo:1992kx}
M.~{Baldo}, J.~{Cugnon}, A.~{Lejeune}, and U.~{Lombardo}.
\newblock {Proton and neutron superfluidity in neutron star matter}.
\newblock \emph{\npa} {\bf 536}:\penalty0 349, 1992.

\bibitem[{Levenfish} and {Yakovlev}(1994)]{Levenfish:1994vn}
K.~P. {Levenfish} and D.~G. {Yakovlev}.
\newblock {Specific heat of neutron star cores with superfluid nucleons}.
\newblock \emph{Astron. Rep.} {\bf 38}:\penalty0 247, 1994.

\bibitem[{Schulze} et~al.(1996){Schulze}, {Cugnon}, {Lejeune}, {Baldo}, and
  {Lombardo}]{Schulze:1996ys}
H.-J. {Schulze}, J.~{Cugnon}, A.~{Lejeune}, M.~{Baldo}, and U.~{Lombardo}.
\newblock {Medium polarization effects on neutron matter superfluidity}.
\newblock \emph{\plb} {\bf 375}:\penalty0 1, 1996.

\bibitem[{Chen} et~al.(1993){Chen}, {Clark}, {Dav{\'e}}, and
  {Khodel}]{Chen:1993ly}
J.~M.~C. {Chen}, J.~W. {Clark}, R.~D. {Dav{\'e}}, and V.~V. {Khodel}.
\newblock {Pairing gaps in nucleonic superfluids}.
\newblock \emph{\npa} {\bf 555}:\penalty0 59, 1993.

\bibitem[{Wambach} et~al.(1993){Wambach}, {Ainsworth}, and
  {Pines}]{Wambach:1993zr}
J.~{Wambach}, T.~L. {Ainsworth}, and D.~{Pines}.
\newblock {Quasiparticle interactions in neutron matter for applications in
  neutron stars}.
\newblock \emph{\npa} {\bf 555}:\penalty0 128, 1993.

\bibitem[{Schwenk} et~al.(2003){Schwenk}, {Friman}, and
  {Brown}]{Schwenk:2003ve}
A.~{Schwenk}, B.~{Friman}, and G.~E. {Brown}.
\newblock {Renormalization group approach to neutron matter: quasiparticle
  interactions, superfluid gaps and the equation of state}.
\newblock \emph{\npa} {\bf 713}:\penalty0 191, 2003.

\bibitem[{Gandolfi} et~al.(2009){Gandolfi}, {Illarionov}, {Pederiva},
  {Schmidt}, and {Fantoni}]{Gandolfi:2009dq}
S.~{Gandolfi}, A.~Y. {Illarionov}, F.~{Pederiva}, K.~E. {Schmidt}, and
  S.~{Fantoni}.
\newblock {Equation of state of low-density neutron matter, and the
  $^{1}$S$_{0}$ pairing gap}.
\newblock \emph{\prc} {\bf  80\penalty0 (4)}:\penalty0 045802, 2009.

\bibitem[{Gezerlis} and {Carlson}(2010)]{Gezerlis:2010fk}
A.~{Gezerlis} and J.~{Carlson}.
\newblock {Low-density neutron matter}.
\newblock \emph{\prc} {\bf 81\penalty0 (2)}:\penalty0 025803, 2010.

\bibitem[{Vigezzi} et~al.(2005){Vigezzi}, {Barranco}, {Broglia}, {Col{\`o}},
  {Gori}, and {Ramponi}]{Vigezzi:2005kx}
E.~{Vigezzi}, F.~{Barranco}, R.~A. {Broglia}, G.~{Col{\`o}}, G.~{Gori}, and
  F.~{Ramponi}.
\newblock {Pairing correlations in the inner crust of neutron stars}.
\newblock \emph{\npa} {\bf 752}:\penalty0 600, 2005.

\bibitem[{Baldo} et~al.(2005){Baldo}, {Lombardo}, {Saperstein}, and
  {Tolokonnikov}]{Baldo:2005vn}
M.~{Baldo}, U.~{Lombardo}, {\'E}.~E. {Saperstein}, and S.~V. {Tolokonnikov}.
\newblock {Self-consistent description of the inner crust of a neutron star
  with allowance for superfluidity effects}.
\newblock \emph{Physics of Atomic Nuclei} {\bf 68}:\penalty0 1812, 2005.

\bibitem[{Margueron} and {Sandulescu}(2012)]{Margueron:2012ys}
J.~{Margueron} and N.~{Sandulescu}.
\newblock {Pairing Correlations and Thermodynamic Properties of Inner Crust Matter}.
\newblock In \emph{Neutron Star Crust},
C.~A. {Bertulani} and J.~{Piekarewicz}, editors. Nova Publisher, 2012.
[arXiv:1201.2774].

\bibitem[{Chamel} et~al.(2012){Chamel}, {Pearson}, and
  {Goriely}]{Chamel:2012zr}
N.~{Chamel}, J.~M. {Pearson}, and S.~{Goriely}.
\newblock {Pairing: from atomic nuclei to neutron-star crusts}.
\newblock In \emph{{50 Years of Nuclear BCS: Pairing in Finite Systems}},
R.~A. {Broglia} and V.~{Zelevinsky}, editors. World Scientific, 2013.
[arXiv:1204.2076].

\bibitem[{Chao} et~al.(1972){Chao}, {Clark}, and {Yang}]{Chao:1972kl}
N.-C. {Chao}, J.~W. {Clark}, and C.-H. {Yang}.
\newblock {Proton superfluidity in neutron-star matter}.
\newblock \emph{\npa} {\bf 179}:\penalty0 320, 1972.

\bibitem[{Takatsuka}(1973)]{Takatsuka:1973qf}
T.~{Takatsuka}.
\newblock {Proton Superfluidity in Neutron-Star Matter}.
\newblock \emph{\ptp} {\bf 50}:\penalty0 1754, 1973.

\bibitem[{Amundsen} and {{\O}stgaard}(1985{\natexlab{b}})]{Amundsen:1985ab}
L.~{Amundsen} and E.~{{\O}stgaard}.
\newblock {Superfluidity of neutron matter (I). Singlet pairing}.
\newblock \emph{\npa} {\bf 437}:\penalty0 487, 1985{\natexlab{b}}.

\bibitem[{Elgar{\o}y} et~al.(1996){Elgar{\o}y}, {Engvik}, {Hjorth-Jensen}, and
  {Osnes}]{Elgaroy:1996cr}
{\O}.~{Elgar{\o}y}, L.~{Engvik}, M.~{Hjorth-Jensen}, and E.~{Osnes}.
\newblock {Model-space approach to $^{1}$S$_{0}$ neutron and proton pairing in
  neutron star matter with the Bonn meson-exchange potentials}.
\newblock \emph{\npa} {\bf 604}:\penalty0 466, 1996.

\bibitem[{Baldo} and {Schulze}(2007)]{Baldo:2007fk}
M.~{Baldo} and H.-J. {Schulze}.
\newblock {Proton pairing in neutron stars}.
\newblock \emph{\prc} {\bf 75\penalty0 (2)}:\penalty0 025802, 2007.

\bibitem[{Cao} et~al.(2006){Cao}, {Lombardo}, and {Schuck}]{Cao:2006uq}
L.~G. {Cao}, U.~{Lombardo}, and P.~{Schuck}.
\newblock {Screening effects in superfluid nuclear and neutron matter within
  Brueckner theory}.
\newblock \emph{\prc} {\bf 74\penalty0 (6)}:\penalty0 064301, 2006.

\bibitem[{Baldo} et~al.(1998){Baldo}, {Elgar{\o}y}, {Engvik}, {Hjorth-Jensen},
  and {Schulze}]{Baldo:1998zr}
M.~{Baldo}, {\O}.~{Elgar{\o}y}, L.~{Engvik}, M.~{Hjorth-Jensen}, and H.-J.
  {Schulze}.
\newblock {$^{3}$P$_{2}$-$^{3}$F$_{2}$ pairing in neutron matter with modern
  nucleon-nucleon potentials}.
\newblock \emph{\prc} {\bf 58}:\penalty0 1921, 1998.

\bibitem[{Zhou} et~al.(2004){Zhou}, {Schulze}, {Zhao}, {Pan}, and
  {Draayer}]{Zhou:2004fv}
X.-R. {Zhou}, H.-J. {Schulze}, E.-G. {Zhao}, F.~{Pan}, and J.~P. {Draayer}.
\newblock {Pairing gaps in neutron stars}.
\newblock \emph{\prc} {\bf 70\penalty0 (4)}:\penalty0 048802, 2004.

\bibitem[{Schwenk} and {Friman}(2004)]{Schwenk:2004ve}
A.~{Schwenk} and B.~{Friman}.
\newblock {Polarization Contributions to the Spin Dependence of the Effective
  Interaction in Neutron Matter}.
\newblock \emph{\prl} {\bf 92\penalty0 (8)}:\penalty0 082501, 2004.

\bibitem[{Page} et~al.(2004){Page}, {Lattimer}, {Prakash}, and
  {Steiner}]{Page:2004zr}
D.~{Page}, J.~M. {Lattimer}, M.~{Prakash}, and A.~W. {Steiner}.
\newblock {Minimal Cooling of Neutron Stars: A New Paradigm}.
\newblock \emph{\apjs} {\bf 155}:\penalty0 623, 2004.

\bibitem[{Page} et~al.(2009){Page}, {Lattimer}, {Prakash}, and
  {Steiner}]{Page:2009qf}
D.~{Page}, J.~M. {Lattimer}, M.~{Prakash}, and A.~W. {Steiner}.
\newblock {Neutrino Emission from Cooper Pairs and Minimal Cooling of Neutron
  Stars}.
\newblock \emph{\apj} {\bf 707}:\penalty0 1131, 2009.

\bibitem[{Page} et~al.(2011){Page}, {Prakash}, {Lattimer}, and
  {Steiner}]{Page:2011ys}
D.~{Page}, M.~{Prakash}, J.~M. {Lattimer}, and A.~W. {Steiner}.
\newblock {Rapid Cooling of the Neutron Star in Cassiopeia A Triggered by
  Neutron Superfluidity in Dense Matter}.
\newblock \emph{\prl} {\bf 106\penalty0 (8)}:\penalty0 081101, 2011.

\bibitem[{Hoffberg} et~al.(1970){Hoffberg}, {Glassgold}, {Richardson}, and
  {Ruderman}]{Hoffberg:1970hc}
M.~{Hoffberg}, A.~E. {Glassgold}, R.~W. {Richardson}, and M.~{Ruderman}.
\newblock {Anisotropic Superfluidity in Neutron Star Matter}.
\newblock \emph{\prl} {\bf 24}:\penalty0 775, 1970.

\bibitem[{Takatsuka}(1972{\natexlab{b}})]{Takatsuka:1972ij}
T.~{Takatsuka}.
\newblock {Energy Gap in Neutron-Star Matter}.
\newblock \emph{\ptp} {\bf 48}:\penalty0 1517, 1972{\natexlab{b}}.

\bibitem[{Zuo} et~al.(2008){Zuo}, {Cui}, {Lombardo}, and {Schulze}]{Zuo:2008zr}
W.~{Zuo}, C.~X. {Cui}, U.~{Lombardo}, and H.-J. {Schulze}.
\newblock {Three-body force effect on $^3$PF$_{2}$ neutron superfluidity in
  neutron matter, neutron star matter, and neutron stars}.
\newblock \emph{\prc} {\bf 78\penalty0 (1)}:\penalty0 015805, 2008.

\bibitem[{Page} and {Applegate}(1992)]{Page:1992nx}
D.~{Page} and J.~H. {Applegate}.
\newblock {The cooling of neutron stars by the direct URCA process}.
\newblock \emph{\apjl} {\bf 394}:\penalty0 L17, 1992.

\bibitem[{Takatsuka} and {Tamagaki}(1997)]{Takatsuka:1997ff}
T.~{Takatsuka} and R.~{Tamagaki}.
\newblock {Nucleon Superfluidity in Neutron Star Core with Direct URCA
  Cooling}.
\newblock \emph{\ptp} {\bf 97}:\penalty0 345, 1997.

\bibitem[{Prakash} et~al.(1992){Prakash}, {Prakash}, {Lattimer}, and
  {Pethick}]{Prakash:1992vn}
M.~{Prakash}, M.~{Prakash}, J.~M. {Lattimer}, and C.~J. {Pethick}.
\newblock {Rapid cooling of neutron stars by hyperons and Delta isobars}.
\newblock \emph{\apjl} {\bf 390}:\penalty0 L77, 1992.

\bibitem[{Chen} et~al.(2008){Chen}, {Li}, {Wen}, and {Liu}]{Chen:2008mi}
W.~{Chen}, B.-J. {Li}, D.-H. {Wen}, and L.-G. {Liu}.
\newblock {$^{3}$P$_{2}$ superfluidity in neutron star matter}.
\newblock \emph{\prc} {\bf 77\penalty0 (6)}:\penalty0 065804, 2008.

\bibitem[{Prakash}(1996)]{Prakash:1996fk}
M.~{Prakash}.
\newblock {The nuclear equation of state and neutron stars}.
\newblock In A.~{Ansari} and L.~{Satpathy}, editors, \emph{{The Nuclear
  Equation of State}}, page 229. World Scientific Publishing Co, 1996.

\bibitem[{Balberg} and {Barnea}(1998)]{Balberg:1998dz}
S.~{Balberg} and N.~{Barnea}.
\newblock {S-wave pairing of {$\Lambda$} hyperons in dense matter}.
\newblock \emph{\prc} {\bf 57}:\penalty0 409, 1998.

\bibitem[Gal and Hungerford(2005)]{Gal05}
A.~Gal and E.~Hungerford.
\newblock Proceedings of the eighth international conference on hypernuclear
  and strange particle physics.
\newblock \emph{\npa} {\bf 754}:\penalty0 1, 2005.

\bibitem[Hashimoto and Tamura(2006)]{Hashimoto06}
O.~Hashimoto and H.~Tamura.
\newblock Spectroscopy of lambda hypernuclei.
\newblock \emph{Prog. Part. Nucl. Phys.} {\bf 57}:\penalty0 564, 2006.

\bibitem[Batty et~al.(1997)Batty, Friedman, and Gal]{Batty97}
C.~J. Batty, E.~Friedman, and A.~Gal.
\newblock Strong interaction physics from hadronic atoms.
\newblock \emph{\prep} {\bf 287}:\penalty0 385, 1997.

\bibitem[Beane et~al.(2012)Beane, Chang, Cohen, Detmold, Lin, Luu, Orginos,
  Parre\~no, Savage, and Walker-Loud]{Beane12}
S.~R. Beane, E.~Chang, S.~D. Cohen, W.~Detmold, H.-W. Lin, T.~C. Luu,
  K.~Orginos, A.~Parre\~no, M.~J. Savage, and A.~Walker-Loud.
\newblock Hyperon-nucleon interactions from quantum chromodynamics and the
  composition of dense nuclear matter.
\newblock \emph{\prl} {\bf 109}:\penalty0 172001, 2012.

\bibitem[{Vida{\~n}a} and {Tol{\'o}s}(2004)]{Vidana:2004qa}
I.~{Vida{\~n}a} and L.~{Tol{\'o}s}.
\newblock {Superfluidity of {$\Sigma$}$^{-}$ hyperons in {$\beta$} -stable
  neutron star matter}.
\newblock \emph{\prc} {\bf 70\penalty0 (2)}:\penalty0 028802, 2004.

\bibitem[{Takatsuka} et~al.(2006){Takatsuka}, {Nishizaki}, {Yamamoto}, and
  {Tamagaki}]{Takatsuka:2006fu}
T.~{Takatsuka}, S.~{Nishizaki}, Y.~{Yamamoto}, and R.~{Tamagaki}.
\newblock {Occurrence of Hyperon Superfluidity in Neutron Star Cores}.
\newblock \emph{\ptp} {\bf 115}:\penalty0 355, 2006.

\bibitem[{Wang} and {Shen}(2010)]{Wang:2010kl}
Y.~N. {Wang} and H.~{Shen}.
\newblock {Superfluidity of {$\Lambda$} hyperons in neutron stars}.
\newblock \emph{\prc} {\bf 81\penalty0 (2)}:\penalty0 025801, 2010.

\bibitem[{Yakovlev} and {Levenfish}(1995)]{Yakovlev:1995ly}
D.~G. {Yakovlev} and K.~P. {Levenfish}.
\newblock {Modified URCA process in neutron star cores.}
\newblock \emph{\aap} {\bf 297}:\penalty0 717, 1995.

\bibitem[{Flowers} et~al.(1976){Flowers}, {Ruderman}, and
  {Sutherland}]{Flowers:1976vn}
E.~{Flowers}, M.~{Ruderman}, and P.~{Sutherland}.
\newblock {Neutrino pair emission from finite-temperature neutron superfluid
  and the cooling of young neutron stars}.
\newblock \emph{\apj} {\bf 205}:\penalty0 541, 1976.

\bibitem[{Voskresenskii} and {Senatorov}(1986)]{Voskresenskii:1986nx}
D.~N. {Voskresenskii} and A.~V. {Senatorov}.
\newblock {Neutrino emission by neutron stars}.
\newblock \emph{Zh. Eksp. Teor. Fiz.}  {\bf 90}:\penalty0 1505, 1986.

\bibitem[{Collins} and {Perry}(1975)]{Collins75}
J.~C. {Collins} and M.~J. {Perry}.
\newblock {Superdense matter: Neutrons or asymptotically free quarks?}
\newblock \emph{\prl} {\bf 34}:\penalty0 1353, 1975.

\bibitem[{Wilczek}(2005)]{Wilczek05}
F.~{Wilczek}.
\newblock {Asymptotic Freedom: From Paradox to Paradigm}.
\newblock \emph{\pnas} {\bf 102\penalty0 (24)}:\penalty0 84038413, 2005.

\bibitem[SIS(2011)]{proc11}
\emph{{Lattice 2011: Proceedings of the XXIX International Symposium on Lattice
  Field Theory}}, 2011. SISSA, {Proceedings of Science}.

\bibitem[{Bodmer}(1971)]{Bodmer71}
A.~R. {Bodmer}.
\newblock Collapsed nuclei.
\newblock \emph{\prd} {\bf 4}:\penalty0 1601, 1971.

\bibitem[{Witten}(1984)]{Witten84}
E.~{Witten}.
\newblock Cosmic separation of phases.
\newblock \emph{\prd} {\bf 30}:\penalty0 272, 1984.

\bibitem[{Farhi} and {Jaffe}(1984)]{Farhi84}
E.~{Farhi} and R.~L. {Jaffe}.
\newblock Strange matter.
\newblock \emph{\prd} {\bf 30}:\penalty0 2379, 1984.

\bibitem[{Alcock} et~al.(1986){Alcock}, {Farhi}, and {Olinto}]{Alcock:1986ys}
C.~{Alcock}, E.~{Farhi}, and A.~{Olinto}.
\newblock {Strange stars}.
\newblock \emph{\apj} {\bf 310}:\penalty0 261, 1986.

\bibitem[{Haensel} et~al.(1986){Haensel}, {Zdunik}, and
  {Schaefer}]{Haensel:1986zr}
P.~{Haensel}, J.~L. {Zdunik}, and R.~{Schaefer}.
\newblock {Strange quark stars}.
\newblock \emph{\aap} {\bf 160}:\penalty0 121, 1986.

\bibitem[Barrois(1977)]{Barrois77}
B.~C. Barrois.
\newblock Superconducting quark matter.
\newblock \emph{\npb} {\bf 129}:\penalty0 390, 1977.

\bibitem[Bailin and Love(1979)]{Bailin79}
D.~Bailin and A.~Love.
\newblock Superfluid quark matter.
\newblock \emph{J. Phys. A: Math. Gen.} {\bf 12}:\penalty0 L283, 1979.

\bibitem[Alford et~al.(1998)Alford, Rajagopal, and Wilczek]{Alford98}
M.~Alford, K.~Rajagopal, and F.~Wilczek.
\newblock QCD at finite baryon density: nucleon droplets and color
  superconductivity.
\newblock \emph{\plb} {\bf 422}:\penalty0 247, 1998.

\bibitem[Rapp et~al.(1888)Rapp, Sch\"{a}fer, Shuryak, and Velkovsky]{Rapp98}
R.~Rapp, T.~Sch\"{a}fer, E.~V. Shuryak, and M.~Velkovsky.
\newblock Diquark bose condensates in high density matter and instantons.
\newblock \emph{\prl} {\bf 81}:\penalty0 53, 1888.

\bibitem[Alford et~al.(2008)Alford, Schmitt, Rajagopal, and
  Sch\"{a}fer]{Alford08}
M.~Alford, A.~Schmitt, K.~Rajagopal, and T.~Sch\"{a}fer.
\newblock Color superconductivity in dense quark matter.
\newblock \emph{\rmp} {\bf 80}:\penalty0 1455, 2008.

\bibitem[Polchinski(2012)]{Polchinski92}
J.~Polchinski.
\newblock Effective field theory and the Fermi surface.
\newblock \emph{arXiv:hep-th/9210046}, 1999.

\bibitem[Shankar(1994)]{Shankar94}
R.~Shankar.
\newblock Renormalization-group approach to interacting fermions.
\newblock \emph{\rmp} {\bf 66}:\penalty0 129, 1994.

\bibitem[Hong(2000)]{Hong00}
D.~K. Hong.
\newblock An effective field theory of {QCD} at high density.
\newblock \emph{\plb} {\bf 473}:\penalty0 118, 2000.

\bibitem[Sch\"{a}fer(2003)]{Schafer03}
T.~Sch\"{a}fer.
\newblock Hard loops, soft loops, and high density effective field theory.
\newblock \emph{\npa} {\bf 728}:\penalty0 251, 2003.

\bibitem[{Nambu} and {Jona-Lasinio}(1961)]{Nambu61}
Y.~{Nambu} and G.~{Jona-Lasinio}.
\newblock {Dynamical Model of Elementary Particles Based on an Analogy with
  Superconductivity. I}.
\newblock \emph{\pr} {\bf 122}:\penalty0 345, 1961.

\bibitem[{Hatsuda} and {Kunhiro}(1994)]{Hatsuda94}
T.~{Hatsuda} and T.~{Kunhiro}.
\newblock {QCD} phenomenology based on a chiral effective lagrangian.
\newblock \emph{\pr} {\bf 247}:\penalty0 221, 1994.

\bibitem[Buballa(2005)]{Buballa05}
M.~Buballa.
\newblock NJL-model analysis of dense quark matter.
\newblock \emph{\prep} {\bf 407}:\penalty0 205, 2005.

\bibitem[Steiner et~al.(2002)Steiner, Reddy, and Prakash]{Steiner02}
A.~W. Steiner, S.~Reddy, and M.~Prakash.
\newblock Color-neutral superconducting quark matter.
\newblock \emph{\prd} {\bf 66}:\penalty0 094007, 2002.

\bibitem['t~Hooft(1976)]{tHooft76}
G.~'t~Hooft.
\newblock Symmetry breaking through {Bell-Jackiw} anomalies.
\newblock \emph{\prl} {\bf 37}:\penalty0 8, 1976.

\bibitem[Buballa and Oertel(1998)]{Buballa98}
M.~Buballa and M.~Oertel.
\newblock Quark droplets in the {NJL} mean field.
\newblock \emph{\npa} {\bf 642}:\penalty0 39, 1998.

\bibitem[Shovkovy and Huang(2003)]{Shovkovy03}
I.~Shovkovy and M.~Huang.
\newblock Gapless two-flavor color superconductor.
\newblock \emph{\plb} {\bf 564}:\penalty0 205, 2003.

\bibitem[Bedaque and Sch\"{a}fer(2002)]{Bedaque02}
P.~F. Bedaque and T.~Sch\"{a}fer.
\newblock High-density quark matter under stress.
\newblock \emph{\npa} {\bf 697}:\penalty0 802, 2002.

\bibitem[Sch\"{a}fer(2004)]{Schafer03b}
T.~Sch\"{a}fer.
\newblock Quark matter.
\newblock In \emph{Quarks and Mesons},
A.~B. {Santra {\it et al.}}, editors. Narosa Publishing House, New Dehli, 2004.
[arXiv:hep-ph/0304281].

\bibitem[Fulde and Ferrell(1964)]{Fulde64}
P.~Fulde and R.~A. Ferrell.
\newblock Superconductivity in a strong spin-exchange field.
\newblock \emph{\pr} {\bf 135}:\penalty0 A550, 1964.

\bibitem[Larkin and Ovchinnikov(1965)]{Larkin65}
A.~I. Larkin and Y.~N. Ovchinnikov.
\newblock Inhomogeneous state of superconductors.
\newblock \emph{Sov. Phys. JETP} {\bf 20}:\penalty0 762, 1965.

\bibitem[Alford et~al.(2001)Alford, Bowers, and Rajagopal]{Alford01}
M.~Alford, J.~A. Bowers, and K.~Rajagopal.
\newblock {Crystalline color superconductivity}.
\newblock \emph{\prd} {\bf 63}:\penalty0 074016, 2001.

\bibitem[{Alford} et~al.(2005){Alford}, {Braby}, {Paris}, and
  {Reddy}]{Alford04}
M.~{Alford}, M.~{Braby}, M.~W. {Paris}, and S.~{Reddy}.
\newblock {Hybrid stars that masquerade as neutron stars}.
\newblock \emph{\apj} {\bf 629}:\penalty0 969, 2005.

\bibitem[{Yakovlev} et~al.(2001){Yakovlev}, {Kaminker}, {Gnedin}, and
  {Haensel}]{Yakovlev:2001dq}
D.~G. {Yakovlev}, A.~D. {Kaminker}, O.~Y. {Gnedin}, and P.~{Haensel}.
\newblock {Neutrino emission from neutron stars}.
\newblock \emph{\physrep} {\bf 354}:\penalty0 1, 2001.

\bibitem[{Voskresensky}(2001)]{Voskresensky:2001cr}
D.~N. {Voskresensky}.
\newblock {Neutrino Cooling of Neutron Stars: Medium Effects}.
\newblock In \emph{Physics of Neutron Star Interiors},
{D.~Blaschke, N.~K.~Glendenning, \& A.~Sedrakian}, editors.
 \emph{{Lecture Notes in Physics}}, vol. 578, page 467. Springer Verlag, 2001.
[arXiv:astro-ph/0101514].

\bibitem[{Lattimer} et~al.(1991){Lattimer}, {Prakash}, {Pethick}, and
  {Haensel}]{Lattimer:1991kx}
J.~M. {Lattimer}, M.~{Prakash}, C.~J. {Pethick}, and P.~{Haensel}.
\newblock {Direct URCA process in neutron stars}.
\newblock \emph{\prl} {\bf 66}:\penalty0 2701, 1991.

\bibitem[{Friman} and {Maxwell}(1979)]{Friman:1979bh}
B.~L. {Friman} and O.~V. {Maxwell}.
\newblock {Neutrino emissivities of neutron stars}.
\newblock \emph{\apj} {\bf 232}:\penalty0 541, 1979.

\bibitem[{Flowers} et~al.(1975){Flowers}, {Sutherland}, and
  {Bond}]{Flowers:1975qf}
E.~G. {Flowers}, P.~G. {Sutherland}, and J.~R. {Bond}.
\newblock {Neutrino pair bremsstrahlung by nucleons in neutron-star matter}.
\newblock \emph{\prd} {\bf 12}:\penalty0 315, 1975.

\bibitem[Iwamoto(1982)]{Iwamoto82}
N.~Iwamoto.
\newblock Neutrino emissivities and mean free paths of degenerate quark matter.
\newblock \emph{Ann. Phys.} {\bf 141}:\penalty0 1, 1982.

\bibitem[{Maxwell} et~al.(1977){Maxwell}, {Brown}, {Campbell}, {Dashen}, and
  {Manassah}]{Maxwell:1977zr}
O.~{Maxwell}, G.~E. {Brown}, D.~K. {Campbell}, R.~F. {Dashen}, and J.~T.
  {Manassah}.
\newblock {Beta decay of pion condensates as a cooling mechanism for neutron
  stars}.
\newblock \emph{\apj} {\bf 216}:\penalty0 77, 1977.

\bibitem[{Brown} et~al.(1988){Brown}, {Kubodera}, {Page}, and
  {Pizzochero}]{Brown:1988ly}
G.~E. {Brown}, K.~{Kubodera}, D.~{Page}, and P.~{Pizzochero}.
\newblock {Strangeness condensation and cooling of neutron stars}.
\newblock \emph{\prd} {\bf 37}:\penalty0 2042, 1988.

\bibitem[{Voskresensky} and {Senatorov}(1987)]{Voskresensky:1987uq}
D.~N. {Voskresensky} and A.~V. {Senatorov}.
\newblock Description of nuclear interaction in the Keldysh diagram technique
  and neutrino luminosity of neutron stars.
\newblock \emph{Sov. J. Nucl. Phys.} {\bf 45}:\penalty0 411, 1987.

\bibitem[{Senatorov} and {Voskresensky}(1987)]{Senatorov:1987ve}
A.~V. {Senatorov} and D.~N. {Voskresensky}.
\newblock {Collective excitations in nucleonic matter and the problem of
  cooling of neutron stars}.
\newblock \emph{\plb} {\bf 184}:\penalty0 119, 1987.

\bibitem[{Schaab} et~al.(1997){Schaab}, {Voskresensky}, {Sedrakian}, {Weber},
  and {Weigel}]{Schaab:1997qf}
C.~{Schaab}, D.~{Voskresensky}, A.~D. {Sedrakian}, F.~{Weber}, and M.~K.
  {Weigel}.
\newblock {Impact of medium effects on the cooling of non-superfluid and
  superfluid neutron stars.}
\newblock \emph{\aap} {\bf 321}:\penalty0 591, 1997.

\bibitem[{Page}(1998)]{Page:1998bh}
D.~{Page}.
\newblock {Thermal Evolution of Isolated Neutron Stars}.
\newblock In \emph{The Many Faces of Neutron Stars}. 
{R.~Buccheri, J.~van Paradijs, \& A.~Alpar}, editors. NATO ASIC Proc. 515, page 539, 1998.
[arXiv:astro-ph/9802171].

\bibitem[{Leinson} and {P{\'e}rez}(2006)]{Leinson:2006fk}
L.~B. {Leinson} and A.~{P{\'e}rez}.
\newblock {Vector current conservation and neutrino emission from
  singlet-paired baryons in neutron stars}.
\newblock \emph{\plb} {\bf 638}:\penalty0 114, 2006.

\bibitem[{Steiner} and {Reddy}(2009)]{Steiner:2009uq}
A.~W. {Steiner} and S.~{Reddy}.
\newblock {Superfluid response and the neutrino emissivity of neutron matter}.
\newblock \emph{\prc} {\bf 79\penalty0 (1)}:\penalty0 015802, 2009.

\bibitem[{Yakovlev} et~al.(1999){Yakovlev}, {Kaminker}, and
  {Levenfish}]{Yakovlev:1999cr}
D.~G. {Yakovlev}, A.~D. {Kaminker}, and K.~P. {Levenfish}.
\newblock {Neutrino emission due to Cooper pairing of nucleons in cooling
  neutron stars}.
\newblock \emph{\aap} {\bf 343}:\penalty0 650, 1999.

\bibitem[{Jaikumar} and {Prakash}(2001)]{Jaikumar:2001zr}
P.~{Jaikumar} and M.~{Prakash}.
\newblock {Neutrino pair emission from Cooper pair breaking and recombination
  in superfluid quark matter}.
\newblock \emph{\plb} {\bf 516}:\penalty0 345, 2001.

\bibitem[{Baym} and {Pethick}(2004)]{Baym:2004nx}
G.~{Baym} and C.~{Pethick}.
\newblock \emph{Landau Fermi-Liquid theory}.
\newblock Wiley-VCH, 2004.

\bibitem[{Romani}(1987)]{Romani:1987kx}
R.~W. {Romani}.
\newblock {Model atmospheres for cooling neutron stars}.
\newblock \emph{\apj} {\bf 313}:\penalty0 718, 1987.

\bibitem[{Page}(1995)]{Page:1995vn}
D.~{Page}.
\newblock {Surface temperature of a magnetized neutron star and interpretation
  of the ROSAT data. I. Dipolar fields}.
\newblock \emph{\apj} {\bf 442}:\penalty0 273, 1995.

\bibitem[{Gudmundsson} et~al.(1983){Gudmundsson}, {Pethick}, and
  {Epstein}]{Gudmundsson:1983kx}
E.~H. {Gudmundsson}, C.~J. {Pethick}, and R.~I. {Epstein}.
\newblock {Structure of neutron star envelopes}.
\newblock \emph{\apj} {\bf 272}:\penalty0 286, 1983.

\bibitem[{Page} et~al.(2006){Page}, {Geppert}, and {Weber}]{Page:2006ly}
D.~{Page}, U.~{Geppert}, and F.~{Weber}.
\newblock {The cooling of compact stars}.
\newblock \emph{\npa} {\bf 777}:\penalty0 497, 2006.

\bibitem[{Lattimer} et~al.(1994){Lattimer}, {van Riper}, {Prakash}, and
  {Prakash}]{Lattimer:1994fk}
J.~M. {Lattimer}, K.~A. {van Riper}, M.~{Prakash}, and M.~{Prakash}.
\newblock {Rapid cooling and the structure of neutron stars}.
\newblock \emph{\apj} {\bf 425}:\penalty0 802, 1994.

\bibitem[{Prakash} et~al.(1988){Prakash}, {Lattimer}, and
  {Ainsworth}]{Prakash:1988oq}
M.~{Prakash}, J.~M. {Lattimer}, and T.~L. {Ainsworth}.
\newblock {Equation of state and the maximum mass of neutron stars}.
\newblock \emph{\prl} {\bf 61}:\penalty0 2518, 1988.

\bibitem[{Page}(2013)]{Page:2012vn}
D.~{Page}.
\newblock {Pairing and the Cooling of Neutron Stars}.
\newblock In \emph{{50 Years of Nuclear BCS: Pairing in Finite Systems}},
R.~A. {Broglia} and V.~{Zelevinsky}, editors. World Scientific, 2013.
[arXiv:1206.5011].

\bibitem[{Potekhin} and {Yakovlev}(2001)]{Potekhin:2001ve}
A.~Y. {Potekhin} and D.~G. {Yakovlev}.
\newblock {Thermal structure and cooling of neutron stars with magnetized
  envelopes}.
\newblock \emph{\aap} {\bf 374}:\penalty0 213, 2001.

\bibitem[{Geppert} et~al.(2004){Geppert}, {K{\"u}ker}, and
  {Page}]{Geppert:2004ly}
U.~{Geppert}, M.~{K{\"u}ker}, and D.~{Page}.
\newblock {Temperature distribution in magnetized neutron star crusts}.
\newblock \emph{\aap} {\bf 426}:\penalty0 267, 2004.

\bibitem[{P{\'e}rez-Azor{\'{\i}}n} et~al.(2006){P{\'e}rez-Azor{\'{\i}}n},
  {Miralles}, and {Pons}]{Perez-Azorin:2006ve}
J.~F. {P{\'e}rez-Azor{\'{\i}}n}, J.~A. {Miralles}, and J.~A. {Pons}.
\newblock {Anisotropic thermal emission from magnetized neutron stars}.
\newblock \emph{\aap} {\bf 451}:\penalty0 1009, 2006.

\bibitem[{Geppert} et~al.(2006){Geppert}, {K{\"u}ker}, and
  {Page}]{Geppert:2006fk}
U.~{Geppert}, M.~{K{\"u}ker}, and D.~{Page}.
\newblock {Temperature distribution in magnetized neutron star crusts. II. The
  effect of a strong toroidal component}.
\newblock \emph{\aap} {\bf 457}:\penalty0 937, 2006.

\bibitem[{Page} et~al.(2007){Page}, {Geppert}, and {K{\"u}ker}]{Page:2007bh}
D.~{Page}, U.~{Geppert}, and M.~{K{\"u}ker}.
\newblock {Cooling of neutron stars with strong toroidal magnetic fields}.
\newblock \emph{\apss} {\bf 308}:\penalty0 403, 2007.

\bibitem[{Aguilera} et~al.(2008){Aguilera}, {Pons}, and
  {Miralles}]{Aguilera:2008qf}
D.~N. {Aguilera}, J.~A. {Pons}, and J.~A. {Miralles}.
\newblock {2D Cooling of magnetized neutron stars}.
\newblock \emph{\aap} {\bf 486}:\penalty0 255, 2008.

\bibitem[{Kaplan} et~al.(2004){Kaplan}, {Frail}, {Gaensler}, {Gotthelf},
  {Kulkarni}, {Slane}, and {Nechita}]{Kaplan:2004nx}
D.~L. {Kaplan}, D.~A. {Frail}, B.~M. {Gaensler}, E.~V. {Gotthelf}, S.~R.
  {Kulkarni}, P.~O. {Slane}, and A.~{Nechita}.
\newblock {An X-Ray Search for Compact Central Sources in Supernova Remnants.
  I. SNRS G093.3+6.9, G315.4-2.3, G084.2+0.8, and G127.1+0.5}.
\newblock \emph{\apjs} {\bf 153}:\penalty0 269, 2004.

\bibitem[{Chang} and {Bildsten}(2004)]{Chang:2004bh}
P.~{Chang} and L.~{Bildsten}.
\newblock {Evolution of Young Neutron Star Envelopes}.
\newblock \emph{\apj} {\bf 605}:\penalty0 830, 2004.

\bibitem[{Page} et~al.(2000){Page}, {Prakash}, {Lattimer}, and
  {Steiner}]{Page:2000kl}
D.~{Page}, M.~{Prakash}, J.~M. {Lattimer}, and A.~W. {Steiner}.
\newblock {Prospects of Detecting Baryon and Quark Superfluidity from Cooling
  Neutron Stars}.
\newblock \emph{\prl} {\bf 85}:\penalty0 2048, 2000.


\bibitem[{Blaschke} et~al.(2000){Blaschke}, {Kl\"ahn}, and {Voskresensky}]{Blaschke:2000kl}
D.~{Blaschke}, T.~{Kl\"ahn}, and D.~N.~{Voskresensky}.
\newblock {Diquark Condensates and Compact Star Cooling}.
\newblock \emph{\apj} {\bf 533}:\penalty0 406, 2000.

\bibitem[{Grigorian} et~al.(2005){Grigorian}, {Blaschke}, and {Voskresensky}]{Grigorian:2005kl}
H.~{Grigorian}, D.~{Blaschke}, and D.~N.~{Voskresensky}.
\newblock {Cooling of neutron stars with color superconducting quark cores}.
\newblock \emph{\prc} {\bf 71\penalty0 (4)}:\penalty0 045801, 2005.

\bibitem[{Hess} et~al.(2011){Hess}, and {Sedrakian}]{Hess:2011kl}
D.~{Hess}, and A.~{Sedrakian}.
\newblock {Thermal evolution of massive compact objects with dense quark cores}.
\newblock \emph{\prd} {\bf 84\penalty0 (6)}:\penalty0 063015, 2011.


\bibitem[{Negreiros} et~al.(2012){Negreiros}, {Dexheimer}, and {Schramm}]{Negreiros:2012kl}
R.~{Negreiros}, V.~A.~{Dexheimer}, and S.~{Schramm}.
\newblock {Quark core impact on hybrid star cooling}.
\newblock \emph{\prc} {\bf 85\penalty0 (3)}:\penalty0 035805, 2012.




\bibitem[{Gusakov} et~al.(2004){Gusakov}, {Kaminker}, {Yakovlev}, and
  {Gnedin}]{Gusakov:2004ys}
M.~E. {Gusakov}, A.~D. {Kaminker}, D.~G. {Yakovlev}, and O.~Y. {Gnedin}.
\newblock {Enhanced cooling of neutron stars via Cooper-pairing neutrino
  emission}.
\newblock \emph{\aap} {\bf 423}:\penalty0 1063, 2004.

\bibitem[{Kaminker} et~al.(2006){Kaminker}, {Gusakov}, {Yakovlev}, and
  {Gnedin}]{Kaminker:2006hc}
A.~D. {Kaminker}, M.~E. {Gusakov}, D.~G. {Yakovlev}, and O.~Y. {Gnedin}.
\newblock {Minimal models of cooling neutron stars with accreted envelopes}.
\newblock \emph{\mnras} {\bf 365}:\penalty0 1300, 2006.

\bibitem[{Heinke} et~al.(2009){Heinke}, {Jonker}, {Wijnands}, {Deloye}, and
  {Taam}]{Heinke:2009dq}
C.~O. {Heinke}, P.~G. {Jonker}, R.~{Wijnands}, C.~J. {Deloye}, and R.~E.  {Taam}.
\newblock {Further Constraints on Thermal Quiescent X-Ray Emission from SAX J1808.4-3658}.
\newblock \emph{\apj} {\bf 691}:\penalty0 1035, 2009.

\bibitem[{Jonker} et~al.(2007){Jonker}, {Steeghs}, {Chakrabarty}, and
  {Juett}]{Jonker:2007cr}
P.~G. {Jonker}, D.~{Steeghs}, D.~{Chakrabarty}, and A.~M. {Juett}.
\newblock {The Cold Neutron Star in the Soft X-Ray Transient 1H 1905+000}.
\newblock \emph{\apjl} {\bf 665}:\penalty0 L147, 2007.

\bibitem[{Lattimer}(2012)]{Lattimer:2012}
J.~M. {Lattimer}.
\newblock The nuclear equation of state and neutron star masses.
\newblock \emph{Annu. Rev. Nucl. Part. Phys.} {\bf 62}:\penalty0 485, 2012.

\bibitem[{Fryer} and {Kalogera}(2001)]{Fryer:2001uq}
C.~L. {Fryer} and V.~{Kalogera}.
\newblock {Theoretical Black Hole Mass Distributions}.
\newblock \emph{\apj} {\bf 554}:\penalty0 548, 2001.

\bibitem[{Ashworth}(1980)]{Ashworth:1980vn}
W.~B. {Ashworth}, Jr.
\newblock {A Probable Flamsteed Observations of the Cassiopeia-A Supernova}.
\newblock \emph{J. Hist. Astron.} {\bf 11}:\penalty0 1, 1980.

\bibitem[{Reed} et~al.(1995){Reed}, {Hester}, {Fabian}, and
  {Winkler}]{Reed:1995kl}
J.~E. {Reed}, J.~J. {Hester}, A.~C. {Fabian}, and P.~F. {Winkler}.
\newblock {The Three-dimensional Structure of the Cassiopeia A Supernova
  Remnant. I. The Spherical Shell}.
\newblock \emph{\apj} {\bf 440}:\penalty0 706, 1995.

\bibitem[{Fesen} et~al.(2006){Fesen}, {Hammell}, {Morse}, {Chevalier},
  {Borkowski}, {Dopita}, {Gerardy}, {Lawrence}, {Raymond}, and {van den
  Bergh}]{Fesen:2006tg}
R.~A. {Fesen}, M.~C. {Hammell}, J.~{Morse}, R.~A. {Chevalier}, K.~J.
  {Borkowski}, M.~A. {Dopita}, C.~L. {Gerardy}, S.~S. {Lawrence}, J.~C.
  {Raymond}, and S.~{van den Bergh}.
\newblock {The Expansion Asymmetry and Age of the Cassiopeia A Supernova
  Remnant}.
\newblock \emph{\apj} {\bf 645}:\penalty0 283, 2006.

\bibitem[{Krause} et~al.(2008){Krause}, {Birkmann}, {Usuda}, {Hattori}, {Goto},
  {Rieke}, and {Misselt}]{Krause:2008hc}
O.~{Krause}, S.~M. {Birkmann}, T.~{Usuda}, T.~{Hattori}, M.~{Goto}, G.~H.
  {Rieke}, and K.~A. {Misselt}.
\newblock {The Cassiopeia A Supernova Was of Type IIb}.
\newblock \emph{Science} {\bf 320}:\penalty0 1195, 2008.

\bibitem[{Willingale} et~al.(2003){Willingale}, {Bleeker}, {van der Heyden},
  and {Kaastra}]{Willingale:2003ys}
R.~{Willingale}, J.~A.~M. {Bleeker}, K.~J. {van der Heyden}, and J.~S.
  {Kaastra}.
\newblock {The mass and energy budget of Cassiopeia A}.
\newblock \emph{\aap} {\bf 398}:\penalty0 1021, 2003.

\bibitem[{Chevalier} and {Oishi}(2003)]{Chevalier:2003ij}
R.~A. {Chevalier} and J.~{Oishi}.
\newblock {Cassiopeia A and Its Clumpy Presupernova Wind}.
\newblock \emph{\apjl} {\bf 593}:\penalty0 L23, 2003.

\bibitem[{van Veelen} et~al.(2009){van Veelen}, {Langer}, {Vink},
  {Garc{\'{\i}}a-Segura}, and {van Marle}]{van-Veelen:2009fv}
B.~{van Veelen}, N.~{Langer}, J.~{Vink}, G.~{Garc{\'{\i}}a-Segura}, and A.~J.
  {van Marle}.
\newblock {The hydrodynamics of the supernova remnant Cassiopeia A. The
  influence of the progenitor evolution on the velocity structure and
  clumping}.
\newblock \emph{\aap} {\bf 503}:\penalty0 495, 2009.

\bibitem[{Young} et~al.(2006){Young}, {Fryer}, {Hungerford}, {Arnett},
  {Rockefeller}, {Timmes}, {Voit}, {Meakin}, and {Eriksen}]{Young:2006bs}
P.~A. {Young}, C.~L. {Fryer}, A.~{Hungerford}, D.~{Arnett}, G.~{Rockefeller},
  F.~X. {Timmes}, B.~{Voit}, C.~{Meakin}, and K.~A. {Eriksen}.
\newblock {Constraints on the Progenitor of Cassiopeia A}.
\newblock \emph{\apj} {\bf 640}:\penalty0 891, 2006.

\bibitem[{Pavlov} et~al.(2004){Pavlov}, {Sanwal}, and {Teter}]{Pavlov:2004kl}
G.~G. {Pavlov}, D.~{Sanwal}, and M.~A. {Teter}.
\newblock {Central Compact Objects in Supernova Remnants}.
\newblock In \emph{Young Neutron Stars and Their Environments},
{F.~Camilo \& B.~M.~Gaensler}, editors.
 \emph{IAU Symposium}, vol. 218 page 239, 2004.
[arXiv:astro-ph/0311526].

\bibitem[{Ho} and {Heinke}(2009)]{Ho:2009fk}
W.~C.~G. {Ho} and C.~O. {Heinke}.
\newblock {A neutron star with a carbon atmosphere in the Cassiopeia A
  supernova remnant}.
\newblock \emph{\nat} {\bf 462}:\penalty0 71, 2009.

\bibitem[{Heinke} and {Ho}(2010)]{Heinke:2010hc}
C.~O. {Heinke} and W.~C.~G. {Ho}.
\newblock {Direct Observation of the Cooling of the Cassiopeia A Neutron Star}.
\newblock \emph{\apjl} {\bf 719}:\penalty0 L167, 2010.

\bibitem[{Shternin} et~al.(2011){Shternin}, {Yakovlev}, {Heinke}, {Ho}, and
  {Patnaude}]{Shternin:2011fu}
P.~S. {Shternin}, D.~G. {Yakovlev}, C.~O. {Heinke}, W.~C.~G. {Ho}, and D.~J.
  {Patnaude}.
\newblock {Cooling neutron star in the Cassiopeia A supernova remnant: evidence
  for superfluidity in the core}.
\newblock \emph{\mnras} {\bf 412}:\penalty0 L108, 2011.

\bibitem[{Yakovlev} et~al.(2011){Yakovlev}, {Ho}, {Shternin}, {Heinke}, and
  {Potekhin}]{Yakovlev:2011kl}
D.~G. {Yakovlev}, W.~C.~G. {Ho}, P.~S. {Shternin}, C.~O. {Heinke}, and A.~Y.
  {Potekhin}.
\newblock {Cooling rates of neutron stars and the young neutron star in the
  Cassiopeia A supernova remnant}.
\newblock \emph{\mnras} {\bf 411}:\penalty0 1977, 2011.

\bibitem[{Shternin} et~al.(2007){Shternin}, {Yakovlev}, {Haensel}, and
  {Potekhin}]{Shternin:2007fk}
P.~S. {Shternin}, D.~G. {Yakovlev}, P.~{Haensel}, and A.~Y. {Potekhin}.
\newblock {Neutron star cooling after deep crustal heating in the X-ray
  transient KS 1731-260}.
\newblock \emph{\mnras} {\bf 382}:\penalty0 L43, 2007.

\bibitem[{Brown} and {Cumming}(2009)]{Brown:2009uq}
E.~F. {Brown} and A.~{Cumming}.
\newblock {Mapping Crustal Heating with the Cooling Light Curves of
  Quasi-Persistent Transients}.
\newblock \emph{\apj} {\bf 698}:\penalty0 1020, 2009.

\bibitem[{Horowitz} et~al.(2009){Horowitz}, {Caballero}, and
  {Berry}]{Horowitz:2009kx}
C.~J. {Horowitz}, O.~L. {Caballero}, and D.~K. {Berry}.
\newblock {Thermal conductivity and phase separation of the crust of accreting
  neutron stars}.
\newblock \emph{\pre} {\bf 79\penalty0 (2)}:\penalty0 026103, 2009.

\bibitem[{Daligault} and {Gupta}(2009)]{Daligault:2009vn}
J.~{Daligault} and S.~{Gupta}.
\newblock {Electron-Ion Scattering in Dense Multi-Component Plasmas:
  Application to the Outer Crust of an Accreting Neutron Star}.
\newblock \emph{\apj} {\bf 703}:\penalty0 994, 2009.

\bibitem[{Blaschke} et~al.(2012){Blaschke}, {Grigorian}, {Voskresensky}, and
  {Weber}]{Blaschke:2012ys}
D.~{Blaschke}, H.~{Grigorian}, D.~N. {Voskresensky}, and F.~{Weber}.
\newblock {Cooling of the neutron star in Cassiopeia A}.
\newblock \emph{\prc} {\bf 85\penalty0 (2)}:\penalty0 022802, 2012.

\bibitem[{Elshamouty}(2013)]{Elshamouty:2013}
K.~G. {Elshamouty}, C.~O. {Heinke}, G.~R. {Sivakoff}, W.~C.~G. {Ho},
P.~S. {Shternin}, D.~G. {Yakovlev}, D.~J. {Patnaude}, and 
L. {David}.
\newblock {Measuring the Cooling of the Neutron Star in Cassiopeia A
with all Chandra X-ray Observatory Detectors}.
\newblock \emph{\apj} {\bf 777}:\penalty0 22, 2013.

\bibitem[{Posselt}(2013)]{Posselt:2013}
B. {Posselt}, G.~G. {Pavlov}, V. {Suleimanov}, and O. {Kargaltsev}.
\newblock {New Constraints on the Cooling of the Central Compact
  Object in Cas A}.
\newblock \emph{\apj} {\bf 779}:\penalty0 186, 2013.

\bibitem[{Pacini}(1968)]{Pacini:1968ly}
F.~{Pacini}.
\newblock {Rotating Neutron Stars, Pulsars and Supernova Remnants}.
\newblock \emph{\nat} {\bf 219}:\penalty0 145, 1968.

\bibitem[{Gunn} and {Ostriker}(1969)]{Gunn:1969ve}
J.~E. {Gunn} and J.~P. {Ostriker}.
\newblock {Magnetic Dipole Radiation from Pulsars}.
\newblock \emph{\nat} {\bf 221}:\penalty0 454, 1969.

\bibitem[{Goldreich} and {Julian}(1969)]{Goldreich:1969qf}
P.~{Goldreich} and W.~H. {Julian}.
\newblock {Pulsar Electrodynamics}.
\newblock \emph{\apj} {\bf 157}:\penalty0 869, 1969.

\bibitem[{Contopoulos} et~al.(1999){Contopoulos}, {Kazanas}, and
  {Fendt}]{Contopoulos:1999cr}
I.~{Contopoulos}, D.~{Kazanas}, and C.~{Fendt}.
\newblock {The Axisymmetric Pulsar Magnetosphere}.
\newblock \emph{\apj} {\bf 511}:\penalty0 351, 1999.

\bibitem[{Spitkovsky}(2008)]{Spitkovsky:2008dq}
A.~{Spitkovsky}.
\newblock {Pulsar Magnetosphere: The Incredible Machine}.
\newblock In \emph{40 Years of Pulsars: Millisecond Pulsars, Magnetars and More},
C.~{Bassa}, Z.~{Wang}, A.~{Cumming}, and V.~M. {Kaspi}, editors.
 \emph{American Institute of Physics Conference Series}, vol. 983 page~20, 2008.

\bibitem[{Manchester} et~al.(2005){Manchester}, {Hobbs}, {Teoh}, and
  {Hobbs}]{Manchester:2005jl}
R.~N. {Manchester}, G.~B. {Hobbs}, A.~{Teoh}, and M.~{Hobbs}.
\newblock {The Australia Telescope National Facility Pulsar Catalogue}.
\newblock \emph{\aj} {\bf 129}:\penalty0 1993, 2005.

\bibitem[{Espinoza}(2012)]{Espinoza:2012dq}
C.~M. {Espinoza}.
\newblock {The spin evolution of young pulsars}.
In \emph{Neutron Stars and Pulsars: Challenges and Opportunities after 80 years},
J. van Leeuwen, editor.
IAUS Proceedings, vol. 291, 2012.
[arXiv:1211.5276].

\bibitem[{Muslimov} and {Page}(1995)]{Muslimov:1995dz}
A.~{Muslimov} and D.~{Page}.
\newblock {Delayed switch-on of pulsars}.
\newblock \emph{\apjl} {\bf 440}:\penalty0 L77, 1995.

\bibitem[{Ho} and {Andersson}(2012)]{Ho:2012tg}
W.~C.~G. {Ho} and N.~{Andersson}.
\newblock {Rotational evolution of young pulsars due to superfluid decoupling}.
\newblock \emph{Nature Physics} {\bf 8}:\penalty0 787, 2012.

\bibitem[{Faucher-Gigu{\`e}re} and {Kaspi}(2006)]{Faucher-Giguere:2006hc}
C.-A. {Faucher-Gigu{\`e}re} and V.~M. {Kaspi}.
\newblock {Birth and Evolution of Isolated Radio Pulsars}.
\newblock \emph{\apj} {\bf 643}:\penalty0 332, 2006.

\bibitem[{Geppert} et~al.(1999){Geppert}, {Page}, and
  {Zannias}]{Geppert:1999ij}
U.~{Geppert}, D.~{Page}, and T.~{Zannias}.
\newblock {Submergence and re-diffusion of the neutron star magnetic field
  after the supernova}.
\newblock \emph{\aap} {\bf 345}:\penalty0 847, 1999.

\bibitem[{Bernal} et~al.(2010){Bernal}, {Lee}, and {Page}]{Bernal:2010bs}
C.~G. {Bernal}, W.~H. {Lee}, and D.~{Page}.
\newblock {Hypercritical accretion onto a magnetized neutron star surface: a
  numerical approach}.
\newblock \emph{\rmxaa} {\bf 46}:\penalty0 309, 2010.

\bibitem[{Bernal} et~al.(2012){Bernal}, {Page}, and {Lee}]{Bernal:2012fv}
C.~G. {Bernal}, D.~{Page}, and W.~H. {Lee}.
\newblock {Hypercritical Accretion onto a Newborn Neutron Star and Magnetic
  Field Submergence}.
\newblock arXiv:1212.0464, 2012.

\bibitem[{Muslimov} and {Page}(1996)]{Muslimov:1996fu}
A.~{Muslimov} and D.~{Page}.
\newblock {Magnetic and Spin History of Very Young Pulsars}.
\newblock \emph{\apj} {\bf 458}:\penalty0 347, 1996.

\bibitem[{Blandford} and {Romani}(1988)]{Blandford:1988qa}
R.~D. {Blandford} and R.~W. {Romani}.
\newblock {On the interpretation of pulsar braking indices}.
\newblock \emph{\mnras} {\bf 234}:\penalty0 57P, 1988.

\bibitem[{Melatos}(1997)]{Melatos:1997kl}
A.~{Melatos}.
\newblock {Spin-down of an oblique rotator with a current-starved outer
  magnetosphere}.
\newblock \emph{\mnras} {\bf 288}:\penalty0 1049, 1997.

\bibitem[{Andersson} et~al.(2012){Andersson}, {Glampedakis}, {Ho}, and
  {Espinoza}]{Andersson:2012tw}
N.~{Andersson}, K.~{Glampedakis}, W.~C.~G. {Ho}, and C.~M. {Espinoza}.
\newblock {Pulsar Glitches: The Crust is not Enough}.
\newblock \emph{\prl} {\bf 109\penalty0 (24)}:\penalty0 241103, 2012.

\bibitem[{Espinoza} et~al.(2011){Espinoza}, {Lyne}, {Stappers}, and
  {Kramer}]{Espinoza:2011oq}
C.~M. {Espinoza}, A.~G. {Lyne}, B.~W. {Stappers}, and M.~{Kramer}.
\newblock {A study of 315 glitches in the rotation of 102 pulsars}.
\newblock \emph{\mnras}, {\bf 414}:\penalty0 1679, 2011.

\bibitem[{Anderson} and {Itoh}(1975)]{Anderson:1975mi}
P.~W. {Anderson} and N.~{Itoh}.
\newblock {Pulsar glitches and restlessness as a hard superfluidity
  phenomenon}.
\newblock \emph{\nat} {\bf 256}:\penalty0 25, 1975.

\bibitem[{Ruderman}(1976)]{Ruderman:1976pi}
M.~{Ruderman}.
\newblock {Crust-breaking by neutron superfluids and the VELA pulsar glitches}.
\newblock \emph{\apj} {\bf 203}:\penalty0 213, 1976.

\bibitem[{Pines} and {Alpar}(1985)]{Pines:1985ff}
D.~{Pines} and M.~A. {Alpar}.
\newblock {Superfluidity in neutron stars}.
\newblock \emph{\nat} {\bf 316}:\penalty0 27, 1985.

\bibitem[{McCulloch} et~al.(1987){McCulloch}, {Klekociuk}, {Hamilton}, and
  {Royle}]{McCulloch:1987fu}
P.~M. {McCulloch}, A.~R. {Klekociuk}, P.~A. {Hamilton}, and G.~W.~R. {Royle}.
\newblock {Daily observations of three period jumps of the VELA pulsar}.
\newblock \emph{Australian J. of Phys.} {\bf 40}:\penalty0 725, 1987.

\bibitem[{Cordes} et~al.(1988){Cordes}, {Downs}, and
  {Krause-Polstorff}]{Cordes:1988lh}
J.~M. {Cordes}, G.~S. {Downs}, and J.~{Krause-Polstorff}.
\newblock {JPL pulsar timing observations. V - MACRO and microjumps in the VELA
  pulsar 0833-45}.
\newblock \emph{\apj} {\bf 330}:\penalty0 847, 1988.

\bibitem[{Link} et~al.(1999){Link}, {Epstein}, and {Lattimer}]{Link:1999ye}
B.~{Link}, R.~I. {Epstein}, and J.~M. {Lattimer}.
\newblock {Pulsar Constraints on Neutron Star Structure and Equation of State}.
\newblock \emph{\prl} {\bf 83}:\penalty0 3362, 1999.

\bibitem[{Chamel}(2012)]{Chamel:2012qo}
N.~{Chamel}.
\newblock {Crustal Entrainment and Pulsar Glitches}.
\newblock \emph{\prl} {\bf 110\penalty0 (1)}:\penalty0 011101, 2012.

\bibitem[{Link}(2013)]{Link:2013il}
B.~{Link}.
\newblock {Thermally-Activated Post-Glitch Response of the Neutron Star Inner Crust and Core}.
\newblock arXiv:1311.2499, 2013.

\bibitem[{Sidery} and {Alpar}(2009)]{Sidery:2009zt}
T.~{Sidery} and M.~A. {Alpar}.
\newblock {The effect of quantized magnetic flux lines on the dynamics of
  superfluid neutron star cores}.
\newblock \emph{\mnras} {\bf 400}:\penalty0 1859, 2009.

\end{thebibliography}


\end{document}